\documentclass[11pt]{JHEP3}
\usepackage{epsfig}

\def\beq{\begin{equation}}
\def\eeq{\end{equation}}
\def\beqa{\begin{eqnarray}}
\def\eeqa{\end{eqnarray}}

\def\PLB#1#2#3{Phys. Lett. B{#1} (19#2) #3}

\def\PRL#1#2#3{Phys. Rev. Lett. {#1} (19#2) #3}

\newcommand{\be}{\begin{equation}}
\newcommand{\ee}{\end{equation}}
\newcommand{\bea}{\begin{eqnarray}}
\newcommand{\eea}{\end{eqnarray}}

\DeclareMathSymbol{\mg}{\mathrel}{symbols}{"1D}

\newcommand{\gb}{\beta}
\newcommand{\gd}{\delta}
\renewcommand{\ge}{\epsilon}

\newcommand{\gl}{\lambda}

\newcommand{\gD}{\Delta}

\newcommand{\cF}{{\cal F}}

\newcommand{\cL}{{\cal L}}
\newcommand{\cM}{{\cal M}}

\newcommand{\cO}{{\cal O}}

\newcommand{\cR}{{\cal R}}

\newcounter{oldcounter}

\preprint{
DAMTP-2002-47\\
FTUAM-02/13,\\
IFT-UAM/CSIC-02-14\\
}
 
\title{
TeV-Scale Z' Bosons from  D-branes}
\author{D.M. Ghilencea$^a$, L.E. Ib\'a\~nez$^b$, 
N. Irges$^{c}$, F. Quevedo$^a$\\
{$^a$ DAMTP, CMS, University of Cambridge, } \\
{\it Wilberforce Road, Cambridge, CB3 0WA, U.K. \\}\\
{$^b$ Departamento de F\'{\i}sica Te\'orica C-XI and Instituto de
F\'{\i}sica Te\'orica C-XVI,} \\
{Universidad Aut\'onoma de Madrid, Cantoblanco, 28049 Madrid, Spain.\\}
\\
{\sl $^c$ Instituto de Estructura de la Materia (CSIC)}\\
\emph{\it Serrano 123, E-28006-Madrid, Spain.\\}}

\abstract{
Generic D-brane string models of particle physics predict the existence of 
extra $U(1)$ gauge symmetries beyond  hypercharge. These
symmetries are not of the $E_6$ class but rather  
include the gauging of Baryon and Lepton numbers as well as
certain Peccei-Quinn-like symmetries. Some of the $U(1)$'s have
triangle anomalies, but they are cancelled by a Green-Schwarz mechanism. 
The corresponding gauge bosons typically
acquire a mass of order  the string scale $M_S$  by  combining
with two-index antisymmetric fields coming from the closed string sector
of the theory. We argue that in string models with a low
string scale $M_S\propto 1-10 $ TeV, the presence of these generic 
$U(1)$'s may be amenable to experimental test.
Present constraints from electroweak precision data already set
important bounds on the mass of these extra gauge bosons. In particular,
for large classes of models,  $\rho$-parameter constraints 
imply $M_S\geq 1.5$ TeV. In the present scheme  some fraction of the 
experimentally measured $Z^0$ mass would  be due not to the Higgs
mechanism, but rather to the mixing with these closed string fields.
We give explicit formulae for  recently constructed 
classes  of intersecting  D6- and D5-brane models  yielding the 
Standard Model (SM)  fermion spectrum.}

\keywords{D-branes, Compactification and String models, Large extra
 dimensions, Beyond Standard Model.}

\begin{document}

\section{Introduction}
One of the aims of string phenomenology is to extract
generic features of realistic string models
 which can be considered
as `predictions' that  could
 eventually be tested
experimentally. Finding one generic property of string models
 is particularly relevant, given the  large degeneracy of string vacua.

At present there are two general classes of chiral  D-brane models 
with realistic properties, corresponding to D-branes at singularities
\cite{bbarmod,aiqu,cuw} and
D-branes intersecting at non-trivial angles
\cite{bgkl,afiru,afiru2,bkl,imr,csu,bklo,pheno,cim1,cim2,cim3}. Both
classes
of models share very
encouraging phenomenological prospects: in both cases it is possible
to construct three-family models resembling very much the structure
of the Standard
Model (SM). The gauge symmetry always arises from a  product of $U(N)$  
groups and the
matter fields transform generically in bi-fundamental representations
of those groups,
giving rise to a rich  structure that can be encoded
in terms of quiver diagrams \cite{cim1,dm}.
Another interesting property of explicit D-brane models is that
the proton  appears to be generically stable.
There are also other characteristics that are
more model dependent:
the models may be non-supersymmetric
\cite{bbarmod,aiqu,bgkl,afiru,afiru2,bkl,imr,csu,bklo,pheno,cim1,cim2,cim3},
 supersymmetric
\cite{aiqu,cuw,csu} or ``quasi''-supersymmetric \cite{cim1,cim2},
the string scale may be as high as the Planck scale and as
low as 1TeV and  gauge and Yukawa couplings may be found to take 
realistic values.

It is then desirable to concentrate on
one generic feature of this class of models. In this paper we will
study in detail the general implications of having extra
$U(1)$ symmetries beyond the Standard Model hypercharge, which is
generic in D-brane models \cite{joe}. 
 The reason for the
appearance of the extra $U(1)$'s is the following.
 The basic structure of D-branes implies the
existence of a $U(1)$ for each D-brane. Having $N$ overlapping
D-branes implies that open string states ending on different D-branes 
become massless when the separation vanishes and the $U(1)^N$ symmetry
is enhanced to $U(N)$.
 Therefore,
whenever we look for a chiral model including the symmetries of the
Standard Model of particle physics, we cannot  obtain just the
symmetries $SU(3)\times SU(2)\times U(1)$ but instead  we will have at least
$U(3)\times U(2)\times U(1)$. 
This  implies that there will  necessarily be additional $U(1)$
symmetries beyond the Standard Model and
also that these extra $U(1)$'s are precisely  defined to be
the ones that complete the $SU(N)$ groups to  $U(N)$.
If the string scale is sufficiently 
low, the associated gauge bosons may be relatively light  and precision
tests of the Standard Model can already set a bound on their masses
which in turn may imply  a bound on the string scale. 
In this context notice that 
signatures of additional $Z'$ bosons can be detected easier than almost any other
 possible signature beyond the Standard
Model \cite{pdg,bkm,hr,cl,leike}.

Most of the extra $U(1)$'s in D-brane models are usually massive and the
mechanism through   which they acquire a mass is well understood \cite{iru,p} 
(for related discussions see  \cite{abd,iq,mk,akt,ka,akr}).
This is due to the existence  of  interaction terms of the form $B\wedge F$,
where $B$ stands for the two-index antisymmetric tensors present in the 
closed string sector of the theory. These terms are crucial for the 
existence of the $D=4$ Green-Schwarz mechanism for the cancellation 
of  $U(1)$ anomalies. 
Upon dualization, these  terms convert into  
Stuckelberg   mass terms for the corresponding gauge
field. Interestingly enough, these  couplings  have
 been found \cite{imr} not only
for `anomalous' $U(1)$'s but also for non-anomalous ones. The general situation
is that only $U(1)$ of hypercharge survives as a massless gauge boson and all
the other extra Abelian gauge bosons
acquire a mass of the order of the string scale. If this scale is close
to 1 TeV then it may be easy to find signatures of  those fields.
In particular, in the present scheme a fraction of the mass of the
$Z^0$ boson would be due not to the standard Higgs mechanism, 
but to the mixing with closed string antisymmetric tensor fields $B$.

The study of explicit models shows that the additional $U(1)$'s may correspond
to physically relevant symmetries: baryon number, lepton number and a
PQ-like  symmetry. The interesting aspect is that these symmetries
may still survive at low energies as exact global symmetries. The
reason for this is the way  those fields acquire a mass, which  is 
not the standard
Higgs mechanism,  since it is not necessarily accompanied by a non 
vanishing vev for
any  scalar field. Thus there is no longer a local invariance 
(there is no massless $U(1)$ boson) but the global symmetry 
remains. The survival of the baryon number as an effective
global symmetry naturally guarantees the proton stability, which
is a serious problem in models with a low string scale.

In this context,  a class of  explicit Type IIA orientifold 
models was recently constructed yielding just the fermions of the SM
at the intersections of D6-branes wrapping a 6-torus \cite{cim1, cim2}.
These brane configurations provide a natural explanation for
properties of the SM like family replication, due to the 
fact that branes wrapped on a compact space typically intersect 
 more than once. More recently,  these constructions have been
extended to the case of intersecting D5-branes \cite{cim3}. These
brane intersection models provide  us for the first time
with explicit realistic  models in which the structure 
of $U(1)$ gauge bosons can be analyzed in detail. In these
models  one can write explicit results for the couplings
of the RR-scalars to the $U(1)$ gauge bosons which
eventually give Stuckelberg masses to the extra $U(1)$'s.
This is the reason why we will concentrate in the numerical analysis
on the specific case of D6- and D5-brane intersection
models.

The paper is organised as follows. In the next chapter  we start 
with  a general  discussion of the mechanism by which two-index 
antisymmetric fields can provide  explicit (Stuckelberg) masses to 
Abelian fields, without the presence of any Higgs mechanism.
Then we review the D-brane intersection scenario in which 
quarks and leptons appear at intersections  of D-branes. 
We show how in this scenario there are in 
general three extra $U(1)$'s, of which  two have  triangle 
anomalies cancelled by a Green-Schwarz mechanism.
We discuss in turn the case of D6- and  D5-brane models and 
provide explicit results for the  couplings of the different  
antisymmetric B-fields to the relevant Abelian fields.

In section \ref{generalmethod} we outline the general approach
we will follow in  the study of the mass matrix of Abelian gauge  
bosons before the electroweak symmetry breaking. This is done for 
some  families  of D6-brane models as well as  D5-brane 
models. We then describe the method to be used for including
the effects of the electroweak symmetry breaking on the $U(1)$ masses.

In Section \ref{specificmodels}  we   detail our analysis 
of  the masses of  $U(1)$ fields before electroweak symmetry breaking
following the strategy of Section \ref{generalmethod}. We
provide explicit results for the masses of $U(1)$ fields as well as the 
constraints  induced on the value of the string  scale $M_S$
for D6 and D5-brane models.
In Section \ref{EWB}  the effects of the electroweak symmetry breaking
are analysed in detail. The neutral gauge bosons acquire now a mass 
from a combination of two sources: mixing with antisymmetric B-fields 
and the standard Higgs mechanism induced by the vevs of electroweak  doublets. 
We show how the presence of the stringy source of mass
for the Abelian gauge bosons has an impact on the $\rho $
parameter. Present $\rho $-parameter constraints imply
sizable bounds on the masses of the extra $Z'$ bosons 
(and hence on the string scale $M_S$). We analyze 
in detail some D6-brane and D5-brane classes of models
and give corresponding bounds.
In Section \ref{conclusions}  some general comments 
and conclusions are presented. The Appendix contains  some
additional formulae referred to in the main text.

\section{Extra U(1)'s  in D-brane Standard Model-like models }\label{extrau1}

In this section  we address the structure 
of gauged Abelian symmetries beyond hypercharge in 
D-brane settings  yielding the SM at low energies.
As mentioned, explicit string D-brane models yielding three generations of 
quarks and leptons have been constructed in the last few
years.  In order to obtain chirality it has been considered the
location of the SM stacks of branes to be at some  (e.g., $Z_N$ 
orbifold) singularity  or/and  settings involving intersecting
branes.  We will concentrate here on models based  on intersecting 
D6- or D5-branes  in which it has been recently shown that models 
with the massless fermion  spectrum of the SM are easy to obtain 
\cite{imr,cim3}, although we think that  much of our results 
are  generalisable to other classes  of D-brane models.

Before proceeding to the description of the intersecting  D-brane 
models  let us review  in more  detail how the $U(1)$ gauge bosons 
acquire a mass from  the presence of $B\wedge F$ couplings, which is a
generic mechanism in string theory models.
We would like to emphasize that there is an important  difference
from  the standard Higgs mechanism in the sense that there is
not necessarily a remnant massive scalar field to play the role of the
corresponding Higgs boson.

\subsection{Green-Schwarz terms and massive $U(1)$'s}

 To understand the basis of the mechanism 
giving masses to the $U(1)$'s 
let us consider the following
Lagrangian coupling an Abelian gauge field $A_\mu$ to an antisymmetric
tensor $B_{\mu\nu}$:

\begin{equation}
\label{dualuno} 
{\cal L}\ =\ -\frac{1}{12} H^{\mu\nu\rho}
H_{\mu\nu\rho}-\frac{1}{4g^2} F^{\mu\nu} F_{\mu\nu} 
+ \frac{c}{4}\ \epsilon^{\mu\nu\rho\sigma} B_{\mu\nu}\ F_{\rho\sigma},
\end{equation}
where 
\beq
\label{definicion}
H_{\mu\nu\rho}=\partial_\mu B_{\nu\rho}+\partial_\rho B_{\mu\nu}
+\partial_\nu B_{\rho\mu}, \qquad  F_{\mu\nu}=\partial_\mu
A_\nu-\partial_\nu A_\mu
\ee
and $g, c$
are arbitrary  constants. This corresponds to
the kinetic term for the fields $B_{\mu\nu}$ and $A_{\mu}$ together
with the $B\wedge F$ term. We will now proceed to dualize this
Lagrangian in two equivalent ways. First we can re-write it in terms of
the (arbitrary) field $H_{\mu\nu\rho}$ imposing the constraint $H=dB$ by the
standard introduction of a Lagrange multiplier field $\eta$ in the
following way:
\begin{equation}
\label{dualdos}
{\cal L}_0=\ -\frac{1}{12} H^{\mu\nu\rho}\
H_{\mu\nu\rho}-\frac{1}{4g^2} F^{\mu\nu}\ F_{\mu\nu} 
- \frac{c}{6}\ \epsilon^{\mu\nu\rho\sigma} H_{\mu\nu\rho}\ A_{\sigma} 
-\frac{c}{6}\eta \epsilon^{\mu\nu\rho\sigma} \partial_\mu H_{\nu\rho\sigma}.
\end{equation}
 Notice that integrating out $\eta$ implies $d^*H=0$ which in turn
implies that (locally) $H=dB$ and then we recover (\ref{dualuno}).  
Alternatively, integrating by parts the last term in (\ref{dualdos}) we
are left with a quadratic action for $H$ which we can solve
immediately to find
\begin{equation}
H^{\mu\nu\rho}= - {c}\ \epsilon^{\mu\nu\rho\sigma}
\left(A_\sigma+\partial_\sigma \eta\right).
\end{equation}
Inserting  this back into (\ref{dualdos}) we find:
\begin{equation}
{\cal L}_{A}\ =\ -\frac{1}{4g^2}\ F^{\mu\nu}\ F_{\mu\nu} -
\frac{c^2}{2} \left(A_\sigma+\partial_\sigma \eta\right)^2
\end{equation}
which is just a mass term for the gauge field $A_\mu$ after ``eating'' 
the scalar $\eta$ to acquire a mass $m^2=g^2 c^2$.
 Notice that this is similar to the St\"uckelberg mechanism
where we do not need a scalar field with a  vacuum expectation value
to give a mass to the gauge boson, nor do we have a massive Higgs-like
field at the end.

Furthermore, we can understand this mechanism in a dual way in which it
is not the gauge field that ``eats'' a scalar, but  the antisymmetric
tensor that ``eats'' the gauge field to gain a mass.
 This can be seen as follows.
Start now with the first order Lagrangian:
\begin{equation}
{\cal L'}_0\ =\ -\frac{1}{12} H^{\mu\nu\rho}\
H_{\mu\nu\rho}-\frac{1}{4g^2} F^{\mu\nu}F_{\mu\nu} 
+ \frac{c}{4}\ \epsilon^{\mu\nu\rho\sigma} B_{\mu\nu}\ F_{\rho\sigma}
+ \frac{c}{4}\ Z_\mu\ \epsilon^{\mu\nu\rho\sigma}\partial_\nu F_{\rho\sigma}.
\end{equation}
where  $F^{\mu\nu}$ is now an arbitrary tensor (not determined by 
(\ref{definicion}))
and $Z_\mu$  is a Lagrange multiplier enabling the condition
$F=dA$. Similar to the previous case, integrating over $Z_\mu$ gives
back the original Lagrangian, (\ref{dualuno}) but integrating by parts
the last term and solving the quadratic equation for $F_{\mu\nu}$ (now
an arbitrary field) gives the dual Lagrangian:
\begin{equation}
{\cal L}_B\ = -\frac{1}{12} H^{\mu\nu\rho}\
H_{\mu\nu\rho} - \frac{g^2 c^2}{4} \left(B_{\mu\nu}+\partial_\mu Z_\nu
\right)^2.
\end{equation}
We can see that this is the Lagrangian for a massive
two-index antisymmetric tensor, which gains a mass after ``eating'' the
vector $Z_\mu$ with the same mass as above $m^2=g^2 c^2$.
 This is completely equivalent to the massive vector, notice that
in four dimensions a massive vector and a {\it massive} two-index
tensor have the same number of degrees of freedom (3). For a general
discussion of massive antisymmetric tensor fields see \cite{qt}.

We would like to emphasize that this mechanism requires the presence  
of the Green-Schwarz term $B\wedge F$  but not necessarily the anomaly
cancellation term ($\eta F\wedge F$). Therefore as long as a $U(1)$
field has a Green-Schwarz coupling $B\wedge F$, it does not have to be
anomalous in order to acquire a mass. In most previous models found in
string theory it was generally the case that it was only the anomalous 
$U(1)$'s that had a Green-Schwarz term
and therefore those were the ones becoming massive. However, in 
recent models \cite{imr,cim1,cim2} there are non anomalous $U(1)$'s that
can also become massive. Those $U(1)$ bosons acquire a  ``topological mass''
induced by the Green-Schwarz term with no  associated massive scalar  field in the
spectrum. Note that this is totally different from 
what happens when a $U(1)$ becomes massive in the standard Higgs
mechanism. Indeed in the latter case in addition to the 
massive gauge boson there is always an explicit scalar field, the
Higgs field. Such a field is not present in the mechanism we discussed.

We finally note that  as emphasized in \cite{imr}, the gauge group is
broken to a global symmetry and therefore symmetries like baryon and
lepton number remain as perturbative global symmetries,
providing a simple explanation for  proton stability in models
with a low string scale.

\subsection{$U(1)$ structure  in intersecting brane SM-like models}
The general structure of intersecting D-brane models is summarized in 
Table \ref{branecontent} and in Fig.\ref{intersecting}. 
%
\TABLE{\renewcommand{\arraystretch}{1.7}
\begin{tabular}{|c|c|c|c|}
\hline
Label & Multiplicity & Gauge Group & Name \\
\hline
\hline
stack $a$ & $N_a = 3$ & $SU(3) \times U(1)_a$ & Baryonic brane\\
\hline
stack $b$ & $N_b = 2$ & $SU(2) \times U(1)_b$ & Left brane\\
\hline
stack $c$ & $N_c = 1$ & $U(1)_c$ & Right brane\\
\hline
stack $d$ & $N_d = 1$ & $U(1)_d$ & Leptonic brane \\
\hline
\end{tabular}
\label{branecontent}
\caption{\small Brane content yielding the SM spectrum.\label{SMbranes}}}
\FIGURE{\epsfxsize=4in
\hspace*{0in}\vspace*{.2in}
\epsffile{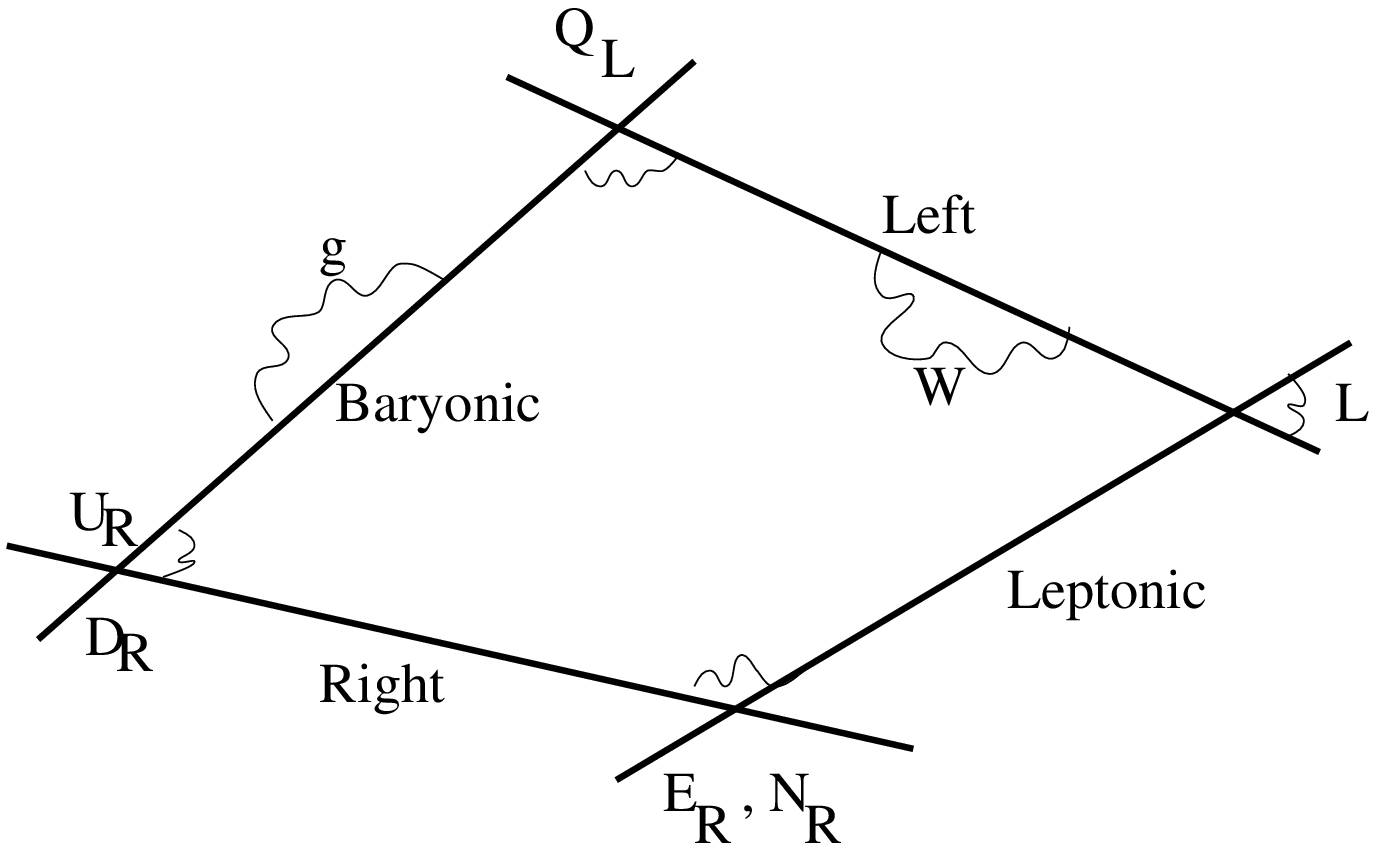}
\caption{\small  Schematic view of an intersecting D-brane standard
model construction. There are four stacks of branes:
{\it Baryonic, Left, Right } and {\it Leptonic}, giving
rise to a gauge group $U(3)_{Baryonic}\times U(2)_{Left}\times
U(1)_{Right}\times U(1)_{Leptonic}$.
Open strings starting and ending on the
same stack of branes give rise to the SM gauge bosons.
Quarks and leptons appear at the intersections of two different stacks
of branes.}
\label{intersecting}}
We have four stacks of branes : the {\it baryonic} stack
contains three parallel branes giving rise to the QCD interactions,
and the {\it left} stack contains two 
parallel branes yielding the electroweak   $SU(2)_L$ SM interactions.
In addition there is the {\it right} and {\it lepton} stacks 
containing each a single brane.
These four stacks of branes intersect in the compact six dimensions
(plus Minkowski) and at the intersections 
chiral fermions with the quantum numbers of the SM appear.
Thus, for example, the right-handed U-quarks occur at three 
different intersections of the {\it baryonic} stack with the {\it right}
stack (see Fig.\ref{intersecting}). 

Each stack of branes
comes along with a unitary gauge group so that
the initial gauge group is $SU(3)_{QCD}\times SU(2)_L
\times U(1)_a\times U(1)_b\times U(1)_c\times U(1)_d$.
A linear combination of these four $U(1)$'s may be 
identified with the standard hypercharge and at some 
level the rest of the $U(1)$'s should become massive. 
In the class of D6-brane models of ref.\cite{imr} 
and D5-brane models of ref.\cite{cim3} the charges 
of quarks and leptons with respect to these 
$U(1)$'s are shown in Table \ref{tabpssm}.
%
\TABLE{\renewcommand{\arraystretch}{1.25}
\begin{tabular}{|c|c|c|c|c|c|c|c|}
\hline Intersection &
 Matter fields  &   &  $q_a$  & $q_b $ & $q_c $ & $q_d$  &  $q_Y$ \\
\hline\hline (ab) & $Q_L$ &  $(3,2)$ & 1  & -1 & 0 & 0 & 1/6 \\
\hline (ab*) & $q_L$   &  $2( 3,2)$ &  1  & 1  & 0  & 0  & 1/6 \\
\hline (ac) & $U_R$   &  $3( {\bar 3},1)$ &  -1  & 0  & 1  & 0 & -2/3 \\
\hline (ac*) & $D_R$   &  $3( {\bar 3},1)$ &  -1  & 0  & -1  & 0 & 1/3 \\
\hline (bd*) & $ L$    &  $3(1,2)$ &  0   & -1   & 0  & -1 & -1/2  \\
\hline (cd) & $E_R$   &  $3(1,1)$ &  0  & 0  & -1  & 1  & 1   \\
\hline (cd*) & $N_R$   &  $3(1,1)$ &  0  & 0  & 1  & 1  & 0 \\
\hline \end{tabular}
 \caption{\small Standard model spectrum and $U(1)$ charges. The
hypercharge
generator is defined as $q_Y = \frac 16 q_a - \frac 12 q_c + \frac 12 q_d$.}
\label{tabpssm}}
%
In this table the asterisk denotes the ``orientifold mirror'' of each given brane,
which must always be present in this type of orientifold constructions
(see \cite{imr} for details).
Note that $U(1)_a$ and $U(1)_d$ can be identified with 
baryon number and (minus) lepton number respectively. Additionally
$U(1)_c$ can be identified with the third component 
of right-handed weak isospin. Finally, $U(1)_b$ is an axial symmetry
with QCD anomalies, very much like a PQ-symmetry.
It is easy to check from the above fermion spectrum that $U(1)_b$
and $3U(1)_a-U(1)_d$ linear combination 
  have triangle anomalies whereas $U(1)_a+3U(1)_d$
and $U(1)_c$ are both anomaly-free.  In fact the standard hypercharge 
may be written as a linear combination of these two symmetries:
\beq
q_Y = \frac 16 q_a - \frac 12 q_c + \frac 12 q_d  \ .
\label{hyperca}
\eeq
The above mentioned anomalies of two linear combinations of $U(1)$'s 
are cancelled by a generalized 4-dimensional Green-Schwarz 
mechanism which may be summarized as follows.
 In Type II string theory 
there are  some  closed string  `Ramond-Ramond'(RR)  modes
coupling to the gauge fields.
In particular in the D6- and D5-brane models here discussed 
there are four 
 RR two-form  fields  $B_i$ with couplings
to  the $U(1)_{\alpha }$ field strengths:
\beq
\sum_i c_i^{\alpha}  \ B_i \wedge  tr (F^{\alpha }),\quad  \ i=1,2,3,4;\quad 
\alpha = a,b,c,d \ 
\label{gsuno}
\eeq
and in addition
there are  couplings of the Poincar\'e  dual scalars 
(representing the same degrees of freedom) $\eta_i$ 
of the $B_i$ fields:
\beq
\sum_i d_i^{\beta } \eta_i tr(F^{\beta }\wedge F^{\beta }),
\label{gsdos}   
\eeq
 where $F^{\beta }$ are the field strengths of any of the gauge groups.
The combination of both couplings,
by tree-level exchange of the RR-fields,  
 cancels the
mixed $U(1)_\alpha $ anomalies  $A_{\alpha \beta }$ 
with any other group $G_\beta $
as: 
\beq
A_{\alpha \beta }\ +\ \sum_i c_i^{\alpha} d_i^{\beta } \ =\ 0   \ .  
\label{gstres}
\eeq
The coefficient  $c_i^{\alpha }$ and $d_i^{\alpha}$ may be
computed  explicitly for each given D-brane configuration 
and we will give later on their values for D6- and D5-brane models.
Note that for given   $\alpha , \beta $,
both   $c_i^{\alpha}$ {\it and }
$d_i^{\beta }$ have to be  non-vanishing for
some $i $, in order to cancel the anomalies. 

As remarked in the previous subsection,  
 the couplings in (\ref{gsuno}) give
masses to some linear 
combinations of $U(1)$'s.
Indeed,
after a duality transformation the $B\wedge F$ couplings turn into   
explicit mass terms for the Abelian gauge bosons given by
the expression:
\begin{equation} 
(M^2)_{\alpha\beta}= g_\alpha g_\beta M_S^2 \sum_{i=1}^{3} c_i^\alpha
c_i^\beta,\qquad \alpha,\beta=a,b,c,d.
\label{masones}
\end{equation}
where the sum runs over the massive RR-fields present in the models
and where $g_\alpha$ is the coupling of $U(1)_\alpha$.
Here we have normalized to unity the gauge boson kinetic functions. 

In principle there could be $4\times 4=16$ $\ $  $c_i^\beta$ coefficients.
However, it turns  out  that in the class of D6- and D5-brane models 
addressed here, there are only seven non-vanishing $c_i^{\beta}$
coefficients and only five of them
are independent . This is due to the particular structure
of the anomalies of this system of four $U(1)$'s. For both 
D6- and D5-brane models there are three RR fields $B_2^i$, $i=1,2,3$
which have the following structure of $B\wedge F^{\alpha} $
couplings \cite{imr,cim3}:
\beqa
B_2^1 &\wedge & \  c_1^b\, F^{b} \nonumber \\
B_2^2 &\wedge  &
 c_2^d\, \left(-3F^a\ +\  F^{d}\right)
\nonumber \\
 B_2^3 &\wedge  &  \
\left[ c_3^a F^a\ +\  c_3^b F^{b} \ +\   
\bigg({1\over 3}c_3^a+c_3^d\bigg) F^c \  +\
 c_3^d F^d \ \right]
\label{bfs}
\eeqa
whereas the fourth antisymmetric field $B_2^0$ does not couple to
any  of the $U(1)$'s. The universality of the 
form of the anomalous $U(1)$'s is the cause for the 
relationship   $c_2^a=-3c_2^d$. In addition
the equality $c_3^c=   ({1\over 3}c_3^a+c_3^d)$ is
required so  that the physical hypercharge
combination  eq.(\ref{hyperca}) does not mix 
with $B_2^3$ and remains massless. 
We thus see that the most general $U(1)$ mass matrix appearing
in this class of models depends  on only five
free parameters: $c_1^b, c_2^d, c_3^a, c_3^b$  and $c_3^d$.
As mentioned, these five coefficients may be computed 
for each model. We now provide their value for 
D6-brane and D5-brane models, respectively.

\subsection{D6-brane orientifold models}\label{d6branemodels}

 The structure of these models is discussed in
ref.\cite{imr}. One  considers
Type IIA  string theory compactified on a six-torus
$T^2\times T^2\times T^2$.
 Now we consider D6-branes containing inside Minkowski
space and wrapping
each of the three remaining dimensions of the
branes on a different torus $T^2$. We  denote by $(n_{\alpha
i},m_{\alpha i})$, i=1,2,3, $\alpha=a,b,c,d$ 
the wrapping numbers\footnote{This notation for $n_{\alpha i}$ is the
equivalent of $n_{\alpha}^i$ used in \cite{imr}.}
 of each brane $D6_\alpha $,
 $n_{\alpha i}$($m_{\alpha i}$) being the
number of times the brane is wrapping around the  x(y)-coordinate of the
$i-th$ torus. In this class of models the coefficients 
$c_i^\alpha $ are given by \cite{imr}:
\beq 
c_i^\alpha \ =\ N_\alpha  n_{\alpha j} \, n_{\alpha k} \, m_{\alpha i} 
\ ;\  i\not= j\not= k\not= i  \ , \ i=1,2,3
\label{cillos}
\eeq
where $N_\alpha$ is the number of parallel branes of type $\alpha $.
The general set of possible wrapping numbers yielding just the 
SM fermion spectrum was provided in ref.\cite{imr} and is 
shown in Table \ref{solution}.
%
\TABLE{\renewcommand{\arraystretch}{2.5}
\begin{tabular}{|c||c|c|c|}
\hline
 $N_\alpha$    &  $(n_{\alpha 1},m_{\alpha 1})$  &
 $(n_{\alpha 2},m_{\alpha 2})$   & $(n_{\alpha 3},m_{\alpha 3})$ \\
\hline\hline $N_a=3$ & $(1/\beta _1,0)$  &  $(n_{a 2},\epsilon \beta_2)$ &
 $(1/\nu ,  1/2)$  \\
\hline $N_b=2$ &   $(n_{b 1},-\epsilon \beta_1)$    &  $ (1/\beta_2,0)$  &
$(1,3\nu /2)$   \\
\hline $N_c=1$ & $(n_{c 1},3\nu \epsilon \beta_1)$  &
 $(1/\beta_2,0)$  & $(0,1)$  \\
\hline $N_d=1$ &   $(1/\beta_1,0)$    &  $(n_{d 2},-\beta_2\epsilon/\nu )$  &
$(1, 3\nu /2)$   \\
\hline \end{tabular}
 \caption{ D6-brane wrapping numbers giving rise to a SM
spectrum.
The general solutions
 are parametrized by a phase $\epsilon =\pm1$, the NS background
on the first two tori $\beta_i=1-b_i=1,1/2$, four integers
$n_{a 2},n_{b 1},n_{c 1},n_{d 2}$ and a parameter $\nu=1,1/3$.
\label{solution} }}
%
From Table \ref{solution} and eq.(\ref{cillos}) one can compute the
relevant  $c_i^\alpha $ coefficients:
\beqa
c_1^b\ &=& \ -{{2\epsilon \beta_1}\over { \beta_2} } \ ;\ 
c_2^d\  =  \ -{{\epsilon \beta_2}\over {\nu  \beta_1} } \nonumber   \\
c_3^a\ &=& \ {{3n_{a 2}}\over { 2  \beta_1} } \ ;\ 
c_3^b\ =  \ {{3\nu n_{b 1}}\over {  \beta_2} }\ ;\ 
c_3^c\  = \ { {n_{c 1}}\over {\beta_2} }\ ;\ 
c_3^d\  =  \ {{3\nu n_{d 2} }\over {2  \beta_1} }
\label{cillos6}
\eeqa
in terms of the free parameters $n_{a 2}$, $n_{d 2}$, $n_{b 1}$, $\epsilon$ 
and $\beta _i$.  To reduce the number of free 
parameters in the numerical study of the next sections 
we will consider  a subclass of models (later called class A and B)
with the simplest Higgs structure \cite{imr}, in which 
the number of free parameters is reduced. 

Another interesting subclass of models  are those in which for certain
choices of the torus moduli all the brane intersections 
become ``approximately'' 
supersymmetric (see section 5.1 in ref.\cite{cim2})). 
In that subclass of models one has:
\beqa
\nu \ & =& \ 1/3 \ ;\ \epsilon\ =\ 1; \nonumber  \\ 
 n_{a 2}\ &=& \  n_{d 2} \ ;\ n_{c 1}\ =\  {{\beta_2}\over {\beta_1}
}n_{a 2} 
\label{qsusy}
\eeqa
so we are left with $n_{a 2},n_{b 1}$ and $\beta_i$ as free parameters.
A specific example of a model with these characteristics (and with $\beta_1=1/2$)
is presented later in the text, see the third 
line in Table \ref{minimal}.

\subsection{D5-brane orientifold models}

These models are obtained from Type IIB  compactification on
an orbifold $T^2\times T^2\times (T^2/Z_N)$
(see ref.\cite{cim3} for details). There are four
stacks a,b,c and d  of D5-branes which are wrapping cycles 
on $T^2\times T^2$ and are located at a $Z_N$ fixed point on 
$T^2/Z_N$. The 6-dimensional world-volume of the D5-branes includes
Minkowski space and the two extra dimensions wrap a different 
2-torus. Thus  each stack of D5-branes are specified by
giving the wrapping numbers $(n_{\alpha i}, m_{\alpha i})$, $i=1,2$ 
as well as the charge with respect to the $Z_N$ symmetry.
For general orientifold models of this type one can compute 
the form of the $c_i^\alpha $ coefficients. 
\TABLE{
\begin{tabular}{|c|c||c|c|}
\hline
$Z_3$-charge &
 $N_\alpha $    &  $(n_{\alpha 1},m_{\alpha 1})$  &  $(n_{\alpha 2},m_{\alpha 2})$  \\
\hline\hline 
1 &  $N_a=3$ & $(-1,\epsilon)$  &  $(3, -\epsilon/2)$   \\
\hline 
$\gamma $ &  $N_b=2$ &   $(1,0)$    &  $ (1 ,-\epsilon/2)$     \\
\hline 
$\gamma $ & $N_c=1$ & $(1,0)$  &  
 $(0,\epsilon)$    \\
\hline 
1 & $N_d=1$ &   $(2,-3\epsilon)$    &  $(1, -\epsilon/2 ) )$   \\
\hline \end{tabular}
\caption{ D5-brane wrapping numbers 
and $Z_3$ charges giving rise to a SM
spectrum in the example discussed in the text.
Here $\gamma=exp(i2\pi /3)$.
\label{solution5}}} 
\noindent
In this case there 
are $2\times(N-1)$ RR-fields participating in the GS mechanism.
In the particularly simple case of a $Z_3$ orbifold the 
number of RR antisymmetric $B_2^i$ fields is again four.
However, unlike the case of D6-brane, the RR-fields 
belong to a twisted spectrum.
Otherwise, the general structure of the $U(1)$ symmetries is 
surprisingly analogous to the case of D6-branes.
One can find specific examples of $Z_N$ symmetries and
wrapping numbers yielding the SM fermion spectrum, 
although there are  no large families of models as in the
D6-brane case. 
The  spectrum and $U(1)$ charges are as in Table  \ref{tabpssm}.  
We will provide here just a specific 
$Z_3$ example extracted from ref.\cite{cim3}.
In this model the wrapping numbers and $Z_3$ charges of the four stacks of branes
are shown in Table \ref{solution5}.

As discussed in ref.\cite{cim3}, in the case of the $Z_3$ orbifold 
with the structure of this example one of the four RR-fields does not
mix with any $U(1)$ (as in the D6-brane case). The other three
have $c_i^\alpha, $  $i=1,2,3$  coefficients given by \cite{cim3}
\beqa
c_1^\alpha \ &= &\,\,\,\, 2C_1 \ N_\alpha^k\, n_{\alpha 1} \, n_{\alpha 2}\, sin(2k\pi/3)
\nonumber \\
c_2^\alpha \ &= &\ -2C_1 \ N_\alpha^k \, m_{\alpha 1} \, n_{\alpha 2} \,
cos(2k\pi/3)\nonumber \\
c_3^\alpha \ &= &\ -2C_1 \ N_\alpha^k\, n_{\alpha 1}\, m_{\alpha 2} \,
cos(2k\pi/3) 
\label{cillos5}
\eeqa
where $C_1=\sqrt{sin(2\pi /3)}$. In these equations we have
 $k=0$ ($k=1$)  for branes with
$Z_3$-charge equal to 1 ($exp(i2\pi /3)$) respectively.
Thus for the example in Table \ref{solution5} one finds
(we take  $\epsilon =1$):
\beqa
c_1^b\ &=&\ 2C_1\sqrt{3} \ ;\  \quad    c_2^a\  = \  -18C_1 \ ;\nonumber \\ 
c_2^d \ & =& \ 6C_1 \ ;\,\,\,\quad \quad   c_3^a \  =  \ -3C_1 \nonumber  \\ 
c_3^b \ &  =& \ -C_1 \ ;  \quad \quad \ c_3^c\  = \ C_1 \ ;\qquad 
c_3^d\ =\ 2C_1 
\label{cillos55}
\eeqa
and the rest of the $c_i^{\alpha }$ vanish.
Note that in a  specific example like this there is just one free 
parameter, which is the overall mass scale related to the string scale. 
We will not give further details of this construction and refer to
\cite{cim3} for further details. We only note that there
are Higgs scalars with charges identical to those 
in the D6-brane models  and hence one can make a unified treatment
of electroweak symmetry breaking in both classes of models.
In this  particular example the Higgs sector is formed by
one set of fields ($H_i$, i=1,2) as will be described later in 
the text, Table \ref{thirdone}.

\subsection{ $U(1)$ gauge coupling constant normalization} 
\label{u1normalisation}

The gauge couplings of the four $U(1)$'s have some
interesting relationships in the D-brane models of particle 
physics.  Consider in particular a $SU(N)$ non-Abelian group arising
from a stack of $N$ parallel branes. Let us   normalize as usual 
the  non-Abelian
gauge coupling $g_N$  so that the quadratic Casimir in the fundamental is 
 $1/2$. Further, the diagonal $U(1)$ field living in the same stack
and giving charges $\pm 1$ to the bi-fundamental fermions at the
intersections will have gauge coupling $g_1=g_N/\sqrt{2 N}$.
For the case of the SM we  will thus have that at the string scale:
\beq
g_a^2 \ = \ {{g_{QCD}^2}\over 6} \ ;\ g_b^2 \ =\ {{g_L^2}\over 4}.
\label{caplillos}
\eeq
where $g_L$ is the $SU(2)_L$ coupling.
In addition, due to eq.(\ref{hyperca}) one has for the 
hypercharge coupling\footnote{This will be derived in detail later 
in the text.}:
\begin{equation}\label{hyperch}
\frac{1}{g_Y^{2}}\ = \ \frac{1}{36g_a^2}+
\frac{1}{4g_c^2}+\frac{1}{4g_d^2}.
\end{equation}
Thus $g_a$ and $g_b$ are determined by eqs.(\ref{caplillos})
whereas $g_c$ and $g_d$ are constrained by eq.(\ref{hyperch}),
with the ratio  $g_d/g_c$ as a free parameter.
In fact if we knew precisely the geometry of the 
brane configuration (torus moduli and shape) and if
the brane configuration was  fully factorizable in the
different tori, one could in principle compute the gauge
coupling constants at the string scale directly, since they
are given by the inverse of the volume that the branes
are wrapping in the tori (see e.g. refs.\cite{afiru2,imr,cim2}).
For the  analysis to follow in the next sections we will keep 
the ratio $g_d/g_c$ as a free parameter\footnote{In the 
numerical analysis we will ignore the
running  between the weak scale and the string scale.
This evolution does not affect much the structure 
of $Z$' masses and would include model-dependence
on the detailed particle content beyond the SM 
in the region  between the weak and the string scales.}.

We mentioned that the $B\wedge F$ couplings of RR-fields
to $U(1)$'s give masses to these gauge bosons  
of the order of the string scale $M_S$. In fact eq.(\ref{masones}) assumes 
a canonical kinetic term for the RR fields which mix
with the $U(1)$'s and give them a mass. However such kinetic terms
are field (radii)- dependent (see e.g., ref.\cite{cim1}) 
and once one re-defines the fields to 
canonical kinetic terms, extra volume factors appear in
eq.(\ref{masones}). In the case of D6-branes one finds that this 
amounts to the replacement
\beq 
c_i^\alpha \ \rightarrow \ \xi ^i c_i^\alpha \ \ ;\ \
\xi ^i\ =\ \sqrt{{R_2^iR_1^jR_1^k}\over {R_1^iR_2^jR_2^k}} \
\ ,\ i\not= j\not= k\not= i
\label{rescaling6}
\eeq
whereas in the case of the $Z_3$ D5-brane  models discussed above 
the appropriate rescaling factors are
\beq
\xi ^1 \ =\ \sqrt{{R_1^1R_1^2}\over {R_2^1R_2^2}} \ ;\
\xi ^2 \ =\ \sqrt{{R_1^1R_2^2}\over {R_2^1R_1^2}} \ ;\ 
\xi ^3 \ =\ \sqrt{{R_2^1R_1^2}\over {R_1^1R_2^2}}.
\label{rescaling5}
\eeq
Note that these volume factors are also relevant to the question
of creating a hierarchy between the string scale $M_S$  (which in the
present paper is assumed to be not far above the electroweak scale)
and the Planck scale $M_p$. In the case of the D5-brane models
\cite{cim3} 
one can take all the radii $R_{1,2}^i$, $i=1,2$ of order one
(in string units) and the volume of the third (orbifold) torus 
very large giving rise to the $M_p\gg M_S$ hierarchy in the standard
way \cite{aadd}. If this is assumed then one can set all $\xi^i=1$ 
and the $U(1)$ masses will be given by  eq.(\ref{masones}).
The case of D6-brane models is more subtle.
As  pointed out in  \cite{bgkl}
in the case of D6-brane models one cannot make some radii  $R_{1,2}^i$ 
very large in order to obtain $M_p \gg M_S$, the reason being that there
are no compact  directions which are  simultaneously transverse to
all the SM branes.  In this D6-brane case an alternative for understanding
the hierarchy could  perhaps be the following. The 
6-torus could be small while being connected to some very large
volume manifold \cite{imr} . For example, one can consider a region of the
6-torus away from the D6-branes, cutting a ball and gluing a throat
connecting it to a large volume manifold.  In this way one would 
obtain a low string scale model without affecting directly
the brane structure.  Alternatively it may be that
the apparent large value of the four dimensional Planck mass
could be associated with the localization of gravity on the branes,
along the lines of \cite{rs}.  
In the numerical analysis to follow we will neglect 
volume effects and set $\xi ^i=1$. The reader should however remember
 that when computing  bounds on the string scale $M_S$, the limits
apply modulo the mentioned volume factors.

\vspace{0.7cm}

\section{Mass eigenstates \& eigenvectors for $U(1)$ fields.}\label{generalmethod}


\noindent
In this section we outline the approach we use in Sections
\ref{specificmodels} and \ref{EWB} to analysing the effects of 
additional $U(1)$ fields in intersecting D-brane models.
The approach may be used for other cases as well, for example models
with D-branes at singularities \cite{bbarmod,aiqu}.
The analysis to follow is applied to  generic D6 and D5-brane models
and makes  little reference to the explicit details of these models.
We  address the mass eigenvalue problem for the $U(1)_\alpha$ fields,
which for  specific  D6 and D5-brane models is thoroughly
investigated in Section \ref{specificmodels} 
and \ref{EWB}.

In the basis ``a,b,c,d'' the kinetic terms of the Abelian fields
$A_{\alpha}$ are all diagonal. However, whether anomalous or not, the 
Abelian fields have mass terms induced not by a Higgs mechanism,
but through their (field strength) couplings to the three RR two-form fields $B_i$
(i=1,2,3). These mass terms are not diagonal, and are given 
by the $4\times 4$ symmetric  mass matrix $M^2_{\alpha\beta}$ of eq.(\ref{masones}).
An orthogonal  transformation (denoted $\cF$) is then introduced 
to ``diagonalise'' $M^2_{\alpha\beta}$. Therefore
\begin{equation}\label{calF}
A_{i}'= \sum_{\alpha=a,b,c,d} \cF_{i\alpha} A_{\alpha}, \qquad i=1,2,3,4.
\end{equation}
where the ``primed'' states correspond to the mass eigenstate basis,
and  $i$ is fixed (i=1,...4) to a value which includes the hypercharge (i=1). 
Since the matrix $\cF$ is orthogonal, gauge kinetic terms
are not affected by this transformation.

This is what happens when electroweak symmetry breaking effects are
neglected. In the presence of the latter, $M^2_{\alpha\beta}$ 
has additional corrections. The $U(1)_\alpha$ fields will then acquire
a mass from two sources: the string mechanism mentioned before, and
the usual Higgs mechanism, with the contribution of the latter
suppressed by $M_Z^2/M_S^2$.

\subsection{$U(1)$ masses before  Electroweak Symmetry Breaking.}

In the absence of the electroweak symmetry breaking, 
the only contribution to the masses  of the $U(1)$ gauge 
bosons is due to  $M^2_{\alpha\beta}$. 
Its eigenvalues $M_i^2\equiv \lambda_i M_S^2$  are  the roots  of
\be \det\Bigl[\lambda M_S^2 I_4-M^2\Bigr]=0,\label{detM}\ee
which can be written as
\be \lambda \bigl(\lambda^3+c_3\lambda^2+c_2\lambda+c_1\bigr)=0,
\label{char4}\ee
with the following coefficients
\begin{eqnarray}\label{lascs} 
c_3&=&-Tr(M^2)<0,   \nonumber \\
c_2&=&-{1\over 2}\Bigl[ Tr (M^4) - (Tr M^2)^2\Bigr]>0, \nonumber \\
c_1&=&-{1\over 3}\Bigl[ Tr (M^6) - (Tr M^2)^3\Bigr]+
{1\over 2}Tr(M^2)\Bigl[ Tr (M^4) - (Tr M^2)^2\Bigr].
\end{eqnarray}
Since the mass matrix $M^2$ is symmetric, its eigenvalues  are  real. 
The roots of (\ref{char4}) have the general form
\begin{eqnarray}
\gl_1&=&0, \nonumber \\
\lambda_2&=&-\frac{c_3}{3}+\frac{2}{3}\sqrt{c_3^2-3c_2}
\cos{\left(\frac{\phi}{3}\right)}, \nonumber \\
\lambda_3&=&-\frac{c_3}{3}+\frac{2}{3}\sqrt{c_3^2-3c_2}
\cos{\left(\frac{\phi}{3}+\frac{2\pi}{3}\right)}, \nonumber \\
\lambda_4&=&-\frac{c_3}{3}+\frac{2}{3}\sqrt{c_3^2-3c_2}
\cos{\left(\frac{\phi}{3}+\frac{4\pi}{3}\right).}\label{eigval4}
\end{eqnarray}
with the definition
\be \cos{\phi}=\pm \bigg[ 
{\frac{(2c_3^3-9c_2c_3+27c_1)^2}{4(c_3^2-3c_2)^3 }}\bigg]^{1/2}
\ee
where the ``+'' (-) sign applies if $(2c_3^3-9c_2c_3+27c_1)<0\,$ $(>0)$.
In realistic models $\lambda_1=0$ is usually the hypercharge,
the massless state before electroweak  symmetry breaking. If $c_1=0$, 
a second massless state exists which may correspond to
$B-L$ (see ref.\cite{cim2}).  Since $c_3 < 0$ and $c_2 > 0$ 
there cannot be a third massless state (unless the matrix is 
identically zero).

For a generic  $D5$-brane model the coefficients  $c^a_{2}$, $c^a_{3}$, 
$c^b_{1}$, $c^b_{3}$, $c^c_{3}$, $c^d_{2}$, $c^d_{3}$ are non-zero, 
see (\ref{cillos55}). If in addition  $c^a_{2}=0$, 
we obtain the $D6$-brane models case, see eq.(\ref{cillos6}). 
In such cases one may show that the coefficient $c_1$ of (\ref{lascs}) 
equals 
\be c_1=-(c^b_{1})^2\Bigl[(c^c_{3})^2 ( (c^a_{2})^2+(c^d_{2})^2)+
( c^a_{2} c^d_{3}-c^d_{2}c^a_{3})^2\Bigr]<0\label{cc11}
\ee
which is thus negative for both $D5$ and $D6$-brane models.
Taking account of the signs of  $c_i$ (i=1,2,3) one can  prove that
the  eigenvalues  $\lambda_i>0$ (i=2,3,4).
This is so because the function 
$f(\gl)\equiv (\gl^3+c_3 \gl^2 +c_2 \gl +c_1) <0$ for $\gl=0$ 
and its second derivative $f''(\gl=0)=2 c_3 <0$.  Thus $\lambda_i$
and the (squared)  masses $M_i^2\equiv \lambda_i M_S^2$  are all 
positive.

After computing the eigenvalues $M_i^2$, the associated 
 eigenvectors $w_i$ (i=1,2,3,4)   with components $(w_i)_\alpha$
($\alpha=a,b,c,d$) may be written as
\begin{equation}
(w_i)_\alpha=- \sum_{\beta=a,b,c} \left( M^2 -\lambda_i \,M_S^2 \,I_4
\right)^{-1}_{\alpha\beta}\, M^2_{\beta d}\, (w_i)_{d},\qquad i=1,...,4,
\,\,\,\alpha=a,b,c.
\end{equation}
The solution can be written more explicitly in terms of
the coefficients $c_i^\alpha$, but the expressions are long and 
not presented here.
As we will see later, there are classes of models\footnote{For this
see Section \ref{classbBEW}.}  for which $U(1)_b$  field does not mix 
with the remaining ones (being a mass eigenstate itself),  
but this depends on the specific
structure of the matrix $M^2_{\alpha\beta}$ and will not be addressed
here. Finally, the matrix $\cF$ of (\ref{calF}) is found
 from eigenvectors  $w_i$ (normalised to unity).

\subsection{$U(1)$ masses after  Electroweak Symmetry Breaking.}\label{generalmethodAEW}

After electroweak symmetry breaking, the mass matrix 
$M^2_{\alpha\beta}$ receives corrections due to mixing 
of initial $U(1)_\alpha$ with  the Higgs sector charged under some
of these symmetries and also from the mixing with $W_3^\mu$ boson of 
$SU(2)_L$ symmetry. These corrections are proportional to
$\eta=<\!\!H\!\!>^2\!\!/M_S^2$ where $<\!\!H\!\!>$ stands for the vev in the 
Higgs sector. The previous mass eigenstates (\ref{eigval4}) will thus 
have additional corrections of this order with a further (non-zero)
mass eigenstate to correspond to Z boson which thus mixes with initial
$U(1)_\alpha$. 
Taking account of such corrections requires one to "diagonalise"   
a $5\!\times\! 5$ matrix in the basis $a,b,c,d$ {\it and} $W_3^\mu$. 
This matrix has the structure 
$\cM^2= M^2+\Delta$ where $M^2$ is just that of (\ref{masones})
extended to a $5\!\times\!5$ one by a fifth line and column 
which have all  entries equal to zero. For illustration we present
below the electroweak corrections $\Delta$ (in the aforementioned
basis)  for $D6$ and $D5$ brane models\footnote{The matrix $\Delta$
will be derived in detail in Section \ref{EWB}. Its exact 
structure for the general discussion of this section is 
irrelevant.}
\be 
{\Delta}  = \pmatrix {0 & 0 & 0 & 0  & 0\cr
0 & \eta g_b^2 & \eta \delta g_bg_c &  0
& -\frac{1}{2}\eta \delta g_b g_L\cr
0 & \eta \delta g_bg_c & \eta g_c^2 
& 0 & -\frac{1}{2}\eta  g_c g_L\cr
0 & 0 & 0 & 0 & 0 \cr
0 & -\frac{1}{2}\eta \delta g_b g_L & 
-\frac{1}{2}\eta g_c g_L & 0 & 
\frac{1}{4}\eta g_L^2}\label{M4C}
\ee
where the Higgs sector was assumed to be charged only under 
$U(1)_b$ and $U(1)_c$ 
(which is the case in the specific models under study) 
while  $\delta$ is just a mixing parameter 
in this sector. The mass eigenvalue equation of ${\cal M}^2$ is now
\begin{equation}
\gl\left( \gl^4+o_3\,\gl^3+ o_2\,  \gl^2+ o_1\, \gl+ o_0\right)
=0,\label{char5}
\end{equation}
with
\begin{eqnarray}\label{llll}
o_i & = & c_i+\eta\, s_i+ \eta^2\, t_i,\qquad i=1,2,3.\nonumber\\
o_0 & = & \eta \,s_0 + \eta^2\, t_0
\end{eqnarray}
to be compared to the one in the absence of electroweak symmetry breaking,
eq.(\ref{char4}). One can compute explicitly 
the  coefficients $s_i$ and $t_i$  
which arise from the matrix  eq.(\ref{M4C}) or other similar 
electroweak corrections.

Eq.(\ref{char5}) shows  that a  massless state $\lambda_1=0$
exists after electroweak 
symmetry breaking (photon).  The remaining eigenvalues $\lambda_i$
may be computed as an expansion 
in the parameter $\eta$ about their value
$\lambda_i$
in the absence of electroweak effects (given in (\ref{eigval4})).
For example
\be 
\gl_i^{EW}=\gl_i+\eta\,\gl_i'+\eta^2\, \gl_i''+\cdots, \qquad i=2,3,4.
\label{aew}
\ee 
Inserting this into (\ref{char5}) and taking account of (\ref{llll})
one can solve for the coefficients
$\gl_i'$ and $\gl''_i$ to find the (electroweak) corrected  
values of the $U(1)$ masses.
Similarly, the mass eigenvalue $M_Z^2=\lambda_5 M_S^2$ 
corresponding to  Z boson  may be written as   an expansion
\be 
\gl_5=\xi_1\,\eta\, \left(1+\eta\, \xi_{21}+\eta^2 \, \xi_{31}+\cdots \right)
\ee
Using this in (\ref{char5}) one computes the coefficients 
$\xi_{1}$, $\xi_{21}$, $\xi_{31}$ to find
\be \xi_1=-{s_0\over {c_1}},\ee
\be \xi_{21}=\frac{1}{c_1^2}(c_2s_0-c_1s_1)+\frac{t_0}{s_0},\ee
\be \xi_{31}=\frac{1}{c_1^4}
\Bigl[ 
2c_2^2s_0^2-c_1s_0(3c_2s_1+c_3s_0)+c_1^2(s_1^2+s_0s_2)
+t_0 \left(2c_1^2c_2-\frac{s_1c_1^3}{s_0}\right)-t_1 c_1^3
\Bigr].\label{x31}\ee
To leading order in $\eta$ one should recover the usual 
mass of Z boson, while
corrections of higher order in $\eta$ 
are due to mixing between the masses
of initial $U(1)_\alpha$ fields and Z boson. 
Such corrections to the mass of Z boson are of string origin
as previously discussed, and they will be extensively
investigated for specific models in the next two sections. 
(the eigenvectors after electroweak symmetry 
breaking may also be computed using this approach, as
outlined in Appendix I for the eigenvector of $Z$ boson).

\vspace{0.7cm}
\section{Analysis of explicit 
D6 and D5  models before EWS breaking.}\label{specificmodels}

Following the steps outlined above, we address in this section
explicit D-brane  models, for which 
the couplings (\ref{gsuno}) of antisymmetric B-fields to
the Abelian gauge bosons are known in detail. This means that 
the  matrix $M^2_{\alpha\beta}$ of (\ref{masones})  (essentially its
entries $c_i^\alpha$) is fully known in terms of the (chosen) 
parameters of the models.
We thus consider explicit constructions of  D6- and D5-brane 
models discussed in Section \ref{extrau1}, in the absence of
electroweak symmetry (EWS) breaking effects.

\subsection{D6-brane models. Masses of $U(1)$ fields and bounds on $M_S$.}
For  D6-brane models the coefficients
$c_i^\alpha $ in the mass matrix of (\ref{masones})
 are given in eq.(\ref{cillos}) and Table \ref{solution}. 
We  analyse a subclass of the D6-brane models,
with parameters as presented in Table \ref{minimal} 
(these models were discussed in ref.\cite{imr}).
%
\TABLE{\renewcommand{\arraystretch}{1.25}
\begin{tabular}{|c|c|c|c|c|}
\hline
Higgs $\sigma$  &  $q_b$   &  $q_c$   & $q_1'\equiv q_y$ & $T_3=1/2\, \sigma_z$ \\
\hline\hline
$h_1$   &    1     & -1       & 1/2              &   +1/2      \\
\hline 
$h_2$   &   -1     &  1       &  -1/2            &   -1/2    \\
\hline\hline
$H_1$   &   -1     & -1       & 1/2              &   +1/2    \\
\hline
$H_2$   &    1     &  1       &  -1/2            &   -1/2     \\
\hline 
\end{tabular}
 \caption{Higgs fields, their U(1)$_{b,c}$
charges and weak isospin with $\sigma_z$  the diagonal Pauli matrix.}
\label{thirdone}}
For these models there are two types of Higgs scalars
denoted $h_i$ and $H_i$ respectively, as shown in Table 
\ref{thirdone}.
For certain choices of integer parameters the number of 
these  Higgs multiplets is minimal, and this defines two classes
of models:   Class  A models which have a single set of Higgs multiplets,
either  $h_i$ or  $H_i$ and Class B models which have
both $h_i$ and $H_i$.  Class A models are defined in 
the first four entries of Table \ref{minimal}. Class  B  models 
are presented by the last four lines in Table \ref{minimal}.

For these specific models we 
investigate the eigenvalue problem for
 $M_{\alpha\beta}^2$  and the transformations 
of couplings and hypercharges of $U(1)$ fields upon ``diagonalising'' 
this mass matrix.

\vspace{1.1cm}
\subsubsection{Class A models.}

The parameters of this class of models are detailed in 
Table \ref{minimal}. 
For generality we keep $\ge,\nu,\beta_1,n_{b1},n_{c1}$
unassigned, to cover simultaneously  all possible cases of this class. 
We however use the definition of $n_{d2}$ (Table \ref{minimal})
ensuring a vanishing mass for the hypercharge state. 
The ratio $g_d/g_c$ and $n_{a2}$ are chosen as free parameters.
The eigenvalues $M_i^2 \equiv \gl_i M_S^2$  are the roots of  
\begin{equation}\label{roots}
\gl^4+c_3 \gl^3 +c_2 \gl^2 +c_1 \gl =0
\end{equation}
with the notation
\begin{eqnarray}\label{c3}
c_1&=&- \frac{4 \ge^4 g_b^2 n_{c1}^2}{\beta_2^2 \nu^2} \left(g_c^2 g_d^2+9
g_a^2(g_c^2+g_d^2)\right)<0     \\
\vspace{0.2cm}\hfill\nonumber\\
c_2&=&\frac{\ge^2}{\beta_1^2 \beta_2^4 \nu^2} 
\left\{\beta_2^2 g_b^2 (9 g_a^2+g_d^2) \left[4 \beta_1^2
\beta_2^2\ge^2+\nu^2\left (\beta_1^2 n_{a2}^2+9 \beta_2^2n_{b1}^2
\right)\right]
\right.\nonumber\\
&&\,\,\,\,\,\,\,\,\,\,\,\,\,\,\,\,\,\,
+\, n_{c1}^2(9 g_a^2 \beta_2^4+4 \beta_1^4\nu^2 g_b^2)(g_c^2+g_d^2)
+\left.\beta_2^4 g_c^2 g_d^2 n_{c1}^2 -4 \beta_1^3\beta_2 g_b^2 
g_d^2 n_{a2} n_{c1}\nu^2
\right\}>0  \nonumber \\
\vspace{0.2cm}\nonumber\\
c_3&=&\frac{-1}{4 \beta_1^2\beta_2^2 \nu^2}\left\{
(9 g_a^2+g_d^2) \left(4 \beta_2^2 \ge^2 +\nu^2n_{a2}^2\right) 
\beta_2^2\right.\nonumber\\
&&\quad\qquad + \left.4 \beta_1\nu^2\left[4 \beta_1^3\ge^2 
g_b^2-\beta_2 g_d^2 n_{a2}
n_{c1}+\beta_1(g_c^2+g_d^2) n_{c1}^2 \right]
+36 \beta_1^2 g_b^2 n_{b1}^2 \nu^4 \right\}<0\nonumber
\end{eqnarray}
\TABLE{\renewcommand{\arraystretch}{1.25}
\begin{tabular}{|c|c|c|c|c|c|c|c|c|}
\hline
 Higgs    &  $\nu
 $  &  $\beta_1$   & $\beta_2$ & $n_{a2}$ & $n_{b1}$
& $n_{c1} $
& $n_{d2}$  & $N_h$ \\
\hline\hline  $n_H=1,n_h=0$  & 1/3  &  1/2  & $\beta_2$ & $n_{a2}$ & -1 & 1&
$\frac{1}{\beta_2}-n_{a2}$   &  $4\beta_2(1-n_{a2})$  \\
\hline  $n_H=1,n_h=0$  & 1/3  &  1/2  & $\beta_2$ & $n_{a2}$ & 1 & -1&
$-\frac{1}{\beta_2}-n_{a2}$   &  $4\beta_2(1-n_{a2})$  \\
\hline  $n_H=0,n_h=1$  & 1/3  &  1/2  & $\beta_2$ & $n_{a2}$ & 1 & 1&
$\frac{1}{\beta_2}-n_{a2}$   &  $4\beta_2(1-n_{a2})-1$  \\
\hline  $n_H=0,n_h=1$  & 1/3  &  1/2  & $\beta_2$ & $n_{a2}$ & -1 & -1&
$-\frac{1}{\beta_2}-n_{a2}$   &  $4\beta_2(1-n_{a2})+1$  \\
\hline\hline  $n_H=1,n_h=1$  & 1   &  1   & $\beta_2$ & $n_{a2}$ & 0  & 1&
$\frac{1}{3}(\frac{2}{\beta_2}-n_{a2})$   &
$\beta_2(8-\frac{4n_{a2}}{3})-\frac{1}{3}$ \\
\hline  $n_H=1,n_h=1$  & 1   &  1   & $\beta_2$ & $n_{a2}$ & 0  & -1&
$\frac{1}{3}(-\frac{2}{\beta_2}-n_{a2})$   &
$\beta_2(8-\frac{4n_{a2}}{3})+\frac{1}{3}$ \\
\hline  $n_H=1,n_h=1$  & 1/3   &  1   & $\beta_2$ & $n_{a2}$ & 0  & 1&
$ \frac{2}{\beta_2}-n_{a2}$   &
$\beta_2(8-{4n_{a2}})-1$ \\
\hline  $n_H=1,n_h=1$  & 1/3   &  1   & $\beta_2$ & $n_{a2}$ & 0  & -1&
$ -\frac{2}{\beta_2}-n_{a2}$   &
$\beta_2 (8-{4n_{a2}})+1$ \\
\hline \end{tabular}
\caption{ Families of D6-brane models with minimal Higgs
 content \cite{imr}. The first four lines correspond to models of Class A,
the remaining ones to models of Class B. $\beta_2=1, 1/2$. 
 The parameters of the models are  $n_{a2}$ and $g_d/g_c$.}
\label{minimal}}
As discussed in the general case, 
$M_i^2$ are real ($M_{\alpha\beta}^2$ symmetric) and one 
may show that $c_1<0$, $c_2>0$ and $c_3<0$, 
irrespective of the value of parameters present  in their definitions
above. Thus all $M_i^2$  are  positive.
Using eq.(\ref{eigval4}) and coefficients $c_i$ of eqs.(\ref{c3}) 
one finds the values of $M_i$ ($\lambda_i$).

\noindent
The eigenvectors may now be computed and are given by:
\begin{eqnarray}\label{eigvecA}
w_{1}&=&\frac{1}{|w_1|} \left\{\frac{g_d}{3 g_a}, 0,
-\frac{g_d}{g_c}, 1\right\}_\alpha,\qquad {\alpha=a,b,c,d}\\
\vspace{0.4cm}
w_{i}
&=&\frac{1}{|w_i|}\left\{w_{i a},w_{i b},w_{i c},1 \right\},
\qquad  \quad i=2,3,4; 
\end{eqnarray}
with the first three  components given by 
\begin{eqnarray}
w_{i a}&=&\frac{3 \beta_2\, g_a \left( 2 \beta_2 \,g_d^2\, n_{c1} \ge^2 - 
 n_{a2} \gl_i \beta_1\nu^2\right)}
{g_d\left[ 18 \beta_2^2 \, g_a^2 n_{c1} \ge^2  + \beta_1 \nu^2 \gl_i (n_{a2}
\beta_2 -2 \beta_1 n_{c1})\right]}, \nonumber \\ 
\nonumber\\
\vspace{0.4cm}\hfill\nonumber\\
w_{ib}&=&-\frac{4 \beta_1^2\,  \beta_2^2 \, \nu^2 \gl_i^2 
+4 \beta_2^2 \ge^2 \left[g_c^2 g_d^2 +9 g_a^2 (g_c^2+g_d^2)\right]n_{c1}^2 }
{6 \nu\, n_{b1}\, g_b\, g_d\left[ 18 \beta_2^2\, g_a^2 n_{c1}\ge^2 +\beta_1
\nu^2 \gl_i (n_{a2} \beta_2 -2 \beta_1 n_{c1})\right] }\nonumber\\   
\vspace{0.2cm}\hfill\nonumber\\
\nonumber\\
&&+\frac{\gl_i\left[\beta_2^2 (9 g_a^2+g_d^2)(4  \beta_2^2 \ge^2 +\nu^2 n_{a2}^2)
-4 \beta_1 \nu^2 n_{c1} (\beta_2  g_d^2 n_{a2} - \beta_1 n_{c1} (g_c^2+g_d^2)\right]}
{6 \nu\, n_{b1}\, g_b\, g_d\left[ 18 \beta_2^2\,  g_a^2\,  n_{c1}\ge^2 +\beta_1
\nu^2 \gl_i (n_{a2} \beta_2 -2 \beta_1 n_{c1})\right] },
\nonumber\\
\vspace{0.4cm}\hfill    \nonumber \\
w_{ic}&=& \frac{2 g_c n_{c1} \left[\beta_2^2 \, \ge^2\, (9 g_a^2+g_d^2) 
-\beta_1^2\nu^2 \gl_i\right]}
{g_d\left[ 18 \beta_2^2 g_a^2 n_{c1} \ge^2  +\beta_1 \nu^2 \gl_i (n_{a2}
\beta_2 -2 \beta_1 n_{c1})\right]}
\label{fi3}
\end{eqnarray}
and where for each eigenvector $w_i$ (i=2,3,4) 
the appropriate eigenvalue $\gl_i$ as given by
eqs.({\ref{eigval4}})  is substituted.
From the above equations we note
that the anomalous  U(1)$_b$ state mixes with the
other states upon diagonalising the mass matrix 
(which is unlike Class B models to follow). The eigenvectors above will be used in 
analysing the ``amount'' of each initial $U(1)_\alpha$  present in
the final mass eigenstates $M_{2,3,4}$.

We have so far presented the eigenvalues and eigenvectors of
$M_{\alpha\beta}^2$ in terms of the parameters of this class of
models. Further, the couplings of the new (mass eigenstates)
$U(1)$ fields also change upon going to the new basis.  
The new couplings $g_i'$ can be expressed in terms of the ``old'' 
couplings $g_\alpha$,  and similar relations exist for the 
transformation of the charges $q_\alpha\rightarrow q_i'$. 
First, the eigenvectors $w_{i}$  of eqs.(\ref{eigvecA}) define the  matrix
$\cF_{i\alpha}=(w_{i})_{\alpha}$ introduced in eq.(\ref{calF}), which 
respects the conditions $\cF \cF^T=\cF^T \cF=1$ and $\cF M^2 \cF^T= \delta_{ij}
M_i^2$. The $U(1)$ charges of the re-defined 
Abelian fields (mass eigenstates) can then be computed by using the 
invariance of the following term which is part of the covariant
derivative
\begin{equation}
\sum_{\alpha=a,b,c,d} q_\alpha g_\alpha A_\alpha
\end{equation}
under  changing the basis from ``a,b,c,d'' to the ``primed'' basis.
The properties of $\cF$  then give  the transformation of $U(1)$ charges:
\begin{equation}\label{q_i}
q_i'=\sum_{\alpha=a,b,c,d} U_{i\alpha}\, q_\alpha,\hspace{1cm}
U_{i\alpha}\equiv  \frac{g_\alpha}{g_i'} \cF_{i\alpha}, \hspace{1cm} i=1,2,3,4. 
\end{equation}
From this equation we notice that the new hypercharges $q_i'$ of associated fields
 depend in general (see definition of $U_{i\alpha}$) on 
the couplings of the model. For the massless state  (hypercharge) we find
\begin{equation}\label{qy}
q_1'=\frac{g_d}{3 g_1' |w_1|}\left(q_a-3 q_c+3 q_d\right),
\qquad (q_1'\equiv q_y,\,\, g_1'\equiv g_y)
\end{equation}
With $q_y=1/6(q_a-3 q_c+3 q_d)$ see Table \ref{tabpssm},
we find  the hypercharge coupling  is given by
\begin{equation}\label{hypercharge1}
\frac{1}{g_1^{'2}}=\frac{|w_1|^2}{4 g_d^2}=\frac{1}{36}\frac{1}{g_a^2}+
\frac{1}{4}\frac{1}{g_c^2}+\frac{1}{4}\frac{1}{g_d^2}
\end{equation}
as anticipated in Section \ref{u1normalisation}.
More generally, the orthogonality of  the matrix $\cF$  gives the following
relation between the ``old'' and ``new'' gauge couplings in terms of
the matrix $U_{i \alpha}$
\begin{equation}
\frac{\delta_{ij}}{g_i^{'} g_j'}=\sum_{\alpha=a,b,c,d}
\frac{U_{i\alpha} U_{j\alpha}}{g_\alpha^2}
\end{equation}
We thus identified the new basis, the mass 
eigenvalues, and the transition between the old and new couplings and 
hypercharges of the system of $U(1)$'s.  With these analytical results, 
we can  compute (lower) bounds on the value of the 
string scale $M_S$ in models of Class A. This is done  by 
analysing the dependence of the eigenvalues eqs.(\ref{eigval4}) 
with coefficients (\ref{c3}) on the parameters of the models.

A thorough analysis of all Class A models parametrised in Table
\ref{minimal}  would require a separate investigation of each of
these models. Although we do not discuss  separately each model, 
the results  are generic for this class.
As an example we present in  Figure \ref{modela}  the dependence of 
the eigenvalues $M_2$, $M_3$ and $M_4$ in  function of $n_{a2}$ and $g_d/g_c$
 for\footnote{All plots of figure \ref{modela} take into account
the correlation (\ref{hypercharge1}) of the couplings of additional $U(1)$'s
such as the hypercharge coupling is fixed  to the experimental value
at the electroweak scale.} a Class A model of Table \ref{minimal} defined by 
$\beta_1=1/2$, $\nu=1/3$, $\beta_2=1$, $n_{c1}=1$, $n_{b1}=-1$.
Figure \ref{modela} shows that the  eigenvalue $M_2$  is 
larger than the string scale ($M_2\geq 8 \,M_S$).
However,  for $M_4$  we have $1.6 M_S\leq M_4\geq 0.4 \,M_S$
but may become smaller for large $n_{a2}$. Finally $M_3$ 
has a mass  between  $0.15 \,M_S\leq M_3\leq 0.32\, M_S$ and hence may 
be small compared  to the string scale, particularly for large values
of $n_{a2}$.

One can draw the generic conclusion that the
masses of the additional $U(1)$ fields are within a factor of $\cO(10)$
from  the  string scale, either larger or smaller.
Given  the current generic  bounds on the masses of 
additional $U(1)$ bosons in the region of 500-800 GeV \cite{pdg}, 
and with the lightest eigenstate satisfying 
$M_3\leq 0.32 M_S$, one concludes that the string scale 
must obey $M_S\geq 1.5-2.4$ TeV.
For $M_3\approx 0.15 M_S$, $M_S\approx  3-4.8$ TeV.
 These lower bounds on $M_S$ will increase
with $n_{a2}$,  when  $M_3$ decreases.
An additional bound on $M_S$ will be derived in Section \ref{EWB} from
taking account of the  electroweak symmetry breaking effects.
The final lower bound on $M_S$ will be provided by the value which
satisfies both of the above constraints.

It is interesting to know which of 
the initial $U(1)_\alpha$ ($\alpha=a,b,c,d$)
dominates in the final mass eigenstates, $M_{2,3,4}$. 
This depends in general  on the parameters of the models of Class A,
as it may be seen from the eigenvectors of eq.(\ref{fi3}).
It turns out however that the result is  not very sensitive to
$n_{a2}$. The only exception is $w_{ib}$ (i.e. $U(1)_b$ presence)
which has a more significant spreading with respect to $n_{a2}$ 
for fixed ratio $g_d/g_c$ than the remaining $w_{i\alpha}$. 
This dependence  is taken into account for this discussion.
The conclusions  will then apply generically to every 
model of Class A of Table \ref{minimal} since the additional dependence of 
the eigenvector components (mixing) on 
$\beta_i$, $n_{c1}$, $n_{b1}$ is rather mild as may be seen by
examining separately every model of Class A.
Therefore the  only significant change of the amount of $U(1)_\alpha$ each 
mass eigenstate has is due to changes of the parameter  $g_d/g_c$
and this is examined below.

Consider first the  mass eigenstate $M_2$ (Figure \ref{modela}). $M_2$ is 
for $g_d/g_c<1$ a mixture of $U(1)_a$ and  $U(1)_d$ where $U(1)_a$
dominates.  The same holds true for the case
$g_d/g_c>1$, with $U(1)_a$ to dominate more for $g_d/g_c$ small, of
of order one,  and less when this ratio increases (top curve) in 
Figure \ref{modela}.(B). In this last case the amount of initial 
$U(1)_a$ and $U(1)_d$ is comparable.
For $M_3$ mass eigenstate,  for  $g_d/g_c<1$  each $U(1)_b$, $U(1)_c$ $U(1)_d$
states contribute with  similar/comparable amount, with $U(1)_b$ and $U(1)_c$ to
dominate for smaller $g_d/g_c$, top curve in figure \ref{modela} (C).
For  $g_d/g_c>1$, $U(1)_b$, $U(1)_c$ and $U(1)_d$ contribute significantly for
lower curves, and $U(1)_b$ dominates for  $g_d/g_c\gg 1$, see Figure
\ref{modela}.(D).
Finally, for  $M_4$ mass eigenstate, the lower curves in Fig.\ref{modela}.(E)
have a $U(1)_b$ dominating component and additional 
comparable amount of $U(1)_c$ and
$U(1)_d$. While decreasing $g_d/g_c$, top curve, $U(1)_b$ dominates
more strongly, with additional amount of $U(1)_a$. 
For the case of $g_d/g_c\gg 1$, Figure \ref{modela} (F)
$U(1)_b$ again dominates for lower curves, while top curves are an equal
mixture of $U(1)_b$ and $U(1)_d$ and a slightly larger $U(1)_a$ component.

To conclude, the anomalous component $U(1)_b$ is manifest mostly in $M_3$
and $M_4$, while $U(1)_a$  dominates in $M_2$ case, with $U(1)_d$  manifest
significantly in all  lower curves ($g_d\approx g_c$) in Figure \ref{modela}.
Generically, the presence of $U(1)_a$  
favours larger (than the string scale) mass
eigenvalues, and $U(1)_b$  together (to a lesser extent) with $U(1)_c$
and $U(1)_d$ is more present in states of mass smaller than the string scale.

\cleardoublepage
\FIGURE[t]{
\begin{tabular}{cc|cr|} 
\parbox{7.1cm}{ 
\psfig{figure=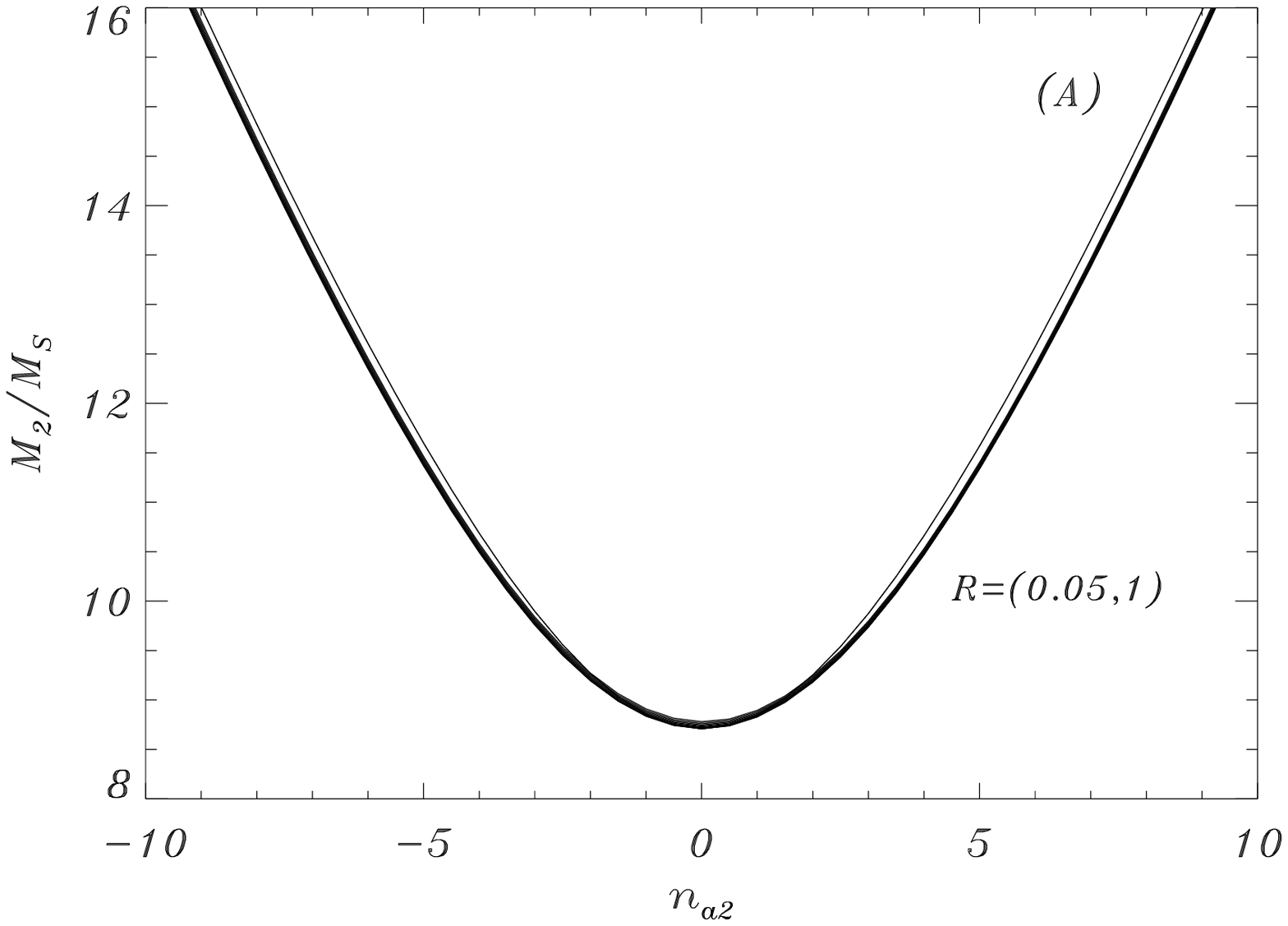,height=6.2cm,width=6.5cm}} 
\hfill{\,} 
\parbox{7.1cm}{ 
\psfig{figure=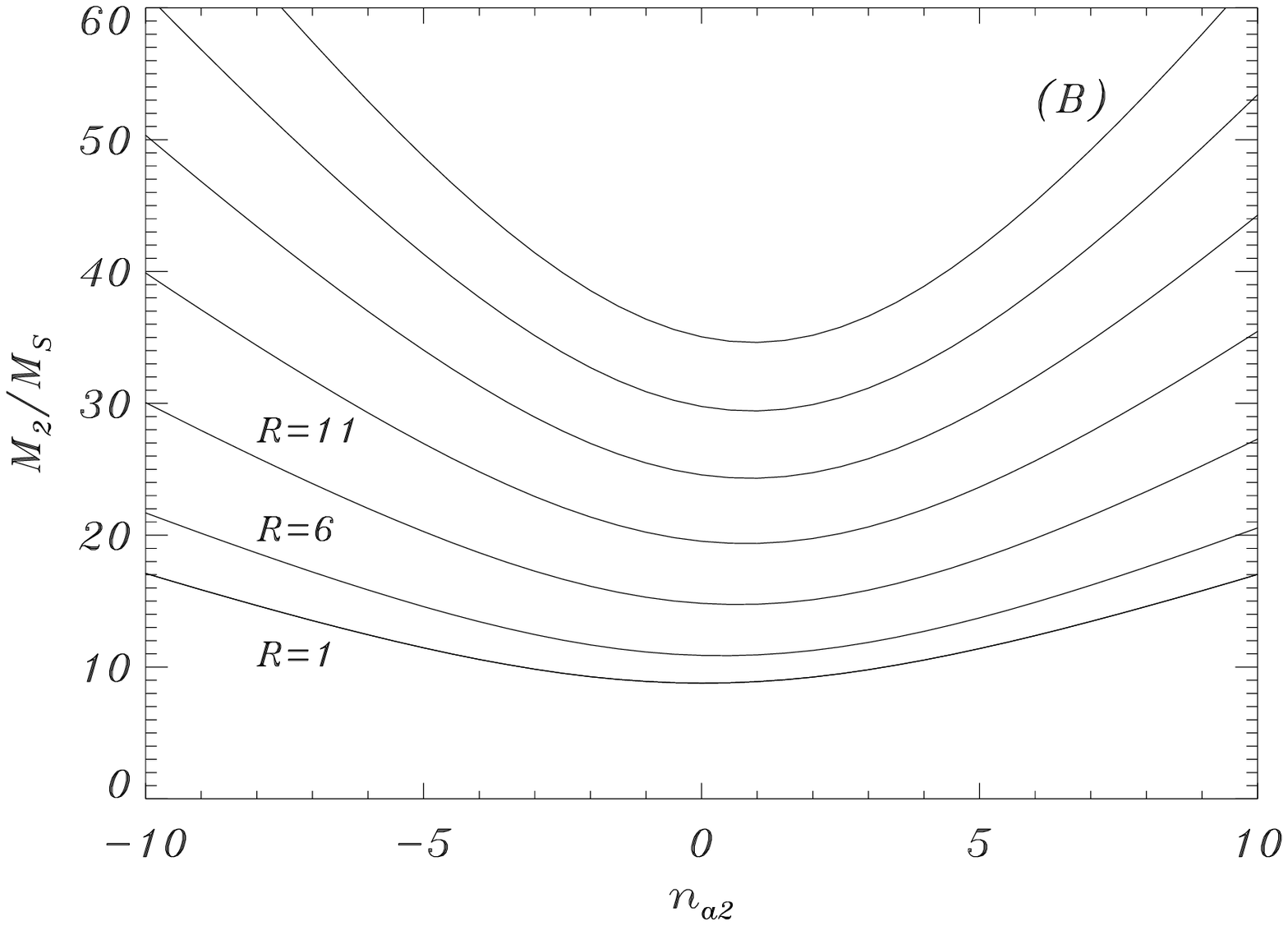,height=6.2cm,width=6.5cm}}  
\end{tabular} 
\begin{tabular}{cc|cr|} 
\parbox{7.1cm}{ 
\psfig{figure=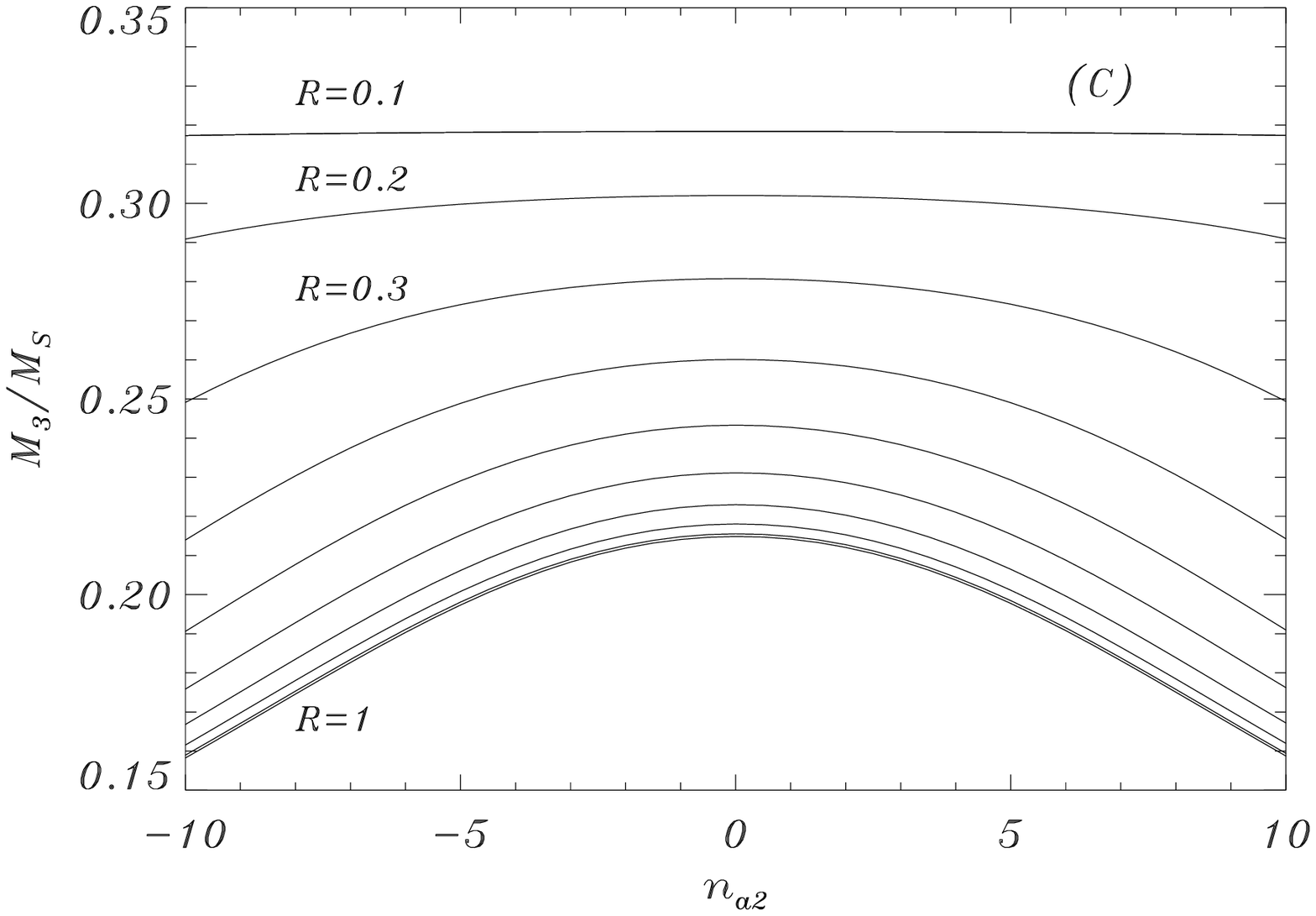,height=6.2cm,width=6.5cm}} 
\hfill{\,} 
\parbox{7.1cm}{ 
\psfig{figure=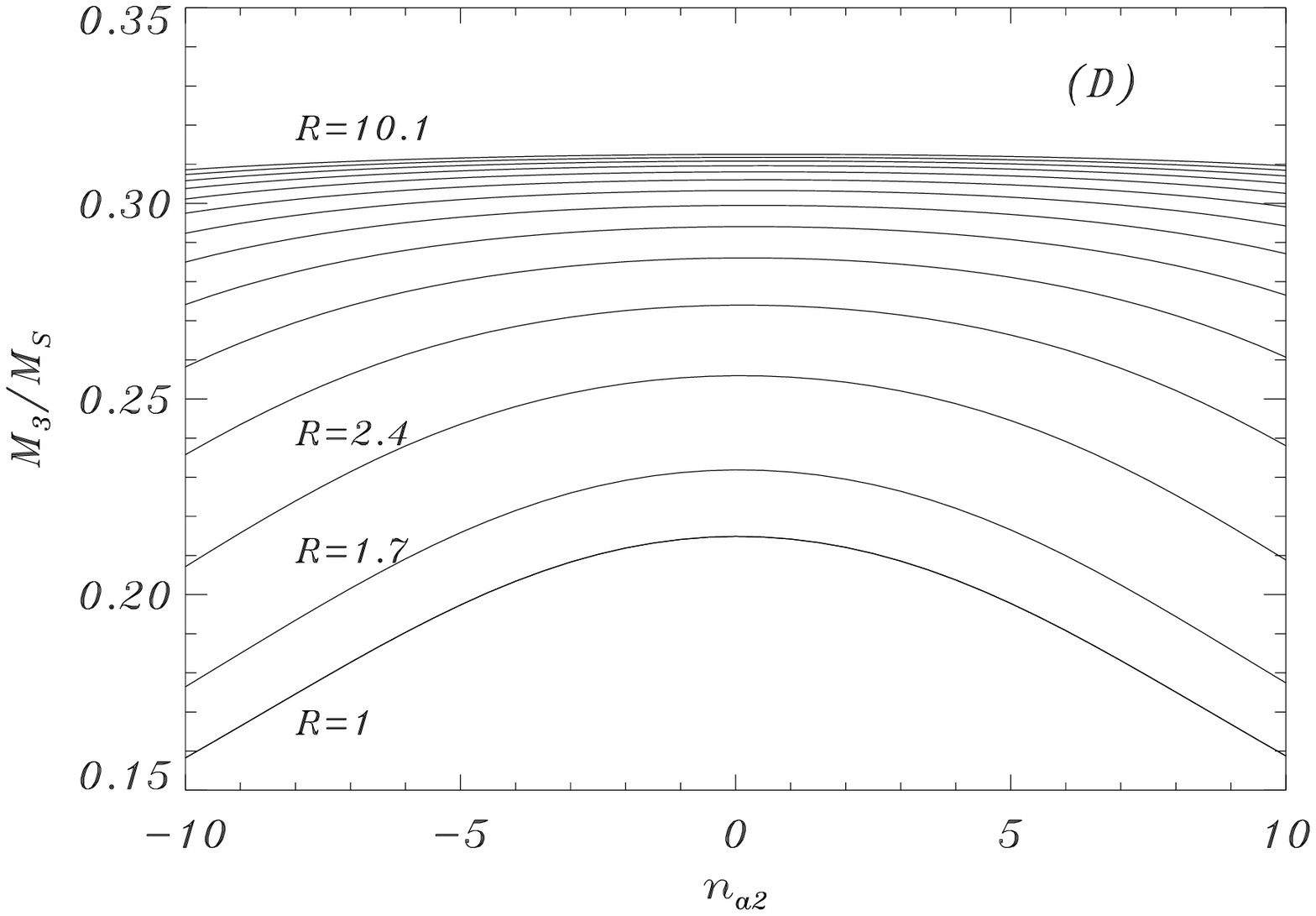,height=6.2cm,width=6.5cm}}  
\end{tabular} 
\begin{tabular}{cc|cr|} 
\parbox{7.1cm}{ 
\psfig{figure=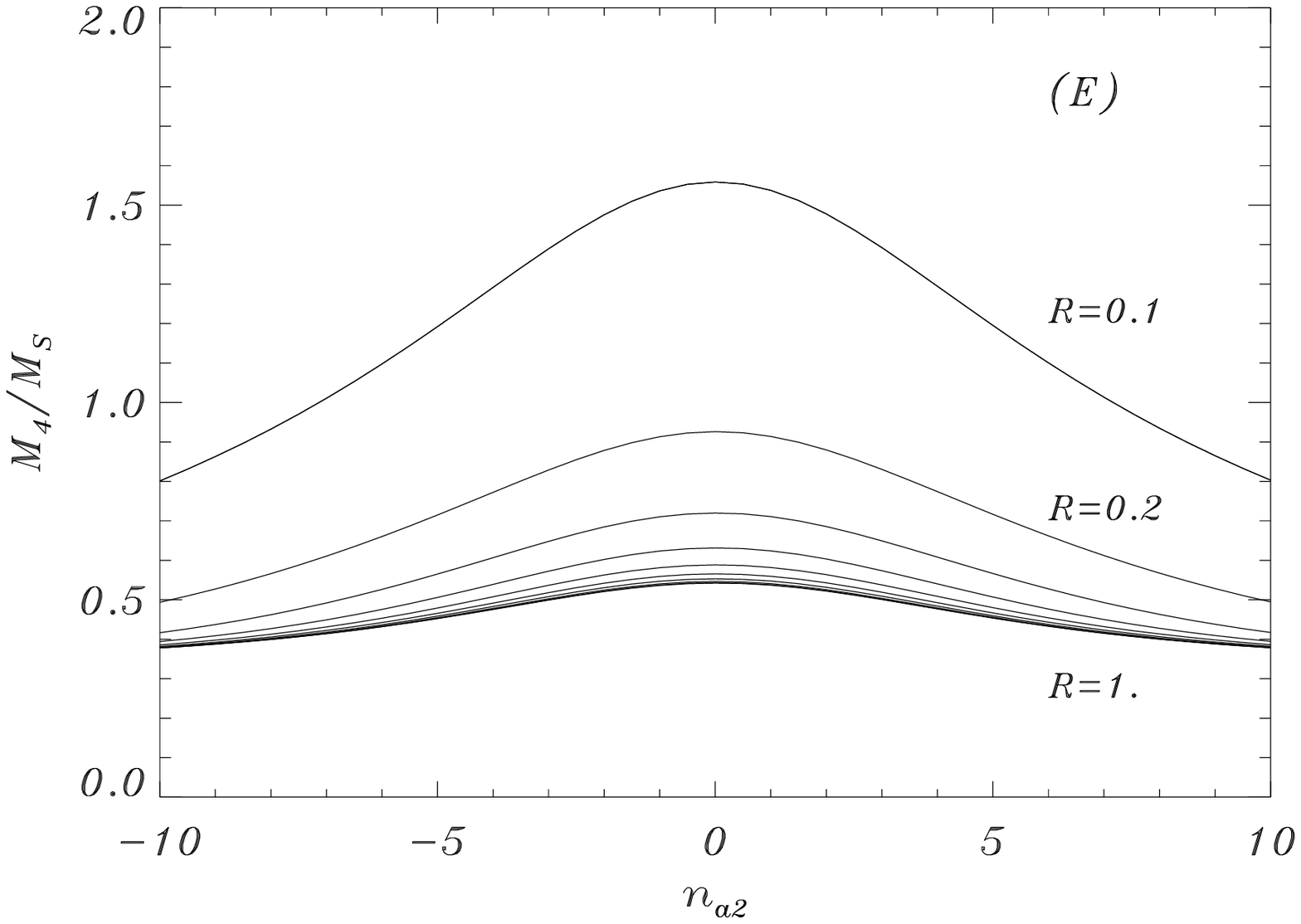,height=6.2cm,width=6.5cm}} 
\hfill{\,} 
\parbox{7.1cm}{ 
\psfig{figure=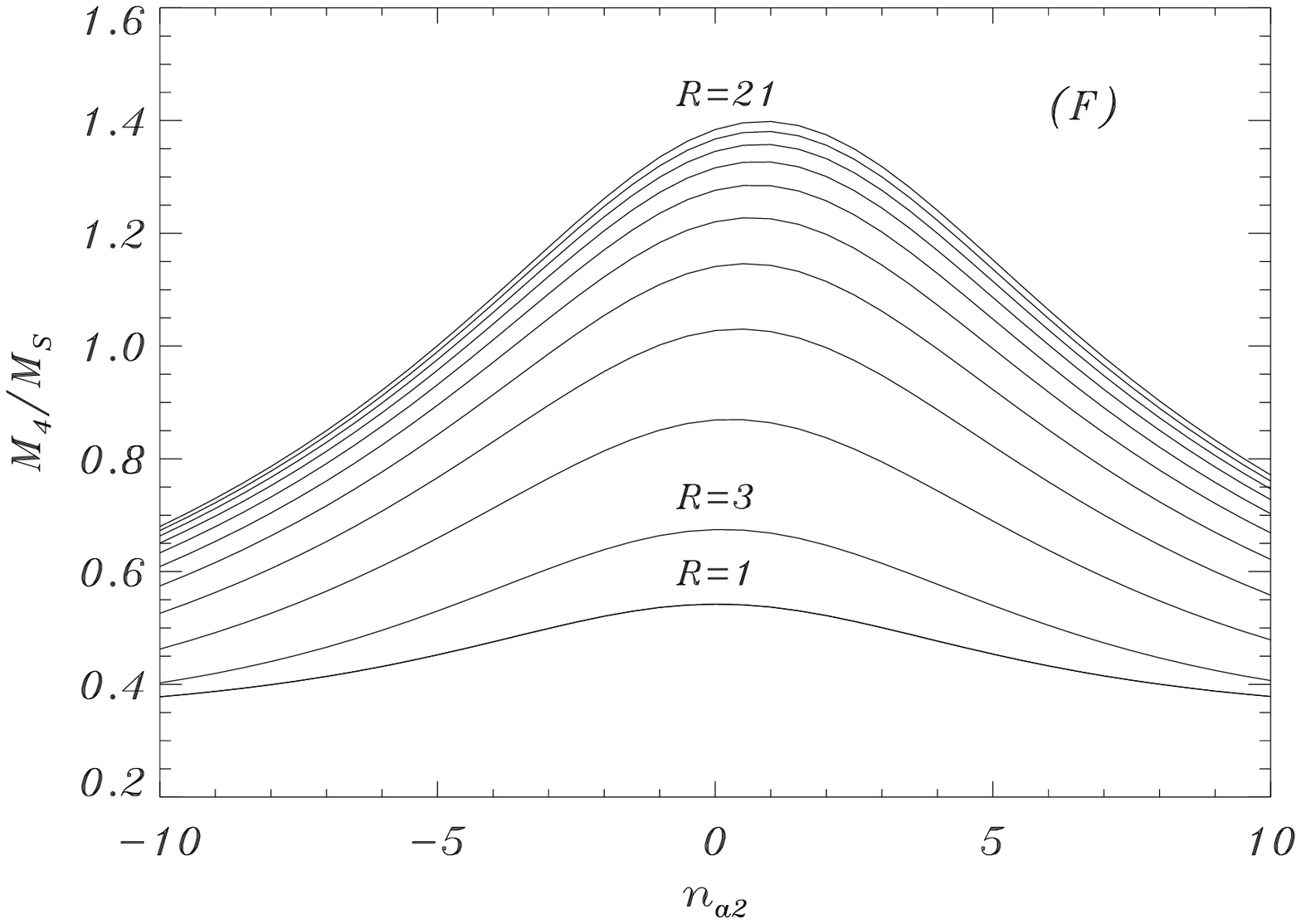,height=6.2cm,width=6.5cm}}  
\end{tabular} 
\caption{\small{D6-brane models: Class A models of Table \ref{minimal}
 (with $\beta_1=1/2$, $\beta_2=1$,
 $\nu=1/3$, $n_{c1}=1$, $n_{b1}=-1$).  $M_{2,3,4}$ (string units) 
in function of $n_{a2}$ for fixed $\cR\equiv g_d/g_c$.
(A): $M_2\!$ for $0.05\!\leq \!\cR\!\leq\! 1$ has small
sensitivity to $\cR$. (B): $M_2$  with $1\!\leq \!\cR\! \leq\! 26$, 
$\cR$ increasing {\it upwards}, 
step 5. (C): $M_3$ with $0.1\! \leq \!\cR\! \leq\! 1$ increasing {\it
downwards}, step 0.1. (D):  $M_3$ with $1\!\leq  \!\cR\! \leq\! 10.1$ increasing
{\it upwards}, step 0.7. (E): $M_4$  with $0.1\!\leq \!\cR\! \leq\! 1.$
increasing {\it downwards} step 0.1. (F): $M_4$ with 
$1\!\leq  \!\cR\! \leq \! 21$ increasing {\it upwards} step 2. 
Electroweak corrections exist for these masses, but they
are significantly suppressed by $M_Z^2/M_S^2$ and are not included here.}}
\label{modela}}
\clearpage
%
%
%
%

\vspace{0.7cm}
\subsubsection{Class B models.}\label{classbBEW}

The parameters of these models are detailed in Table \ref{minimal}.
In this case, $n_{b1}=0$. One also has $\ge=\pm1$,
$\nu=1,1/3$, $\,\beta_1=1$, $\,n_{c1}=\pm 1$, $\beta_2=1,1/2$,
although we keep  all these unassigned, to analyse simultaneously all
possible combinations.
The free parameters of the model are again chosen as $g_d/g_c$ and $n_{a2}$.
Compared to the general case,  the mass matrix $M_{\alpha\beta}^2$ is
significantly simpler (since $n_{b1}=0$). Its  eigenvalues $M_i^2$  are
\begin{eqnarray}\label{eigvalB}
M_1^2&=&0, \nonumber \\
M_2^2&=&(x+y)\, M_S^2,\nonumber \\
M_3^2&=&\left(\frac{2 \beta_1}{\beta_2}\right)^2 g_b^2 \ge^2 M_S^2,
\nonumber \\
M_4^2&=&(x-y)\, M_S^2
\end{eqnarray}
with the notation
\begin{eqnarray}\label{spect}
x&=&\frac{1}{8\beta_1^2 \beta_2^2\nu^2}\left\{
\left[9g_a^2+g_d^2\right]\left[4
\beta_2^4\ge^2+\nu^2\beta_2^2n_{a2}^2\right]+4\beta_1
n_{c1}\nu^2 \left[\beta_1(g_c^2+g_d^2) n_{c1}-\beta_2
g_d^2 n_{a2}\right]\right\} \nonumber \\
y&=&\left\{x^2-\frac{\ge^2 n_{c1}^2}{\beta_1^2\nu^2}
\left[g_c^2 g_d^2+9 g_a^2(g_c^2+g_d^2)\right]\right\}^{1/2}
\end{eqnarray}
The massless eigenvalue corresponds to the observed  hypercharge. 
An interesting consequence
of (\ref{eigvalB}), (\ref{spect}) is that one may in principle  
have a very light state
$M_4$,  without the need for a Higgs mechanism. In fact
\begin{equation}\label{product}
M_2 \, M_4 =\frac{|\ge n_{c1}|}{\beta_1 \nu}
\left[g_c^2 g_d^2+9 g_a^2 \left(g_c^2+g_d^2)\right)\right]^{1/2} M_S^2
\end{equation}
With couplings of  fixed value,  $M_4$ may be significantly 
small(er than the string scale) 
if $M_2$ is made large enough (larger than the string scale) 
by a large choice for the integer (wrapping number) $n_{a2}$.

The (normalised) orthogonal eigenvectors associated with 
the eigenvalues of eq.(\ref{eigvalB}) are 
\begin{eqnarray}\label{eigvecB}
w_{1}&=&\frac{1}{|w_1|} \left\{\frac{g_d}{3 g_a}, 0,
-\frac{g_d}{g_c}, 1\right\}_\alpha,\hspace{4cm}{\alpha=a,b,c,d} \nonumber \\
w_{2}&=&\frac{1}{|w_2|}\left\{\frac{3 g_a}{g_d}\frac{1}{2
\omega_1}\left(\omega_2-8 \beta_1^2\beta_2^2\nu^2 y\right), 0,
\frac{g_c}{g_d}\frac{1}{2\omega_1}\left(\omega_3- 8
\beta_1^2\beta_2^2\nu^2 y\right),1\right\}_\alpha,\nonumber \\
w_{3}&=&\left\{0,1,0,0\right\}_\alpha, \nonumber \\
w_{4}&=&\frac{1}{|w_4|}\left\{\frac{3 g_a}{g_d}\frac{1}{2
\omega_1}\left(\omega_2+8 \beta_1^2\beta_2^2\nu^2 y\right), 0,
\frac{g_c}{g_d}\frac{1}{2\omega_1}\left(\omega_3+ 8
\beta_1^2\beta_2^2\nu^2 y\right),1\right\}_\alpha
\end{eqnarray}
with the norm of the vector
$$|w_i|=\bigg[\sum_{\alpha} (w_{i})_{\alpha}^2\bigg]^{1/2}$$ 
and with
\begin{eqnarray}
\omega_1&=&36 \beta_2^4 \ge^2 g_a^2+\nu^2\left(\beta_2n_{a2}-2 \beta_1
n_{c1}\right)\left(9 \beta_2 g_a^2 n_{a2}+2 \beta_1 n_{c1} g_c^2\right),
 \\
&&\vspace{0.3cm}\hfill\nonumber\\
\omega_2&=&-\left\{\beta_2^2(9 g_a^2-g_d^2)(4 \beta_2^2\ge^2+\nu^2
n_{a2}^2)+4 \nu^2 \beta_1n_{c1}(g_c^2+g_d^2)(n_{a2} \beta_2-\beta_1
n_{c1})\right\}, \nonumber \\
&&\vspace{0.3cm}\hfill\nonumber \\
\omega_3&=&\left\{\beta_2(9 g_a^2+g_d^2)(4 \beta_2^3\ge^2+\nu^2\beta_2
n_{a2}^2-4 \nu^2\beta_1 n_{a2} n_{c1})
+4\nu^2\beta_1^2(g_d^2-g_c^2) n_{c1}^2\right\}\nonumber
\end{eqnarray}
The first eigenvector 
in eq.(\ref{eigvecB}) is that corresponding to the hypercharge.
\FIGURE{\psfig{figure=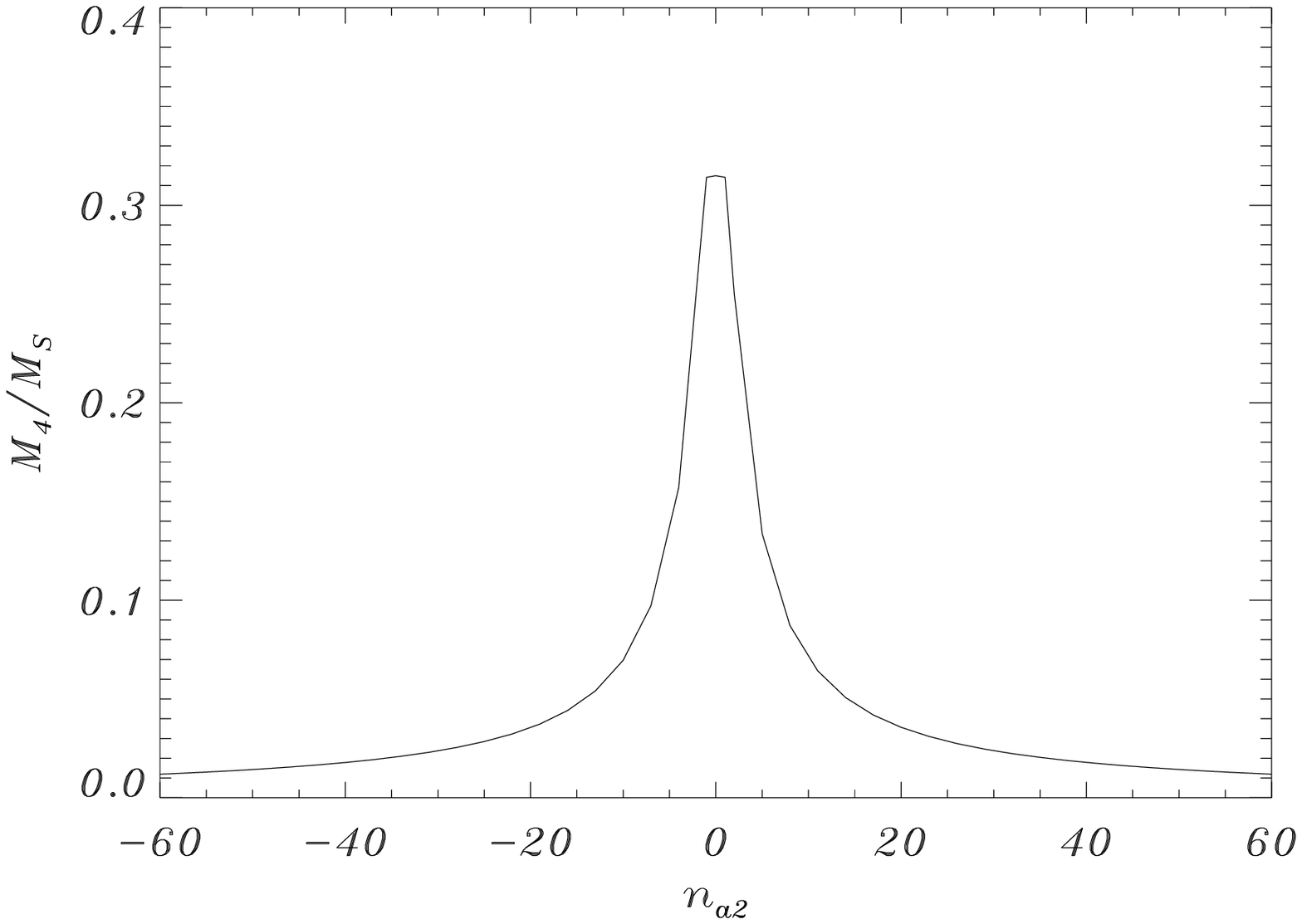,height=6.8cm,width=7.cm}
\caption{\small{
Class B: The curve of minimal values of $M_4$ (string units) in
function of  $n_{a2}$, is realised for $g_d/g_c=1$. $M_4$ decreases 
for larger integer $n_{a2}\approx 60$ 
to $1/100 M_S$. The mass of this state has 
a stringy origin rather than a standard  Higgs mechanism.  It has
corrections from the  usual Higgs mechanism, 
suppressed by the ratio $M_Z^2/M_S^2$. Similar state exists in Class A
models, see lower curves in Fig.\ref{modela}.(C), (D), (E), (F) at
large $n_{a2}$.}}
\label{caseBsingle}}
It is interesting to notice that unlike  Class A models,
U(1)$_b$ (anomalous) does not mix
with the rest of the Abelian fields and remains massive. Of the
remaining  eigenvectors ($w_2$, $w_4$), one linear combination is anomalous 
and one is anomaly free.  The eigenvectors $w_{2,3,4}$ can be used
as in Class A to investigate which original $U(1)_\alpha$ dominates in 
every mass eigenstate (this is done in the last part of this section).

The eigenvectors $w_{i}$   define the  matrix
$\cF_{i\alpha}=(w_{i})_{\alpha}$, with
 $\cF \cF^T=\cF^T \cF=1$ and with $\cF M^2 \cF^T= \delta_{ij}
M_i^2$ where $M^2$ is the initial mass matrix, eq.(\ref{masones}).  
The new $U(1)$ charges, including hypercharge (and 
their normalisation) of the re-defined (mass eigenstates) Abelian 
fields can  be computed as in Class A. 
We  find again  ($g_1'\equiv g_y$)
\begin{equation}\label{hyper1}
\frac{1}{g_1^{'2}}=\frac{|w_1|^2}{4 g_d^2}=\frac{1}{36}\frac{1}{g_a^2}+
\frac{1}{4}\frac{1}{g_c^2}+\frac{1}{4}\frac{1}{g_d^2}
\end{equation}
with similar relations to hold for the remaining $g_i'$ and $q_i'$.
Using (\ref{hyper1}) we can also re-write (\ref{product}) 
in terms of the chosen parameter, the ratio $\cR\equiv g_d/g_c$ 
\begin{equation}
M_2 \, M_4 = 
\frac{|\ge n_{c1}|}{\beta_1 \nu} \frac{54 g_a^3\, M_S^2 }{g_y (36
g_a^2/g_y^2-1)}\left[\cR+\frac{1}{\cR}\right] \nonumber\\
\end{equation}
where $g_a$ is related to QCD coupling, (\ref{caplillos}).
Thus the  string scale is  approximately the geometric mean
of  $M_2$ and $M_4$. This also  shows that at fixed $g_d/g_c$,
$M_4$ may become  very light, even  of the  order of $M_Z$ scale
for (fixed) string scale $M_S\approx \cO(1-10\, TeV)$ (see Figure
\ref{caseBsingle}), without a  Higgs mechanism. (This happens 
if $n_{a2}$ (i.e. $M_2$) is made  large enough). 
This is an interesting mechanism, regardless of the phenomenological 
viability of such a light boson. Alternatively, its (light) mass can 
imply stringent bounds on the string scale by the requirement this
light state be massive enough to avoid current experimental bounds.

As mentioned, the $U(1)_b$ gauge boson with eigenvector $w_{3}$ 
(mass eigenstate $M_3$) does not mix with  the others and one can
check  that is 
lighter than $M_S$, see eq.(\ref{eigvalB}).  
From this eq. for  $\beta_1=\beta_2=1$, together with
$M_3>500-800$ GeV one finds $M_S > (500-800)/g_L$ i.e. 
lower bound of $M_S \approx 800-1350 $ GeV.
Additional higher  (lower) bounds  on $M_S$ are provided 
from the analysis of the remaining mass  eigenstates.

Further  numerical results for Class B models are presented in 
Fig. \ref{modelb}. These figures, (although somewhat generic for all
models of this class) correspond to the first example of
class B in Table \ref{minimal} (fifth row)
defined by $\beta_1=1$, $\nu=1$, $\beta_2=1$, $n_{c1}=1$.
The figures present the dependence of the eigenvalues $M_4$
and $M_2$ in string units in function of $n_{a2}$ and 
$g_d/g_c$. Generically, the masses are within a
factor of $\cO(10)$ from the string scale, either larger or smaller.
With bounds on the gauge bosons of 500-800 GeV, this implies  
$M_S> 5-8$ TeV. However, for large $n_{a2}$, the bound on $M_S$ can further
increase. (This is the case of Figure \ref{caseBsingle}
showing the value of $M_4/M_S$ in function of $n_{a2}$
for $g_d=g_c$, yielding the minimal values for $M_4/M_S$ which 
together with the constraint $M_4>500-800$ GeV, give the highest
(lower) bound for  $M_S$). Electroweak corrections to $M_{2,3,4}$
are suppressed by $M_Z^2/M_S^2$ and small, even for low string scale.
Additional bounds on  the string scale can be
achieved by constraining the correction to $M_Z$ due to new
physics (U(1)'s) effects be within the current experimental error on the
$\rho$ parameter,  and this is the purpose of Section \ref{EWB}.

Finally, we analyse the amount of
mixing of the initial $U(1)_\alpha$ ($\alpha=a,b,c,d$), which is
encoded by the eigenvector components  in (\ref{eigvecB}). 
As for Class A, we do not address every specific model of 
Class B, but rather consider the pattern that emerges after investigating 
each of them separately.  For class B models we thus address
the ``amount'' $w_{i\alpha}$ $(\alpha=a,c,d)$
of original $U(1)_\alpha$  present in each mass eigenstate 
other than the $U(1)_b$ which does not mix. 
For the hypercharge
state, the amount of initial $U(1)_\alpha$ is somewhat clear, 
see $w_{1\alpha}$, and only depends on the parameter $g_d/g_c$.
For the remaining $w_{2\alpha}$ and $w_{4\alpha}$, their 
dependence  on $n_{a2}$ for fixed ratio $g_d/g_c$ 
(with correlation (\ref{hyper1})  to {\it fixed} $g_y$ included)
is stronger than in Class A models. Still, one can say that 
in $M_4$ of Figure \ref{modelb} (E) upper curves are dominated by $U(1)_a$ and
$U(1)_c$ while lower curves correspond to states  
with contributions from $U(1)_c$, $U(1)_d$ and $U(1)_a$. If
$g_d/g_c>1$ Figure \ref{modelb} (F), lower curves correspond to 
$U(1)_c$, $U(1)_d$ and $U(1)_a$, while upper curves are dominated by
$U(1)_a$ and with a less  but comparable ``amount'' of  
 $U(1)_d$. The cases of Figures \ref{modelb} (E) and (F) 
have a correspondent in Class A models in Figures \ref{modela} (E) and
(F) as far as the size of the mass is concerned.
For the plots in Figure \ref{modelb} (A) and (B) the situation is
to some extent similar to the case of $M_2$ of Figure \ref{modela}(A)
and (B).
Lower curves for $M_2$ with  $g_d/g_c\leq 1$ are dominated by $U(1)_a$ and $U(1)_d$
(to a less extent) while upper curves by $U(1)_{a,c,d}$ in a comparable ``amount''. 
In Fig.\ref{modelb} (B) lower curves are dominated by $U(1)_a$ (and
some less amount of $U(1)_d$) while the 
upper curves   contain a comparable amount of $U(1)_{a,d}$.

As in Class A models,  $U(1)_a$ is mostly present in cases 
with masses larger than the string scale.  $U(1)_d$ is manifest in 
all lower curves with $g_d \approx g_c$ and $U(1)_b$ is present 
un-mixed, as a mass eigenstate with $M_3$ smaller than the string 
scale. The analysis shows  patterns similar to Class A models.

\cleardoublepage
\FIGURE[p]{
\begin{tabular}{cc|cr|} 
\parbox{7.1cm}{ 
\psfig{figure=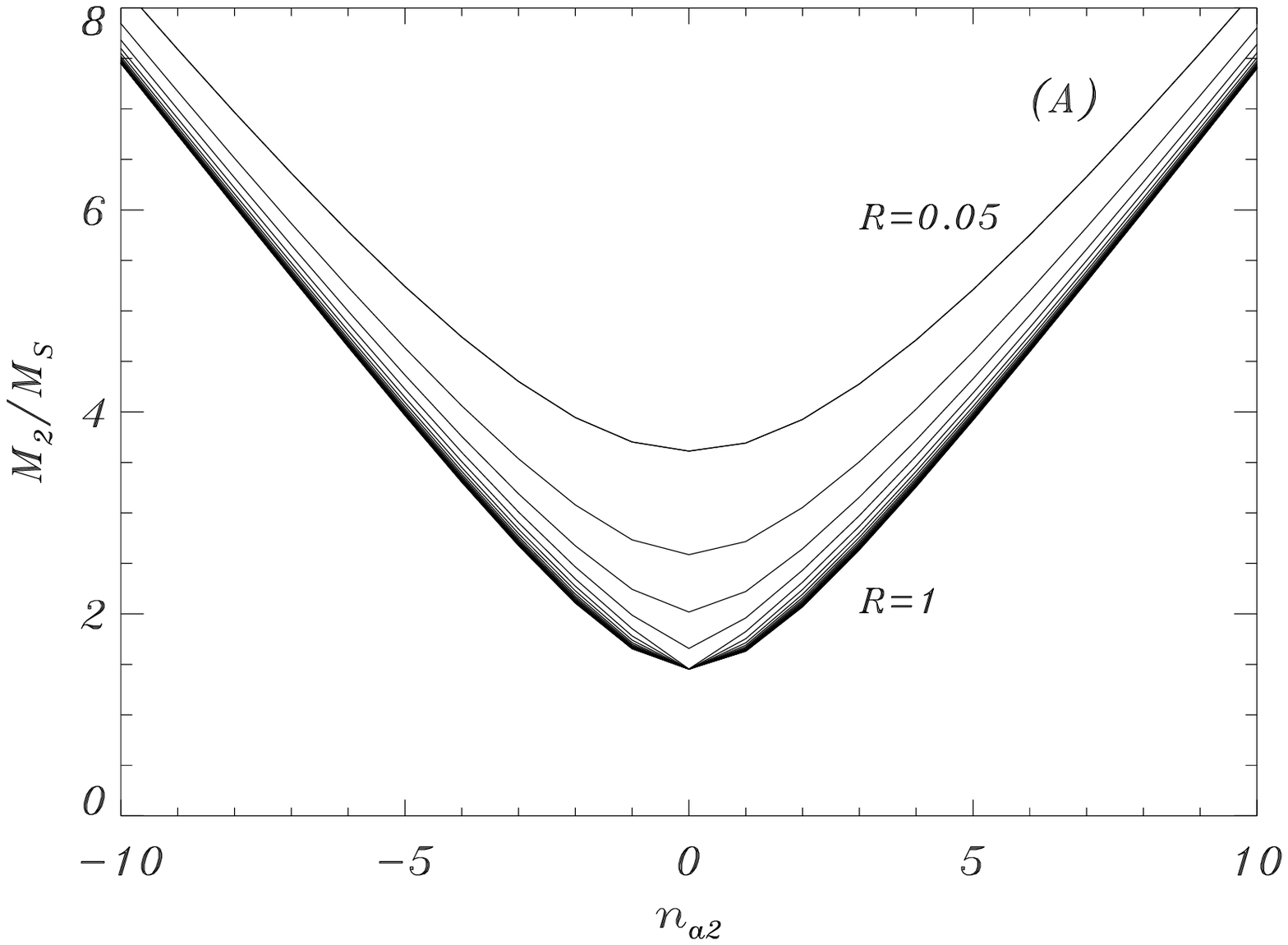,height=6.2cm,width=6.5cm}} 
\hfill{\,} 
\parbox{7.1cm}{ 
\psfig{figure=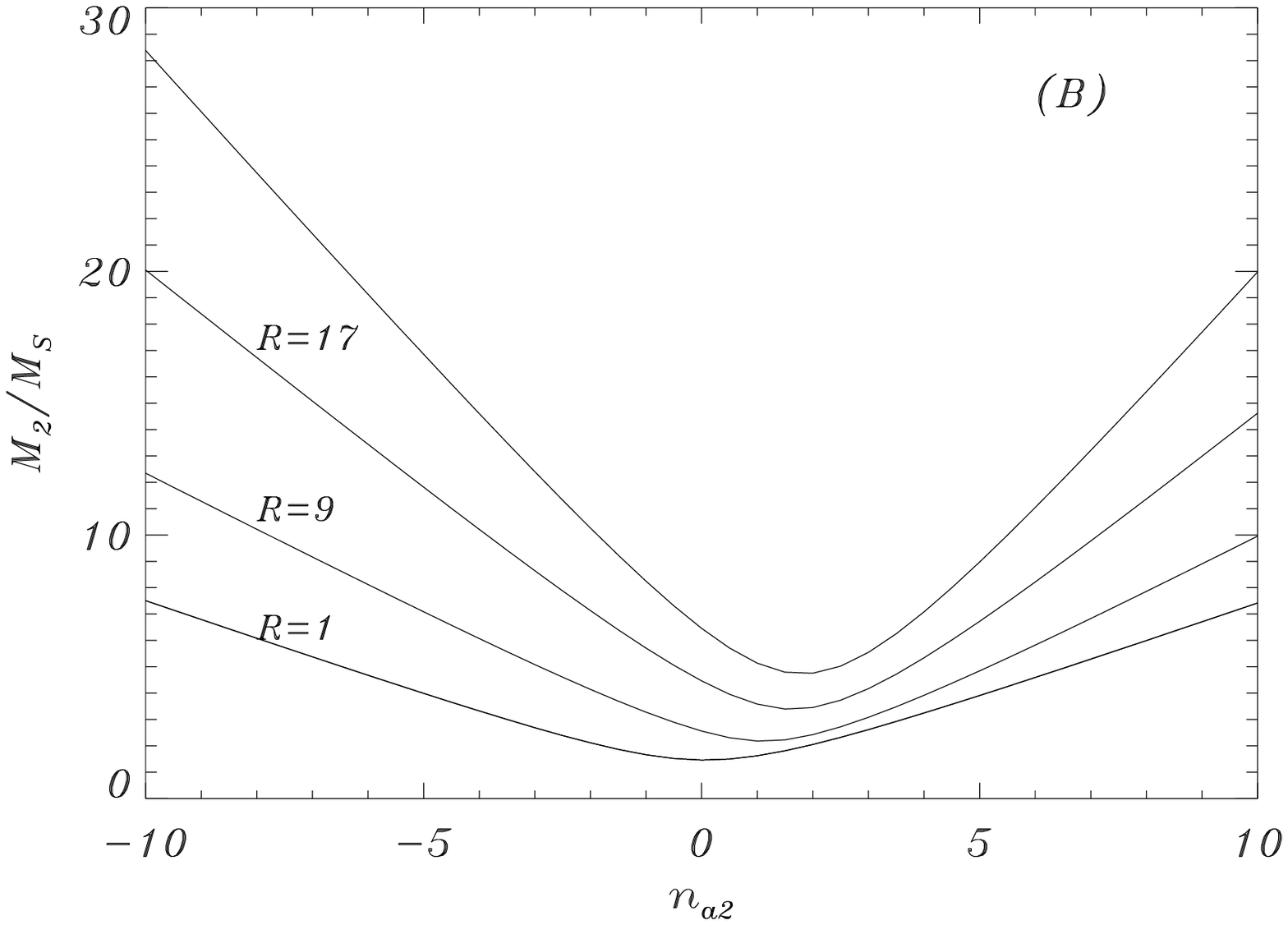,height=6.2cm,width=6.5cm}}  
\end{tabular} 
\begin{tabular}{cc|cr|} 
\parbox{7.1cm}{ 
\psfig{figure=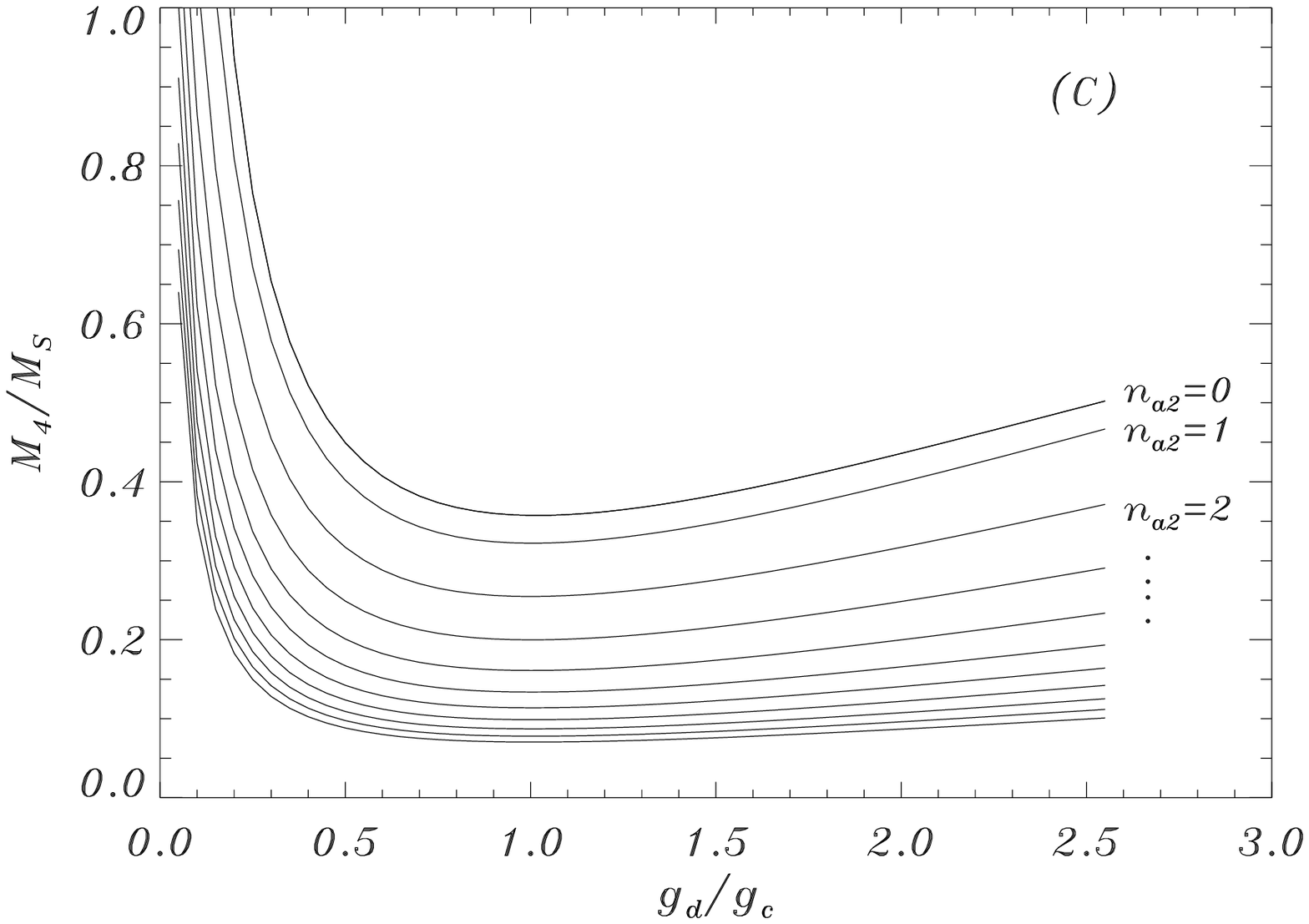,height=6.2cm,width=6.5cm}} 
\hfill{\,} 
\parbox{7.1cm}{ 
\psfig{figure=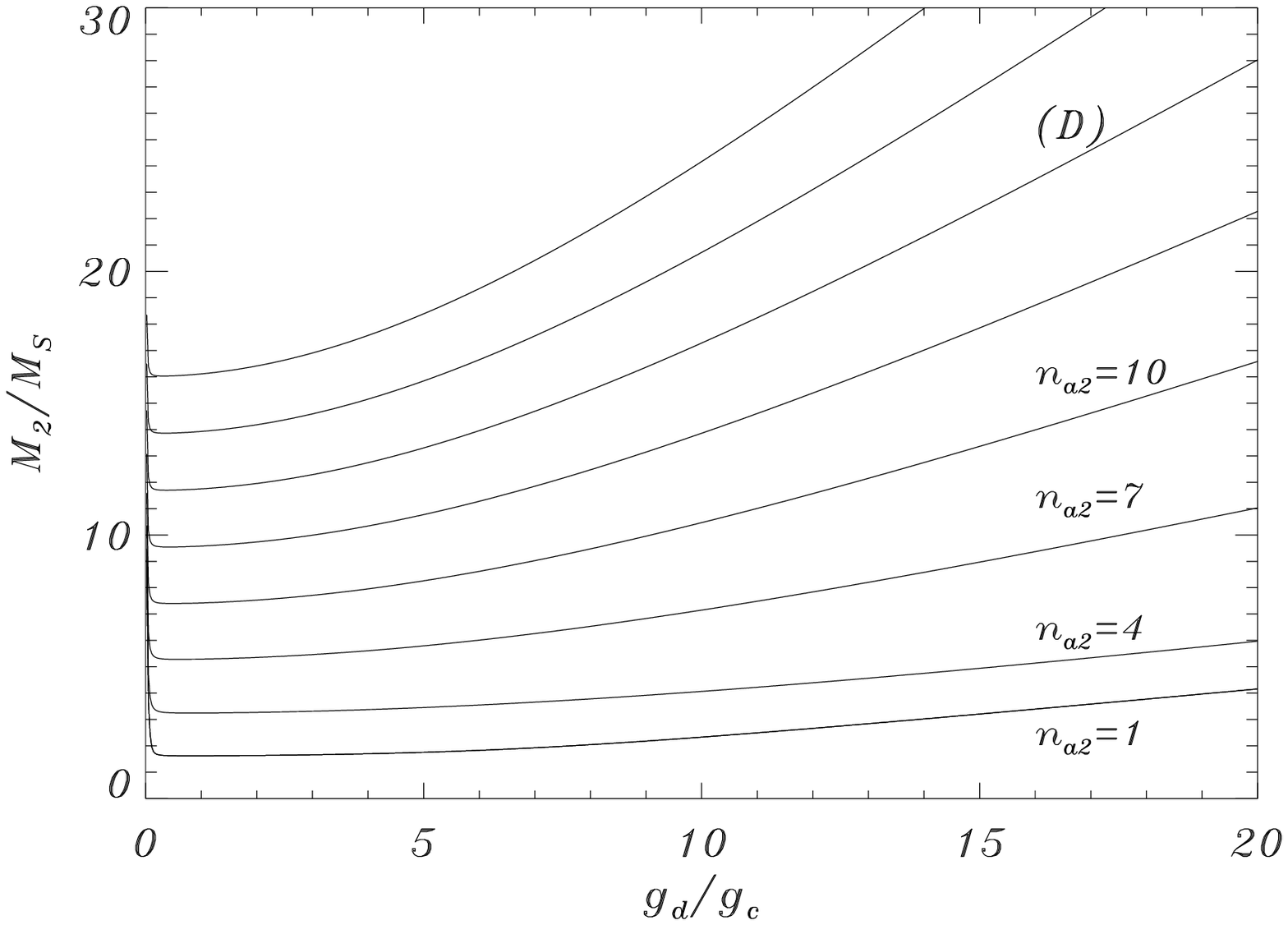,height=6.2cm,width=6.5cm}} 
\end{tabular} 
\begin{tabular}{cc|cr|} 
\parbox{7.1cm}{ 
\psfig{figure=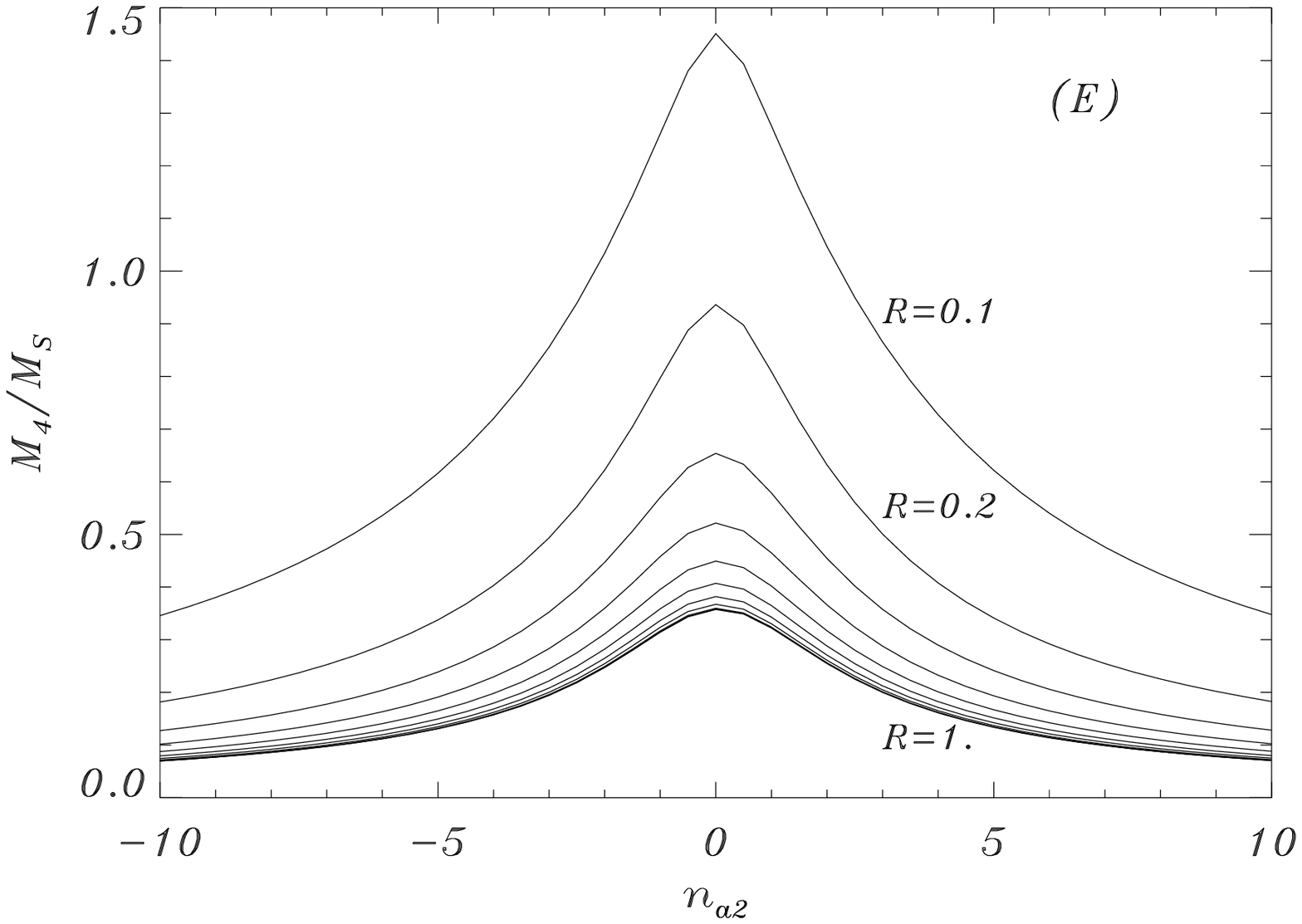,height=6.2cm,width=6.5cm}} 
\hfill{\,} 
\parbox{7.1cm}{ 
\psfig{figure=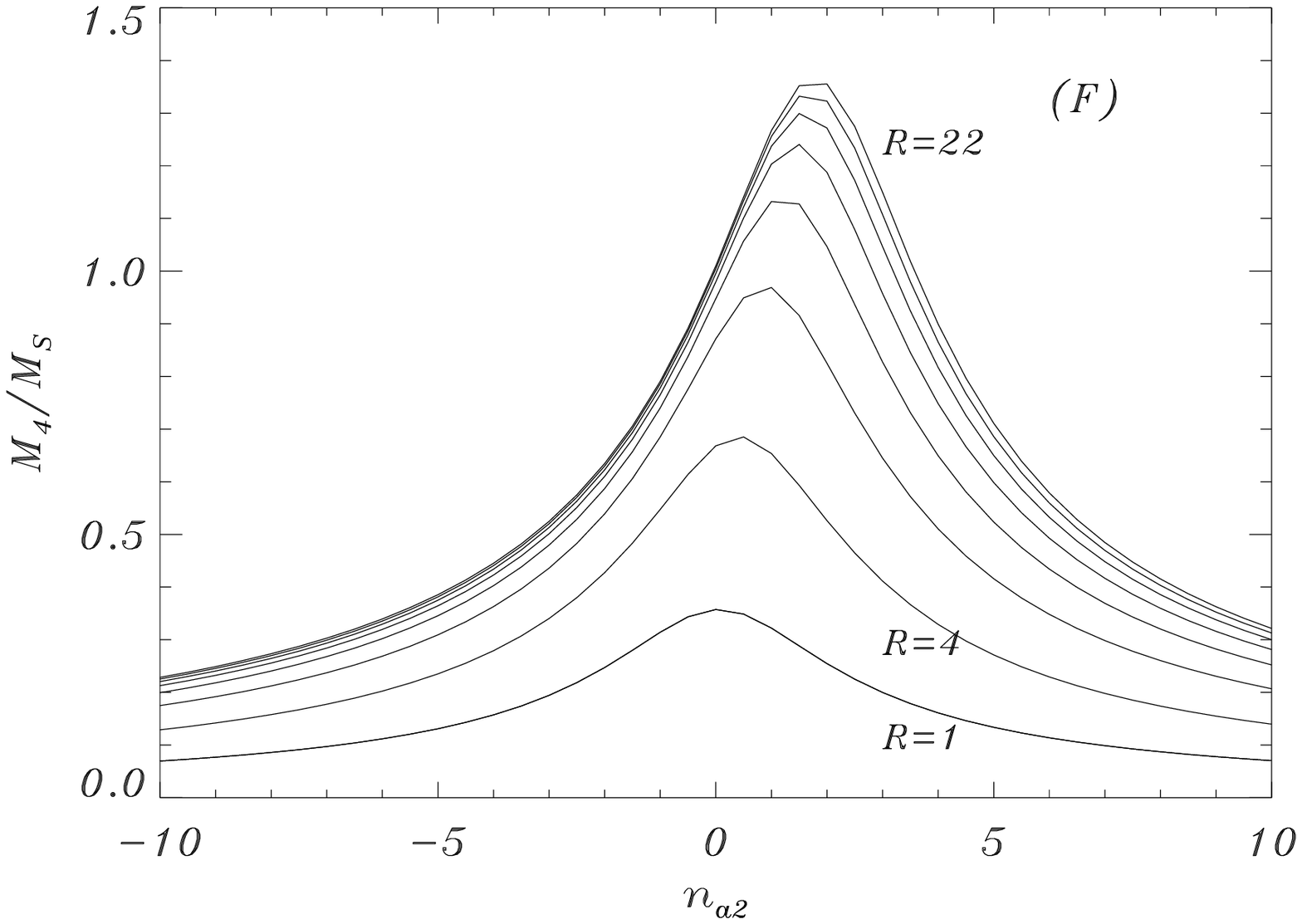,height=6.2cm,width=6.5cm}}  
\end{tabular} 
\caption{\small D6-brane models: Class B models of Table \ref{minimal} 
(with $\beta_1=\beta_2=1$, $\nu=1$, $n_{c1}=1$).  $M_{2,4}$ (string units), in
function of $n_{a2}$ for fixed  $\cR\equiv g_d/g_c$ (A,B,E,F), or in
function  of $\cR$ for fixed  $n_{a2}$ (C,D).
(A): $M_2$    for   $0.05\!\leq \! \cR\! \leq \!1$ increasing {\it downwards} 
step 0.02. 
(B): $M_2$   for   $1\!\leq  \! \cR\! \leq \!25$ 
increasing {\it upwards} step 8.
(C): $M_4$  in function of $\cR$,  $n_{a2}$  increases {\it downwards}
 step 1. 
(D): $M_2$  in function of $\cR$ for $n_{a2}$  increasing {\it upwards}  step 3.
(E): $M_4$   for   $0.1\!\leq \!\cR\! \leq\! 1$ increasing {\it downwards}  step 0.1.
(F): $M_4$  for  $1\!\leq \! \cR\!\leq \!22$ increasing {\it upwards}  step 3.}
\label{modelb}}
\clearpage

\vspace{0.7cm}
\subsection{D5-brane models. Masses of $U(1)$ fields and 
bounds on $M_S$.} \label{D5BEW}

Explicit D5 brane models that we briefly address in this section are defined 
by the mass matrix of eq.(\ref{masones}), with  coefficients $c_i^\alpha$
presented in eq.(\ref{cillos5}). In this case an analysis similar to that for D6
brane models applies.  In the following we choose the simple  example  of
eqs.(\ref{cillos55})  which is rather representative of the SM's 
considered in ref.\cite{cim3}  and 
has only one parameter, the ratio $g_d/g_c$. 
This allows  us to make more definite predictions for the value 
of the string scale or boson masses. 
The mass eigenvalues are as usual  the roots of the equation
\begin{equation}
\gl^4+c_3 \gl^3 +c_2 \gl^2 +c_1 \gl =0
\end{equation}
with coefficients
\begin{eqnarray}\label{ciD5}
c_1& =& -162\,\sqrt3\,g_b^2\,[g_c^2\,g_d^2 + 9\,g_a^2\,(g_c^2 + g_d^2)]<0\nonumber\\
c_2& = & 9\,[3\,g_c^2\,g_d^2 + g_b^2\,(g_c^2 + 43\,g_d^2) + 
     9\,g_a^2\,(40\,g_b^2 + 3\,(g_c^2 + g_d^2))]>0\nonumber\\
c_3& =& -\sqrt3\,(333\,g_a^2 + 13\,g_b^2 + g_c^2 + 40\,g_d^2)/2<0
\end{eqnarray}
With these coefficients and (\ref{eigval4}) one finds the
 (real, positive)  eigenvalues
for this case. 
\FIGURE[b]{
\psfig{figure=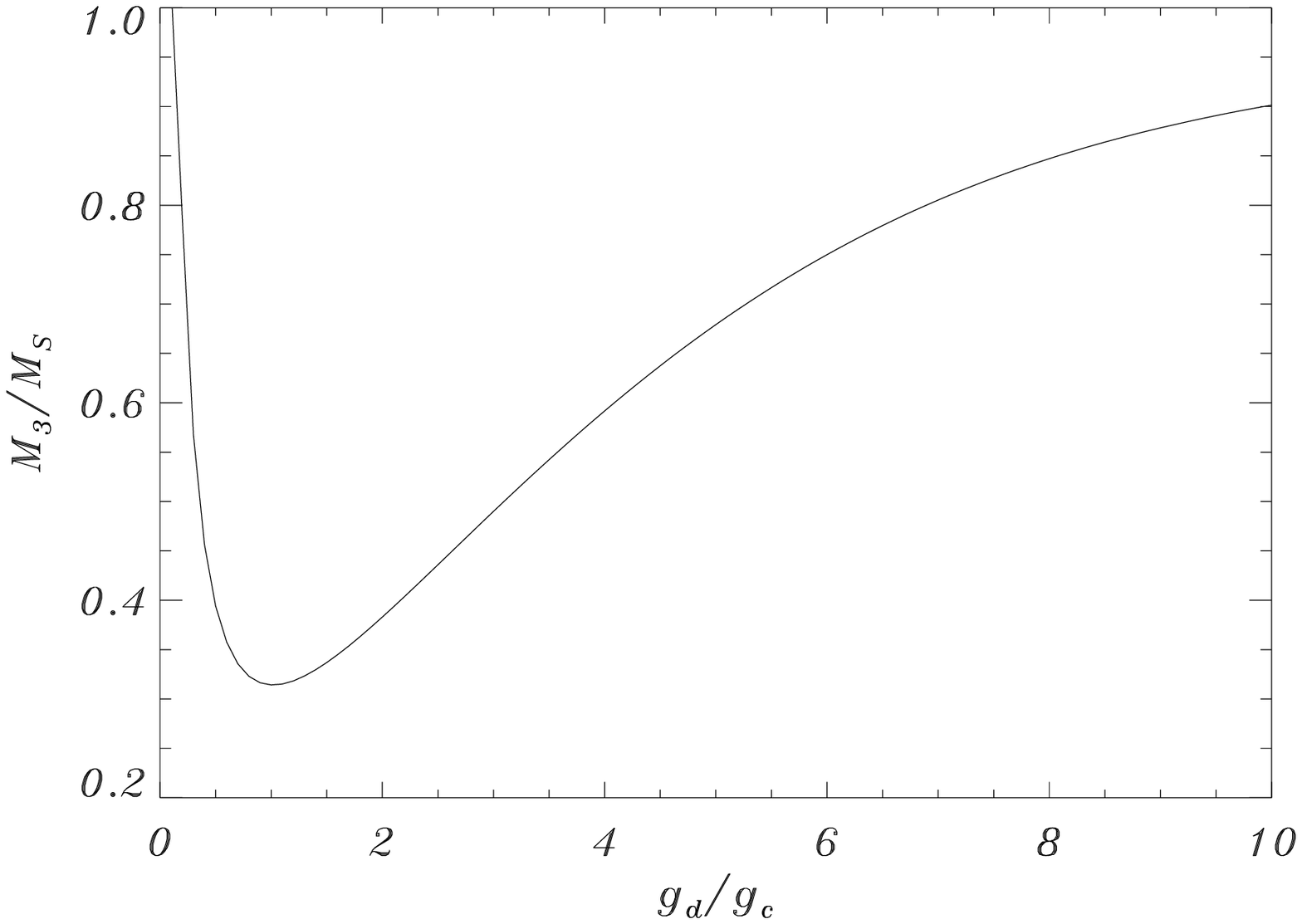,height=6.8cm,width=7.cm}
\caption{\small{The lightest mass $M_3$ (string units) of the D5 brane
model addressed in section \ref{D5BEW} in function of  
$g_d/g_c$ is reached when $g_d\approx g_c$. The remaining 
(non-vanishing) mass eigenvalues (not shown) are $M_2\approx
(8-13.5)\times M_S$ and  $M_4\approx M_S$ (for same range of $g_d/g_c$).}}
\label{d5mass}}
One root is again massless, 
$\lambda_1=0$ corresponding  to  the hypercharge field $U(1)_Y$.
Its associate eigenvector is identical  to the corresponding 
one of the D6 brane models (Class A  and B) 
and leads to the same value for the hypercharge coupling $g_y$ 
in terms of  $g_\alpha$, ($\alpha=a,..,d$). Hence
\begin{eqnarray}\label{hyperD5}
\frac{1}{g_y^2}=\frac{1}{36}\frac{1}{g_a^2}+\frac{1}{4}\frac{1}{g_c^2}+
\frac{1}{4}\frac{1}{g_d^2}
\end{eqnarray}
This  value of $g_y$
is compatible with the  hypercharges
of the SM fields as given in Table \ref{tabpssm}. The constraint
(\ref{hyperD5}) is again  used for evaluating the mass spectrum of the U(1)
fields.  Here we only discuss  the lightest of  these states which is 
$M_3=\sqrt\lambda_3 M_S$  in terms of the ratio $g_d/g_c$.
From Figure \ref{d5mass} we note the smallest value of $M_3$ 
(in string units) exists when the couplings $g_d$ and $g_c$ have 
close values,  and is $M_3\approx 0.3 M_S$. This value has further
corrections from electroweak symmetry 
breaking, but these are significantly suppressed (by the
factor $M_Z^2/M_S^2$) even for a low string scale and are neglected in
this discussion. The remaining mass
eigenvalues  (not presented) have the 
values: $M_2\approx 8 M_S$ for $g_d\approx 0.1 g_c$ with mild variation
(increase) up to $M_2\approx 13.5 M_S$ for $g_d=10 g_c$. Also  
$M_4\approx M_S$  for most values of  $g_d/g_c$ between (0.05,20).  

Generic constraints  on the masses of additional bosons  
of about 500-800 GeV give the strongest constraints on the value of $M_S$ 
when they are 
 applied to the lightest (in string units) state $M_3$ rather than to (larger)
$M_{2,4}$.    Taking into account 
the minimal value for $M_3\approx 0.3 M_S$,  one predicts 
$M_S \geq 1.5-2.4$ TeV. This lower bound on $M_S$  is rather 
similar (although somewhat smaller)
to the previous case 
 of D6 brane models that we addressed.
The effects of electroweak symmetry breaking in this D5 model
and consequences for $M_S$ values are addressed in
Section \ref{D5AEW}.

\vspace{0.7cm}
\section{Constraints on explicit D6 and D5 
models from the $\rho $-parameter. \label{EWB}}

We have so far  investigated the masses of the additional $U(1)$
fields in the absence of electroweak symmetry breaking.    
At low energies, the effects of electroweak scale physics (EW) become  
important and the action term 
below due to (\ref{masones}) receives further corrections. 
Such corrections induced onto the masses of the extra $U(1)$'s 
of the previous section  may be computed using the procedure outlined 
in Section \ref{generalmethodAEW}, eq.(\ref{aew}),
 as a perturbative expansion in $\eta$  about 
their values in the  absence of EWS breaking. Since these effects are 
relatively small, they  will not be addressed. In this section we
instead compute the corrections induced on the mass of $Z$ boson 
by the presence of the additional $U(1)$'s, as outlined in Section
\ref{generalmethodAEW}. Such corrections, induced 
by the mixing of the Higgs state with some of the initial $U(1)$'s
(massive) are of stringy origin  since  the masses of the additional 
$U(1)$  are themselves of this nature. Therefore the corrections to 
$M_Z$ we compute are {\it not} due to a Higgs mechanism. 
Regardless of the phenomenological viability of our models, we find
this mechanism very interesting  and worth further attention.
Finally, we  use the corrections to $M_Z$ that we compute  for D6 
(Class A and B) and D5 brane models  to impose (lower) bounds on the 
value of $M_S$, by making use of the  current constraints on the value 
of  $\rho$ parameter.

We start with the following term in the action
\begin{equation}\label{massterm}
\cL_1=\frac{1}{2}\sum_{\alpha,\beta} M^2_{\alpha\beta} A_\alpha A_\beta\equiv
  \frac{1}{2}\sum_{\alpha,\beta} 
\left[g_\alpha g_\beta M_S^2 \sum_{i=1}^{3} c_i^\alpha
c_i^\beta\right] \, A_\alpha A_\beta,\,\,\,\,\,\,\,
\alpha,\beta=a,b,c,d.
\end{equation}
Including electroweak symmetry breaking effects on the masses of 
$U(1)$ fields requires a careful examination of the Higgs sector.
We remind the reader that the Higgs sector is charged
under both $U(1)_b$ and $U(1)_c$, as presented in Table
\ref{thirdone}, thus $A_\alpha$ ($\alpha=b,c$)  
fields   will have couplings at the electroweak scale to 
$\sigma=H_i$ or $h_i$ and also to $W_3^\mu$. 
All the couplings of $U(1)_\alpha$ fields  
can be read off from  the following additional term in the Lagrangian
\begin{eqnarray}
\cL_2&= & (D^\mu \sigma)^\dagger (D_\mu \sigma),\\
D^\mu \sigma&=& 
\left[\partial^\mu +i g_L {\vec T}.{\vec W}^\mu +i \sum_{\alpha=a,b,c,d}
g_\alpha q_\alpha A_\alpha^\mu\right] \sigma
= \left[\partial^\mu +i g_L {\vec T}.{\vec W}^\mu +i \sum_{i=1}^{4}
g'_i q'_i A_{i}^{'\mu}\right] \sigma \nonumber 
\end{eqnarray}
In the last step, following the previous approach,
we made explicit the change of basis performed in previous section.
The field $\sigma$ stands for any of the Higgs fields in 
Table \ref{thirdone}. The above equation also shows
 the relative  normalisation of the couplings (also $g_L=2 g_b$) in  the 
$SU(2)_L$ and $U(1)$ sectors, in agreement with $U(1)_\alpha$ charge
normalisations/assignments  in  Tables \ref{tabpssm}, \ref{thirdone}.

The mass eigenvalue problem for  $U(1)$ fields with electroweak symmetry breaking 
reduces to diagonalising the  $5\! \times\! 5$  mass matrix
$\cM^2_{\gamma\gamma'}$ with $\gamma,\gamma'=\{a,b,c,d,e\}$ which we
present below in the basis of $U(1)_\alpha$ ($\alpha=a,..,d$) and $W_3^\mu$. 
We denote by $v_i/\sqrt{2}$ (i=1,2) the v.e.v. of 
a Higgs field $\sigma_i$ with $\sigma_i=H_i$ (Class A, $n_H=1, n_h=0$) 
or $\sigma_i=h_i$  (Class A, $n_H=0, n_h=1$). 
For Class B   the mass matrix can again
be read from the one written below. We  have
\begin{eqnarray}\label{5dim}
\cM^2_{\gamma\gamma'}&=&M^2_{\gamma\gamma'}+\gD_{\gamma\gamma'},
\hspace{1cm} \gamma,\gamma'=a,b,c,d,e \nonumber \\
M^2_{e\gamma}&=&M^2_{\gamma e}\, = \, 0, \hspace{1.8cm} \gamma=a,b,c,d,e
\end{eqnarray}
with $M_{\alpha\beta}^2$  ($\alpha,\beta=a,...,d$) as in (\ref{masones}) and with
\begin{eqnarray}
\Delta_{bb}&=& 
g_b^2 \left[q^2_{b_{\sigma_1}} v_1^2+q^2_{b_{\sigma_2}} v_2^2\right]=g_b^2\,
\left(v_1^2+v_2^2\right),\hspace{2cm}<\phi>^2=v_1^2+v_2^2. \nonumber \\
\Delta_{bc}&=& \Delta_{cb}\,=\, 
 g_b g_c \left[ q_{b_{\sigma_1}} q_{c_{\sigma_1}}v_1^2 + q_{b_{\sigma_2}}
 q_{c_{\sigma_2}} v_2^2\right]
= \gd\, g_b g_c\, (v_1^2+v_2^2), \nonumber \\
\Delta_{cc}&=&g_c^2\left[ q_{c_{\sigma_1}}^2 v_1^2+q_{c_{\sigma_2}}^2
v_2^2\right]=g_c^2 (v_1^2+v_2^2)  \nonumber \\
\Delta_{be}&=&\Delta_{eb}\,=\, 
 g_b g_L \left[ T_3 (\sigma_1)\, q_{b_{\sigma_1}} v_1^2+T_3(\sigma_2)\,
q_{b_{\sigma_2}} v_2^2\right]=- \gd\, g_b g_L/2  \,(v_1^2+v_2^2),
\nonumber \\
\Delta_{ce}&=&\Delta_{ec}\,=\,
 g_c g_L \left[ T_3 (\sigma_1)\, q_{c_{\sigma_1}} v_1^2+T_3(\sigma_2)\,
q_{c_{\sigma_2}} v_2^2\right]=\,\,\,\,\,\,- g_c g_L/2 \, (v_1^2+v_2^2),
\nonumber \\
\Delta_{ee}&=& g_L^2\left[ T_3^2(\sigma_1)\, v_1^2+T_3^2(\sigma_2)\, 
v_2^2\right]\,=\, g_L^2/4\, (v_1^2+v_2^2).\label{5dime}
\end{eqnarray}
with $\gd=+1$ if $\sigma_i=H_i$ (Class A, $n_H=1, n_h=0$) and 
$\gd=-1$ if $\sigma_i=h_i$ (Class A, $n_H=0, n_h=1$) respectively. 
Class B of D6 models has  a Higgs sector which is 
a superposition of both possibilities above, with appropriate
choice for the v.e.v. of the Higgs fields involved.
For Class B one formally replaces
$\delta\rightarrow \cos(2\theta)$ with $\theta$ the mixing between
$h_i$ and $H_i$ sectors, to be detailed later (formally
$\theta=0,\pi/2$ for Class A).  For the D5 brane model discussed
before, the electroweak sector is that of class A models with $\delta=+1$.
 All matrices ($\cM^2, \gD$)  are symmetric, with remaining 
entries of $\gD$ not defined above,  equal to zero.
The mass eigenvalue equation for $\cM^2$ gives (see also (\ref{char5}))
\begin{equation}\label{soll}
\gl(\gl^4+o_3 \gl^3 +o_2 \gl^2 +o_1 \gl+o_0) =0
\end{equation}
Thus  the matrix $\cM^2_{\gamma\gamma'}$ has a zero mass eigenvalue 
corresponding to the
photon eigenstate. Its eigenvector is a simple  extension of that of
hypercharge eigenvector $w_1$ given in eqs. (\ref{eigvecA}) and (\ref{eigvecB}):
\begin{equation}
e_1=\frac{1}{|e_1|}\left\{\frac{g_d}{3 g_a}, 0, -\frac{g_d}{g_c},1, 
-\frac{g_d}{g_b}\right\}_{\alpha}
\end{equation}
This confirms that the anomalous $U(1)_b$ field is not part of the photon
eigenstate, as expected.

We now compute the root of eq.(\ref{soll}) corresponding to $M_Z$.
The remaining roots which are electroweak corrections to  (stringy) masses of $U(1)$
fields of previous sections can be computed in exactly the same way.  
Although the method we use to computing $M_Z$   is similar to D6 (Class A and B) 
and D5 brane models as well, we consider  these cases  
separately. This is done  because  the Higgs sector is different
and because it is our intention to compare separately the 
results of D6, D5
models with their corresponding cases before  EWS breaking.


\vspace{1.1cm}
\subsection{D6-brane models. 
Stringy Corrections  to $M_Z$ and Bounds on $M_S$.} 
\vspace{0.2cm}
\subsubsection{Class A models.} 
For Class A models of Table \ref{minimal} one may 
compute the roots of eq.(\ref{soll}) explicitly. 
Analytic formulae for these
exist, but they are long and not very enlightening.
Instead one can compute these solutions  as expansions in 
$\eta$ about their values in the absence of the electroweak symmetry
breaking.  We take  account that  the 
coefficients $o_0,\,o_1,\,o_2,\,o_3$ of (\ref{soll})  contain
corrections linear in $\eta$ in addition to their values 
in the absence of the  EWS breaking (represented by 
the coefficients in eqs.(\ref{c3})). In 
Class A models the coefficients   $o_i$ are
\begin{eqnarray}
o_0&=&\eta\, s_0,\qquad \eta \equiv <\phi>^2/M_S^2\nonumber\\
o_i&=&c_i+ \eta\, s_i \qquad i=1,2,3.\label{coef2}
\end{eqnarray}
$s_0,s_i$ are independent on $<\phi>$ ($o_i\rightarrow c_i$ 
if $<\phi>\rightarrow 0$).
\begin{eqnarray}
s_0&=&\frac{4\, {\ge^4}\,{g_b^2}\,{n_{c1}^2}}{\beta_2^2\,\nu^2}
\left\{ {g_b^2}\,{g_c^2}\,
         {g_d^2} + 9\,{g_a^2}\, \left[ {g_c^2}\,{g_d^2} + 
         {g_b^2}\, \left( {g_c^2} + {g_d^2} \right) 
          \right]  \right\} \nonumber \\
\vspace{0.2cm}\nonumber\\
s_1&=&\frac{\ge^2}{\beta_1^2\beta_2^4 \nu^2}
      \left\{\gb_1^2 g_b^2 \nu^2\left[4\,{\beta_1}\,{\gb_2}\, 
             \left({g_b^2} + {g_c^2} \right) \,
       {g_d^2}\,n_{a2}\,{n_{c1}} - 4\,{\beta_1^2}\,
       \left( {g_c^2}\,{g_d^2} + {g_b^2}\, \left( {g_c^2} +
          {g_d^2} \right) \right) \,{n_{c1}^2}\right]
\right. \nonumber \\
&&\hspace{1cm}
+\,\beta_2^4\, \left[6 g_b^2 g_c^2 (9 g_a^2+g_d^2) n_{b1} n_{c1} \nu \,\delta
-n_{c1}^2\left(2 g_b^2 g_c^2 g_d^2 +9 g_a^2
\left(g_c^2 g_d^2+2 g_b^2 (g_c^2+g_d^2)\right) \right)\right]\nonumber\\
&&\hspace{1cm} - \left. \,{\beta_2^2}\,g_b^2
          \left({g_b^2} + {g_c^2} \right) \,
          \left( 9\,{g_a^2} + {g_d^2} \right) \,
          \left( 4\,\beta_1^2 {\beta_2^2}\,{\ge^2} + 
            {n_{a2}^2}\,{\nu^2}\beta_1^2+ 9 \beta_2^2 \nu^2 n_{b1}^2
\right)\right\}  \nonumber \\
\vspace{0.2cm}\nonumber\\
s_2&=&\frac{1}{4 \beta_1^2 \beta_2^2 \nu^2}
      \left[ \left(9 g_a^2+g_d^2\right)\left(4 \beta_2^2
\ge^2+n_{a2}^2\nu^2\right)\beta_2^2 -4 \beta_1 \beta_2 g_d^2 n_{a2}n_{c1}
      \nu^2\right] \left(2 g_b^2+g_c^2\right) + 
\frac{g_c^2 g_d^2 n_{c1}^2}{\beta_2^2} \nonumber\\
&&\,\,\,\,\,\,\,\,\,\,
+\,\frac{g_b^2}{\beta_2^2}\left[
  2 (g_c^2+g_d^2) n_{c1}^2-6 g_c^2 n_{b1} n_{c1} \nu \,\gd 
 +\left(4 \beta_1^2\ge^2 +9 n_{b1}^2 \nu^2\right) 
\left(g_b^2+g_c^2\right)\right] \nonumber \\
\vspace{0.2cm}\nonumber\\
s_3&=&-2 g_b^2 - g_c^2
\end{eqnarray}
One eigenvalue in (\ref{soll}) is  $\gl_0=0$ with  the remaining ones 
denoted by $\gl_i$ with $i=2,3,4,5$. In the limit  $\eta\rightarrow
0$, $\gl_{2,3,4}$ are  equal to those in the absence of the
electroweak symmetry breaking (and non-zero) see eqs.(\ref{eigval4})
(\ref{c3}). Their electroweak corrections are
suppressed by powers of $\eta$  
relative to their non-vanishing value in the absence of EWS breaking,
are very small and we will not address them further. 
For our  purpose, we are  interested in the remaining $\gl_5$
which is the {\it only} one to vanish {\it in the limit} $\eta\rightarrow 0$,
and may  thus be approximated as  an expansion in positive 
powers  of $\eta$ (no constant term). This is just the mass 
of the $Z$ boson. We thus search for a solution to $M_Z$ of the type
\begin{eqnarray}\label{mz1}
M_Z^2
\equiv \gl_5 M_S^2 &=&\left [\eta\, \xi_1+ \eta^2\,\xi_2 + \eta^3 \, \xi_3 + \eta^4
 \, \xi_4 +\,\cdots\right] M_S^2 \nonumber \\
&=&\left[1 +\eta\, \xi_{21} +  \eta^2 \, \xi_{31}+ \eta^3 \, \xi_{41}+ 
\cdots\right] \xi_1 <\phi>^2\label{mz2}
\end{eqnarray}
To compute $\xi_1$, $\xi_{21}$, $\xi_{31}$ 
  we use eqs.(\ref{soll}), (\ref{coef2}),
(\ref{mz1})  together
with $c_i$ of (\ref{c3}).   Discarding terms of order  $\cO(\eta^5)$
in (\ref{soll}) we  find\footnote{
The definition of $\xi_1$, $\xi_{21}$, $\xi_{31}$
 in terms of $c_i$ and $s_i$ was outlined in (\ref{x31}).}
\begin{eqnarray}\label{xi1}
\xi_1&=& \left\{ g_b^2 g_c^2 g_d^2 + 9 g_a^2 \left[g_c^2 g_d^2 +g_b^2
(g_c^2+g_d^2)\right] \right\}
\left[ g_c^2 g_d^2 +9 g_a^2 (g_c^2 +g_d^2)\right]^{-1} \nonumber \\
&=& \frac{1}{4} (4 g_b^2+g_y^2)
\end{eqnarray}
where in the last step  we used the hypercharge coupling definition 
eq.(\ref{hypercharge1}). For $\xi_{21}$ we find 
\begin{eqnarray}\label{xi21}
\xi_{21}&=&- \left\{4 \beta_2^4 \beta_1^2 \epsilon^2 g_c^4
(9g_a^2+g_d^2)^2 + \beta_2^4 \left[\left (g_c^2 g_d^2+9
g_a^2(g_c^2+g_d^2)\right) 
n_{c1}-3 g_c^2 n_{b1}\nu \, \delta \, (9 g_a^2 +g_d^2)
\right]^2\right.
\nonumber\\
&&\!\!\!\!\!\!\!\!\!\!
\left. + \beta_1^2 g_c^4 \nu^2 
\left[\beta_2 n_{a2}(9 g_a^2+g_d^2)- 2 \beta_1 g_d^2 n_{c1}\right]^2\right\}
\left\{4 \beta_1^2 \beta_2^2 \ge^2 \left[g_c^2 g_d^2 +9 g_a^2
(g_c^2+g_d^2)\right]^2n_{c1}^2\right\}^{-1}\nonumber
\end{eqnarray}
Since $\xi_{21}<0$ the mass $M_Z$ (\ref{mz2}) will be smaller than 
that of the SM Z boson. 
Note that $\xi_{21}$ is similar for all models with same $n_{a2}$, $\beta_2$
 of Class A defined in
Table \ref{minimal} (except the term in the 
second square bracket depending on $n_{c1}$).
The first (squared)  square bracket in  $\xi_{21}$  is similar for all
models of Class A, since $n_{c1} n_{b1} \nu=-1$ is invariant 
for all models.  This will imply similar behaviour for all 
models of Class A that we address.

In terms of the chosen  parameters of Class A models, the ratio
$\cR=g_d/g_c$ and $n_{a2}$, $\xi_{21}$ may be re-written as
\begin{eqnarray}
\xi_{21} & = & -\left\{
\beta_1^2 \left[2 \beta_1 g_y^2 \,\nu\, n_{c1} (1+\cR^2)-(36 g_a^2+g_y^2
\cR^2)\beta_2 n_{a2} \nu\right]^2+
4 \beta_1^2 \beta_2^4 \epsilon^2 (36 g_a^2+g_y^2 \cR^2)^2\right.\nonumber\\
&&\!\!\!\!\!\!\!\!\!\!\!\!\!\!\!\!\!
+\left. 9\beta_2^4\left[ g_y^2 n_{b1} \,\nu\, \cR^2 +12 g_a^2 
[ 3 n_{b1}\,\nu\,-n_{c1}(1+\cR^2)\delta]\right]^2
\right\} 
\left[5184\beta_1^2 \beta_2^2 \epsilon^2 g_a^4 n_{c1}^2
(1+\cR^2)^2\right]^{-1}
\end{eqnarray}
and this will be used in the last part of this section.
Finally, the expression of $\xi_{31}$ is 
\begin{equation}
\xi_{31}=
\left(16 \beta_1^4
\beta_2^4 \ge^4 g_b^4 a_3^4 n_{c1}^4\right)^{-1}
\left[2 a_0^2 a_1^2 + 3 a_0 a_1 a_2 a_3 
-\beta_1^2 \beta_2^4 g_b^2 a_0^2 a_3 a_4
+a_3^2 (\beta_1^2 \beta_2^4 g_b^2 a_0 a_5+a_2^2 ) \right]
\label{xi31}
\end{equation}
where  the coefficients $a_i$ (i=1,..5) are given in Appendix II.
Unlike $\xi_{21}$, $\xi_{31}$ has no definite 
sign\footnote{For a Class A model with
 $\beta_1=1/2$, $\nu=1/3$, $\beta_2=1$, $n_{c1}=1$, $n_{b1}=-1$,
$\delta=+1$ numerical investigations show that
if $g_d/g_c\approx 0.7$ or less, and with $n_{a2}$
taking values smaller than 20, $\xi_{31}>0$ and  has an opposite
effect to $\xi_{21}$, to increase $M_Z$. For  $g_d/g_c>0.7$,
$\xi_{31}<0$, and its effects add to those of $\xi_{21}$ to decrease
$M_Z$.  Generically  $\xi_{21}=\cO(1)$ up to $\cO(10)$ and the 
ratio  $\xi_{31}/\xi_{21}$ is of order $\cO(1)$, 
except  cases with $g_d/g_c<1$ for large $n_{a2}$ when the ratio may
become larger, up to $\cO(10)$.}. 
Numerical investigations for all models of this class 
show that the effects of $\xi_{31}$ are
very small relative to those induced by $\xi_{21}$ and this ensures
our procedure is rapidly convergent. 

We therefore find that  
\begin{eqnarray}
M_Z^2&=& \frac{1}{4}\left(4 g_b^2+g_y^2\right) <\phi>^2 \left[1+\eta\, \xi_{21} 
+ \eta^2\, \xi_{31} +\cdots\right]\nonumber\\
&=& M_{Z, 0}^{2} \left[1+\eta\, \xi_{21} 
+ \eta^2\, \xi_{31} +\cdots\right]\label{ZZ}
\end{eqnarray}
which in the lowest order in $\eta$ recovers the usual mass formula of
$Z$ boson ($4 g_b^2=g_L^2$) induced by the Higgs mechanism alone.
$\xi_{21}$ does not depend on the coupling $g_b (g_L)$ 
as this (electroweak) dependence is factorised into the first
term in the expansion of $M_Z$. Higher corrections ($\xi_{31}$) 
depend on $g_b$ since Higgs state was charged under $U(1)_b$, $U(1)_c$
(which also mix) see Table \ref{thirdone}.
The difference between models of Class A with 
$n_H=1, n_h=0$ or with $n_H=0, n_h=1$ is marked in eqs.(\ref{xi21}),
(\ref{xi31}) defining $\xi_{21}$ and $\xi_{31}$ by the presence of
$\delta=+1$ and  $\delta=-1$ respectively.

The above correction to $M_Z$  is due to the initial presence of
additional $U(1)_\alpha$'s and their mass in the absence of EWS breaking, 
eq.(\ref{massterm}) induced by  the couplings of their field strength
tensor to the RR two form fields, $F \wedge B$. Thus the correction (\ref{ZZ})
is {\it not} due to the Higgs mechanism. 
Given the string origin of the couplings $F \wedge B$,
the additional correction to $M_Z$, as a result of mixing with the
$U(1)_\alpha$ bosons,  is itself of string nature and 
there is no Higgs particle associated with it. This
mechanism is  interesting in itself, regardless of the phenomenological
viability of the models.

From equation (\ref{ZZ}) we  can now extract (lower) bounds on the value of the
string scale for all possible models  of Class A, Table  \ref{minimal}.
This is done by using the  definition of the
$\rho$ parameter
\begin{equation}
\rho=\frac{M_W^2}{M_Z^2 \cos\theta_W}
\end{equation}
with $\rho=\rho_0$ for SM case. Experimental constraints
give that  for a Higgs mass\footnote{We use the value of the {$\rho$}
parameter which complies with this value for the Higgs mass. One can
ignore this constraint, and consider the value of 
$\rho= 1.0012 +0.0023/-0.0014$ \cite{pdg} whose implications (of
decreasing $M_S$ by $\approx 1.96$)
for our findings may easily be recovered from eq.(\ref{mstring}).}
at 115 GeV \cite{pdg}
\begin{equation}\label{limits}
\frac{\Delta \rho}{\rho_0}= \frac{\pm 0.0006}{1.0004}
\end{equation}
and this will be used in the following to impose bounds on $M_S$.
From eq.(\ref{ZZ}) for corrected $Z$ mass we find that
\begin{equation}\label{bound}
\frac{\Delta \rho}{\rho_0}=-1+\frac{1}{1+\eta\,
\xi_{21}+\eta^2\, \xi_{31}+{\cal O}(\eta^3)}
\end{equation}
which may be solved for $\eta$ in terms of $\Delta\rho/\rho_0$.
From (\ref{bound}), the (lower) bound on $M_S$ is given by
\begin{equation}\label{msnew}
M_S^2= <\phi>^2 (- \xi_{21}) \left[1+\frac{\rho^0}{\Delta\rho}\right]
\left[1-\, 2\, \frac{\xi_{31}}{\xi_{21}^2}
\frac{\Delta\rho/\rho^0}{1+\Delta\rho/\rho^0}\right]^{1/2}
\end{equation}
Since  the correction to $M_S$  due to $\xi_{31}$ 
(relative to the case when it is ignored) 
is very small (less than $0.1 \%$) we can safely neglect it (also
$\rho_0/\Delta\rho\gg 1$) and 
keeping the  lowest order in $\eta$ one finds
\begin{equation}\label{mstring}
M_S^2\approx\, <\phi>^2\, (-\xi_{21}) \, \frac{\rho_0}{\Delta\rho}
\end{equation}
Since  $\xi_{21}<0$, $M_Z$ is decreased from its SM value, so
$\Delta\rho =\rho-\rho_0>0$. Therefore the positive correction 
in (\ref{limits}) is considered.
Eq. (\ref{mstring}) shows that the (lower) bound on $M_S$ 
is raised  as the uncertainty   $\Delta\rho$ of measuring 
$\rho_0$ is  reduced. With our choice for 
the $\rho$ parameter consistent with $M_H=115$ GeV
eq. (\ref{mstring}) gives  
\begin{equation}
M_S\approx 10046.7 \times \vert \xi_{21}\vert^{\frac{1}{2}} \quad GeV.
\end{equation}
Using expression (\ref{msnew})  or (\ref{bound}) 
one can derive bounds on $M_S$ in terms of the parameters of the
 model (ratio $g_d/g_c$ and $n_{a2}$).
These are in very good agreement ($<2\%$ for $M_S$) with the one derived 
using a full numerical approach to finding the value of $M_Z$ from
$\cM^2_{\alpha\beta}$ of (\ref{soll}) and from that constraints on $M_S$.
\FIGURE[t]{\begin{tabular}{cc|cr|} 
\parbox{7.1cm}{ 
\psfig{figure=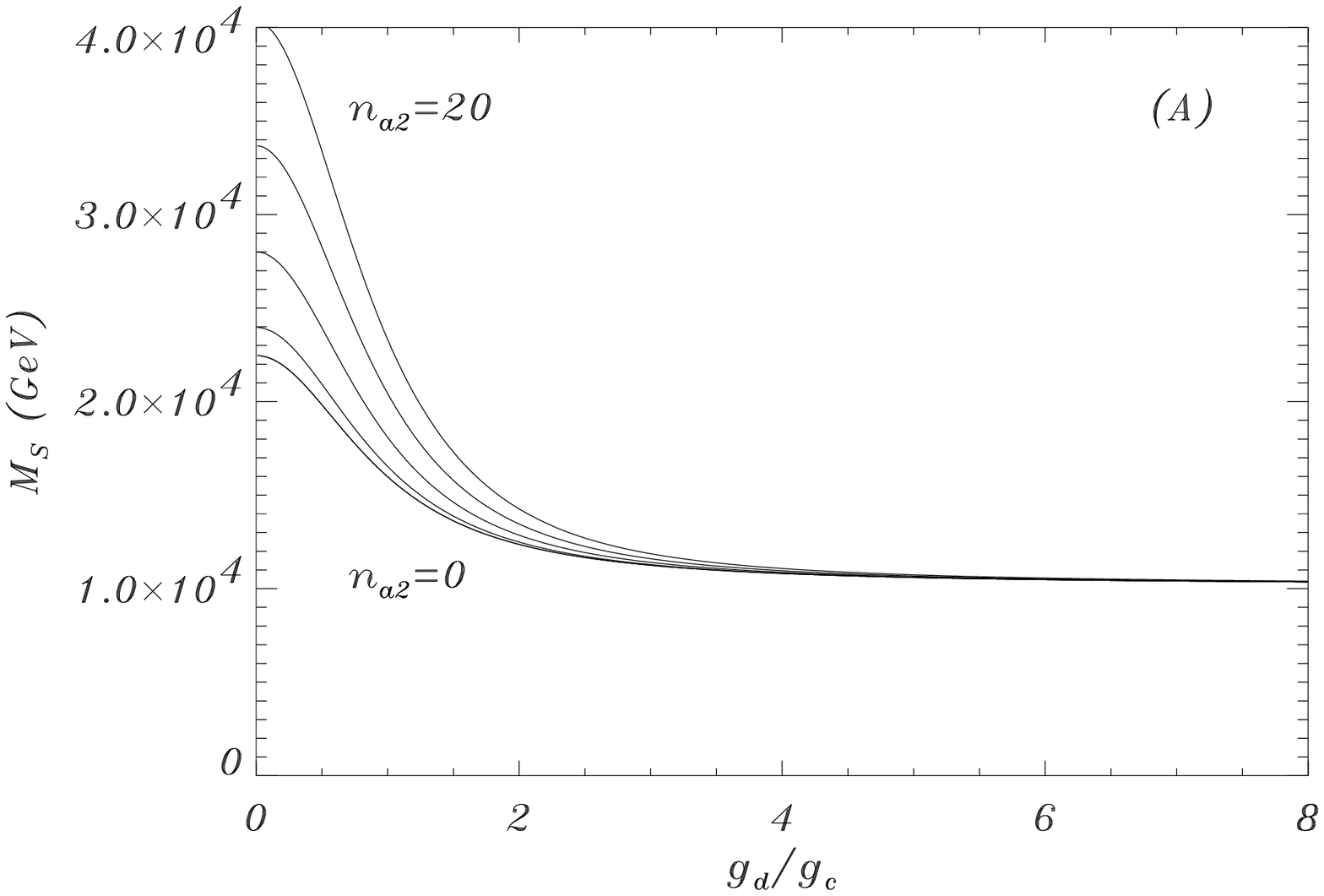,height=6.8cm,width=6.8cm}} 
\hfill{\,\,\,\,} 
\parbox{7.1cm}
{\psfig{figure=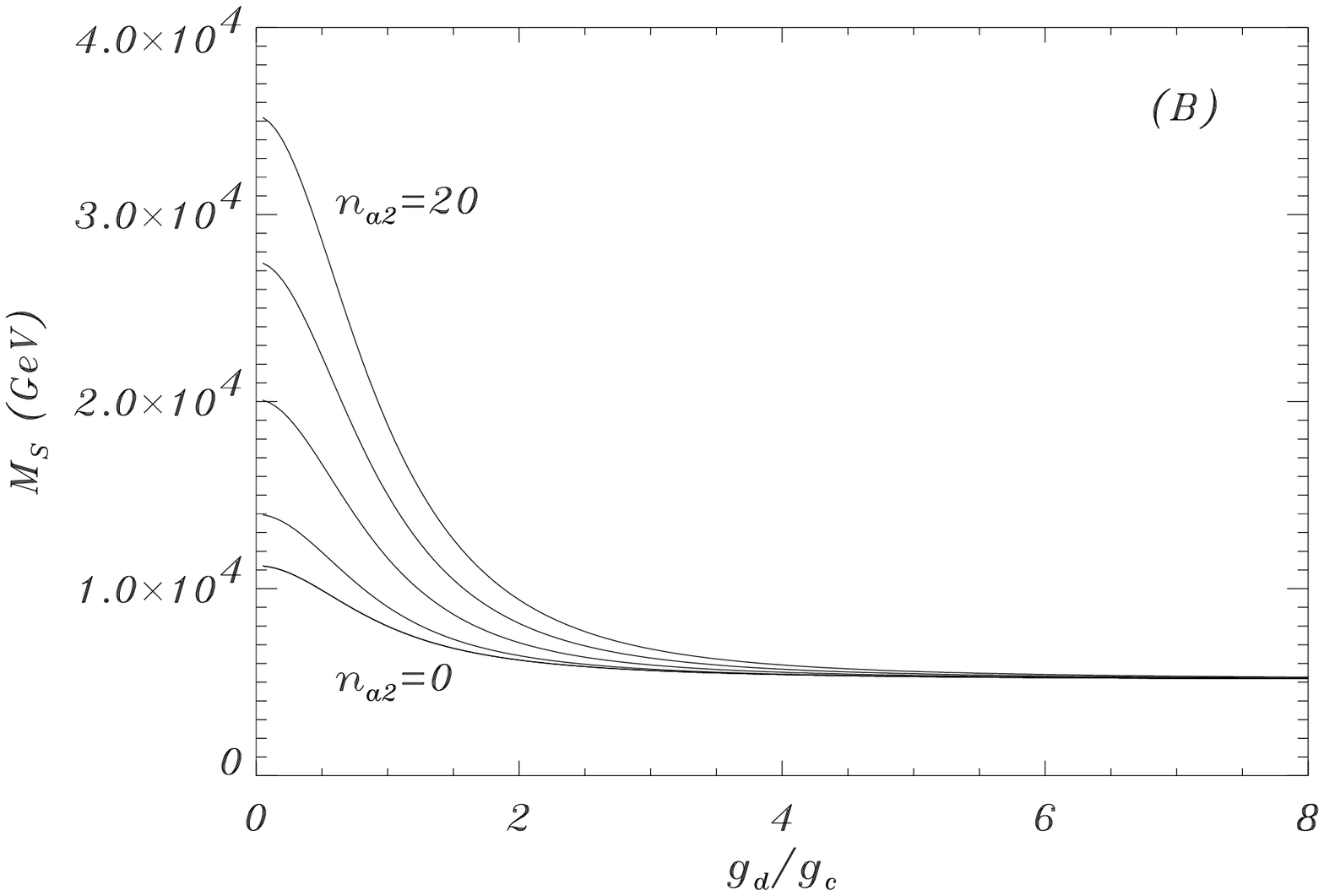,height=6.8cm,width=6.8cm}}
\end{tabular}
\caption{\small{D6 brane models: Class A models.
\newline\noindent
(A): Lower bounds on the string scale, $M_S$  (GeV), 
in function of the ratio $g_d/g_c$, with 
$n_{a2}$ fixed for each curve. $n_{a2}$ is
 increasing upwards (step 5) from 
$n_{a2}=0$ (lowest curve) to $n_{a2}=20$.  
The plots correspond to a model of Class A defined by
$\beta_1=1/2$, $\nu=1/3$, $\beta_2=1$, $n_{c1}=1$, $n_{b1}=-1$,
$\delta=+1$.  Lowest $M_S$ is 10 TeV. 
\newline\noindent
(B): As for (A) but $\beta_2=1/2$ (also  $\nu=1/3$,
$\beta_1=1/2$, $n_{c1}=1$, $n_{b1}=-1$,
$\delta=+1$). The lowest allowed  $M_S$ can be 
as small as 5 TeV. This value and that in (A) may further be
decreased, by a factor of up to $1.96$ when relaxing the constraints on
$M_H$, see text.}}
\label{mstringa}}
%
%
\FIGURE{\psfig{figure=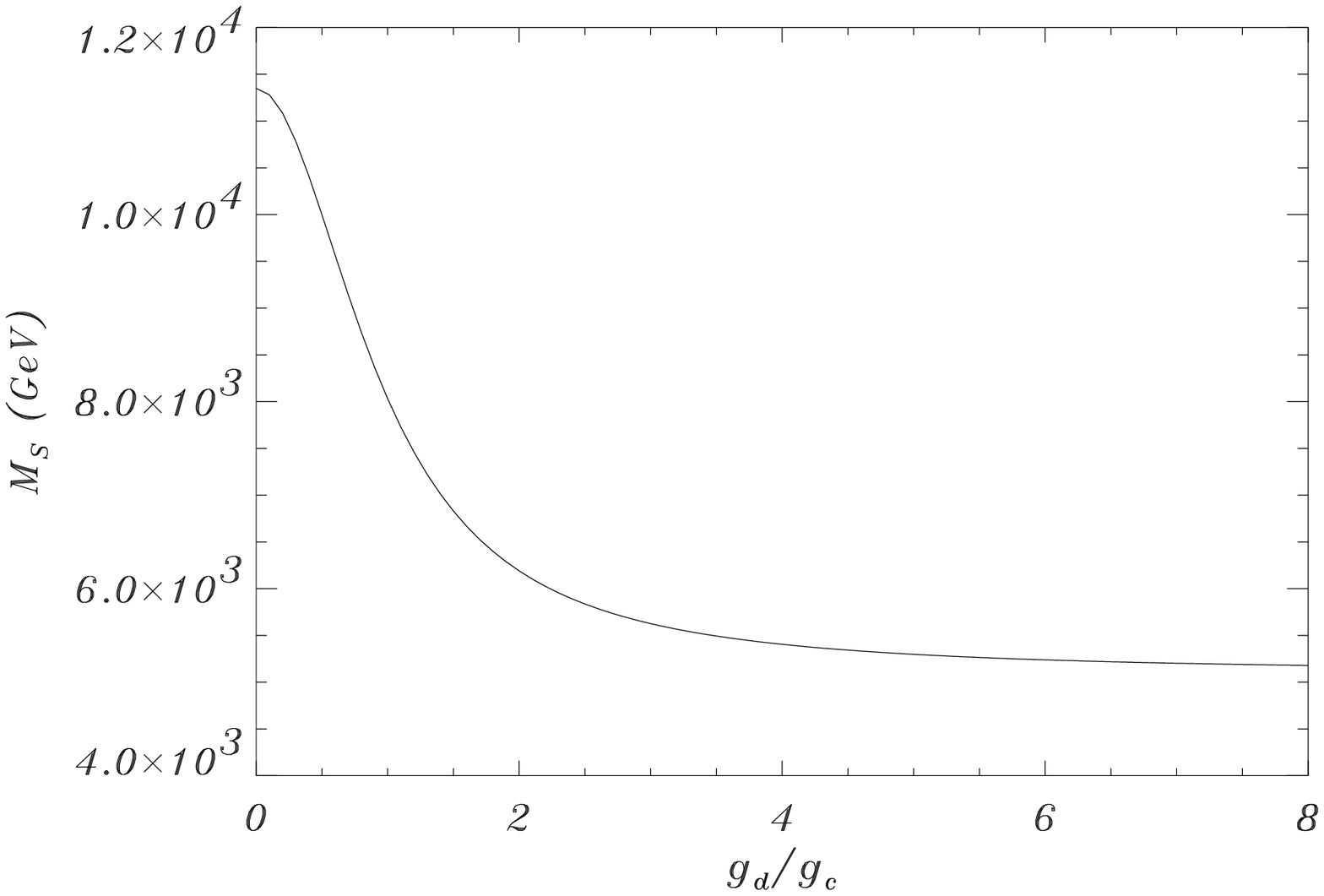,height=6.8cm,width=7cm}
\caption{\small{D6 brane models: The string scale $M_S$ (GeV), in function of the
 ratio $g_d/g_c$, corresponding to a model of Class A, line 3 of the Table
 \ref{minimal}, parametrised by $\nu=1/3$,  $\beta_1=1/2,\, \beta_2=1/2$,
 $n_{a2}=n_{b1}=n_{c1}=n_{d2}=1$.  }}
\label{qsusyfig}}

We start with the first example of Class A
models of Table \ref{minimal}. For this particular model with: 
$\delta=+1$, $\nu=1/3$, $\beta_1=1/2$, $\beta_2=1$, $n_{b1}=-1$, $n_{c1}=1$,
Figure \ref{mstringa}.(A) presents  the  numerical dependence of the
string scale in function of $g_d/g_c$  for various $n_{a2}$.
The plot takes into account that $g_d$ and $g_c$ are not
independent, but correlated to $g_y$ (via eq.(\ref{hypercharge1})) which
is kept fixed. The lower bounds on $M_S$ are in the region of 10 TeV
for large  (parameter) ratio $g_d/g_c$.
These bounds increase significantly up to a range of $22-40$ TeV, 
as $n_{a2}$ is increased from $0$ to 20,
for regions with the ratio of  couplings $g_d/g_c<1$.
All models of Class A (Table \ref{minimal}) which differ
from this example only by $n_{c1}$, $ n_{b1}$, $\delta$, 
with $\beta_1=1/2$, $\beta_2=1$, $\nu=1/3$,
have very similar dependence with respect to $g_d/g_c$ (for given
$n_{a2}$) and the same bounds on $M_S$, as in Figure
\ref{mstringa} (A), (B).
This similarity  may be noticed from a careful analysis of 
$\xi_{21}$ which  sets the behaviour of $M_S$.

We  note  the large spreading with respect to $n_{a2}$ 
for fixed  $g_d/g_c<1$  (thus the wide range of values for $M_S$)
in Figure \ref{mstringa}. 
For similar ratios of $g_d/g_c$ and range of $n_{a2}$ values as in
this figure, the amount of
$U(1)_b$ in the mass eigenstates $M_{4}$ has a significant
effect (spread) with respect to $n_{a2}$
(similar case for $M_2$ but less significant), 
while in $M_3$ this effect is small. Note that very approximately 
$M_S$ is  proportional to the  product of  $M_2$ and $M_4$. The 
spreading/sensitivity of $M_S$  with respect to $n_{a2}$ for fixed $g_d/g_c$
in Figure \ref{mstringa} is then  induced via $M_4$.
As we increase the ratio $g_d/g_c$,  the spreading  with respect to 
$n_{a2}$ disappears, and $U(1)_b$ dominates in $M_3$ but not in
$M_{2,4}$.

As another example 
we show in Fig.\ref{qsusyfig} the constraints  for a model
corresponding  to the third line of Table \ref{minimal}, defined
by  $\nu=1/3$,  $\beta_1=1/2$,
$\beta_2=1/2$, $n_{a2}=n_{b1}=n_{c1}=n_{d2}=1$. As mentioned in 
Section \ref{d6branemodels} such a model may be ``approximately''
supersymmetric in the  sense discussed in \cite{cim2}. With 
all these parameters fixed,
only $g_d/g_c$ remains  a free parameter and we can plot 
the limit on the value of the string scale $M_S$ 
as a  function of $g_d/g_c$. In this case we must have 
$M_S\geq 5$ TeV in order to be within the experimental limits for $\rho$.

As a general conclusion one finds that  constraints
from the $\rho$-parameter are somewhat stronger than those obtained from direct
searches. Very often the constraints imply the need of rather high 
(of order 10-20 TeV) values for the string scale.  Still, 
we find  that the string scale can have  values
as low as 5 TeV, Figure \ref{mstringa}.(B), while still  accommodating 
the constraints from the $\rho$ parameter.
These values of the string scale may further 
decrease by a factor of $\approx 1.96$   if the uncertainty in the $\rho$
parameter that  we used (+0.0006) throughout this analysis and in
Fig.\ref{qsusyfig} and Fig.\ref{mstringa} was relaxed to 
(+0.0023) \cite{pdg} which does not take account of the 
constraint $M_H=115 $ GeV.  This would lead to $M_S$ as low as 
2.5 TeV for (some) Class A  models.

%
%

\vspace{0.7cm}
\subsubsection{Class B models.}

\noindent
For class B models of Table \ref{minimal} we use an approach similar 
to Class A models to find lower bounds on the value of the string scale. 
However, eq.(\ref{coef2}) has a different
structure, as corrections  $\cO(\eta^2)$ to $c_i$ are  now present in
the definition of $o_i$ and care must be taken in finding the
analytical expansion in $\eta$ for $M_Z$. For this one may
use the expansion outlined in eq.(\ref{x31}).

In Class B models there are two Higgs sectors, $H_i$ (i=1,2) with vev's
 $v_i/\sqrt2$  and $h_i$ (i=1,2) with vev's ${\overline v_i}/\sqrt{2}$. 
We therefore need to introduce a further parameter (in addition to
$g_d/g_c$ and $n_{a2}$),  the angle $\theta$ 
defined as the mixing between the two Higgs sectors ($H_i$
 and $h_i$ respectively):
\begin{eqnarray}\label{mixing}
<\phi>^2\equiv v_1^2+v_2^2=<\Phi>^2 \cos^2 \theta\nonumber\\
<{\overline{\phi}}>^2\equiv {\overline v}_1^2+ 
{\overline v}_2^2=<\Phi>^2\sin^2\theta
\end{eqnarray}
As in Class A one can compute  the corrections to $M_Z$ as an expansion
in the parameter $<\Phi>^2\!\!/M_S^2$.
Using eqs.(\ref{5dim}), (\ref{5dime}), and (\ref{mixing}) one finds 
\begin{eqnarray}\label{ZZZ}
M_Z^2&=& \left[1+\eta' \,\xi_{21}+ \eta'^2\, \xi_{31}+\cdots\right]
\xi_1 <\Phi>^2,\qquad\qquad \eta'=<\Phi>^2/M_S^2 \nonumber \\
&=& \frac{1}{4} \left(4 g_b^2+g_y^2\right) <\Phi>^2 \left[1+\eta'\, \xi_{21} 
+\eta'^2 \, \xi_{31} +\cdots\right] \nonumber \\
&\equiv& M_{Z, 0}^{2} \left[1+\eta'\, \xi_{21} 
+\eta'^2 \, \xi_{31} +\cdots\right]
\end{eqnarray}
In (\ref{ZZZ}) we used eq.(\ref{hyper1}) to re-write $\xi_1$ in terms
of the hypercharge coupling:
\begin{equation}
\xi_1= \left\{ g_b^2 g_c^2 g_d^2 + 9 g_a^2 \left[g_c^2 g_d^2 +g_b^2
(g_c^2+g_d^2)\right] \right\}
\left[ g_c^2 g_d^2 +9 g_a^2 (g_c^2 +g_d^2)\right]^{-1}= 
\frac{1}{4} (4 g_b^2+g_y^2)
\end{equation}
We also have that
\begin{eqnarray}\label{bxi21}
\xi_{21}&=&- \left\{4 \beta_2^4\beta_1^2 \epsilon^2 g_c^4 (9
g_a^2+g_d^2)^2
+\beta_2^4 n_{c1}^2\left(g_c^2 g_d^2+9 g_a^2 (g_c^2+g_d^2)\right)^2
\cos^2 (2 \theta)\right.\\
&&\!\!\!\!\!\!\!\!
+\left.\beta_1^2 g_c^4 \nu^2 \left[\beta_2 (9 g_a^2+g_d^2)
n_{a2}-2 \beta_1 g_d^2 n_{c1}\right]^2 \right\}
\left\{4 \beta_1^2 \beta_2^2 \epsilon^2
\left[g_c^2 g_d^2 +9 g_a^2 (g_c^2 +g_d^2)\right]^2 n_{c1}^2
 \right\}^{-1} < 0\nonumber
\end{eqnarray}
which has an expression very similar to that of Class A, see 
eq.(\ref{xi21}). As in Class A,  $\xi_{21}<0$ thus $M_Z$ is  
decreased from its SM expression. 
$\xi_{21}$ may be re-written in terms of the (chosen) parameters 
($\cR=g_d/g_c$, $n_{a2}$, $\theta$) as
\begin{eqnarray}
\xi_{21} &=&
- \left\{\beta_1^2 \left[2 \beta_1 g_y^2 \,\nu\, n_{c1} (1+\cR^2)-(36 g_a^2+g_y^2
\cR^2)\beta_2 n_{a2} \nu\right]^2
+4 \beta_1^2 \beta_2^4 \epsilon^2 (36 g_a^2+g_y^2 \cR^2)^2\right.\nonumber\\
&+&\left. 1296 \beta_2^4 g_a^4 n_{c1}^2 (1+\cR^2)^2 \cos^2(2\theta)
\right\}
\times 
\left[5184\beta_1^2 \beta_2^2 \epsilon^2 g_a^4 n_{c1}^2
(1+\cR^2)^2\right]^{-1}
\end{eqnarray}
which will be used in the following.
Finally, for the coefficient $\xi_{31}$ we have
\begin{eqnarray}\label{newxi31}
\xi_{31}&=&g_c^2 \left[  b_4 +b_7-b_5+b_6 \cos(4
\theta)\right]\left( 32 \beta_1^4 \beta_2^4 \epsilon^4 g_b^2 
n_{c1}^4 b_1^4\right)^{-1}
\end{eqnarray}
where the coefficients $b_i$ $(i=1,..7)$ are presented in Appendix 
II. Unlike $\xi_{21}$ which is of definite sign, 
$\xi_{31}$ is either positive or negative in function of the ratio
$g_d/g_c$ and the value of $n_{a2}$. Considerations similar to those
made in Class A for the expansion in $\eta'$ apply here as well.
The accuracy of this analytical approach compared to the full 
numerical one to computing 
(the corrections to) $M_Z$ from  (\ref{soll})) and from these the
bounds on $M_S$,  is within $2 \%$ for the string scale prediction.

Using the same procedure as in Class A, we find a similar expression for
the string scale bound
\begin{equation}\label{newmstring}
M_S^2= <\Phi>^2 (- \xi_{21}) \left[1+\frac{\rho_0}{\Delta\rho}\right]
\left[1-\, 2\, \frac{\xi_{31}}{\xi_{21}^2}
\frac{\Delta\rho/\rho_0}{1+\Delta\rho/\rho_0}\right]^{1/2}
\approx\, <\Phi>^2\, (-\xi_{21}) \, \frac{\rho_0}{\Delta\rho}
\end{equation}
and where $\xi_{21}$ also depends on the angle $\theta$, the direction
of the vev solution.
\FIGURE[t]{
\begin{tabular}{cc|cr|} 
\parbox{7.1cm}{ 
\psfig{figure=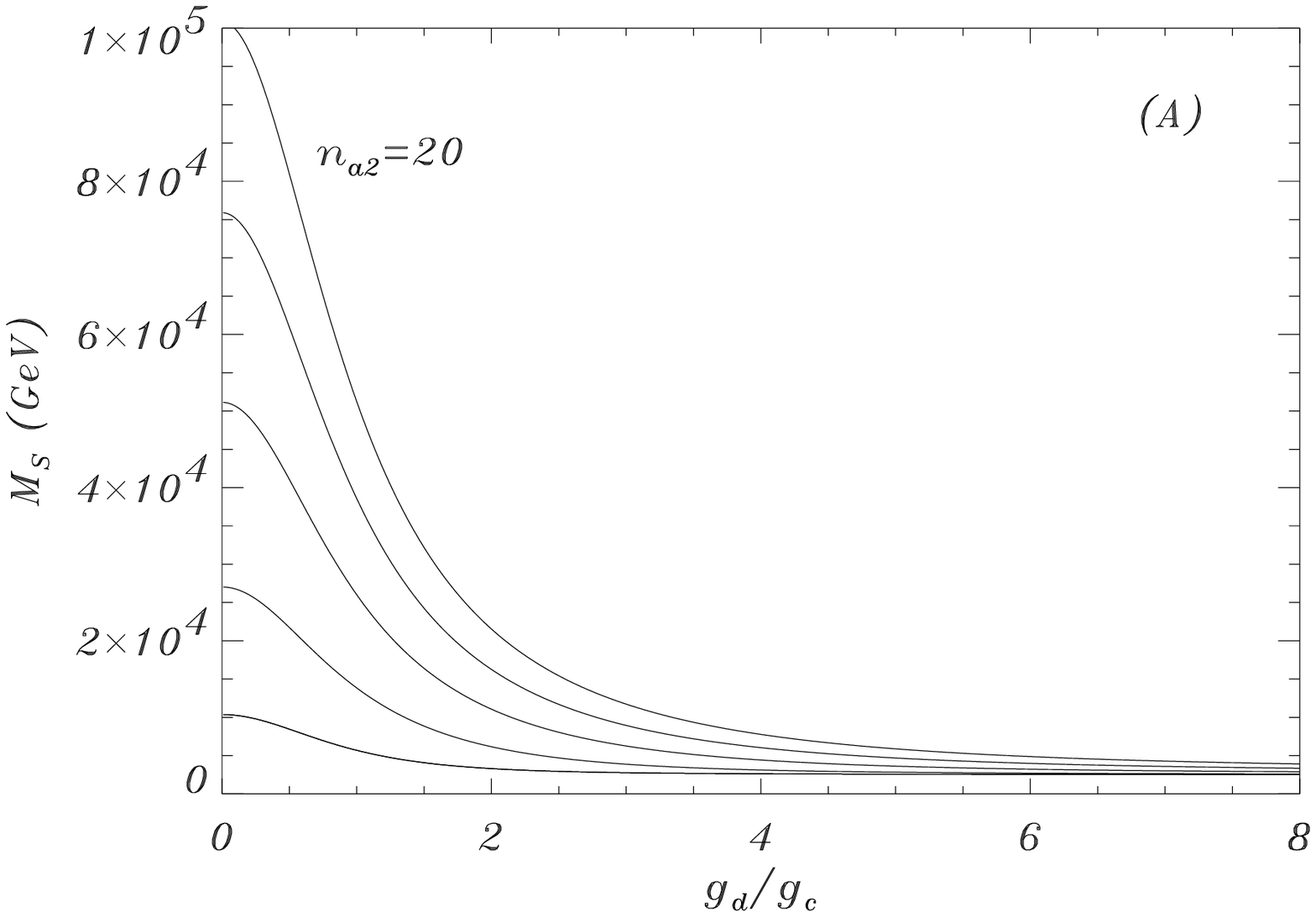,height=6.8cm,width=6.8cm}} 
\hfill{\,\,\,\,} 
\parbox{7.1cm}{  
\psfig{figure=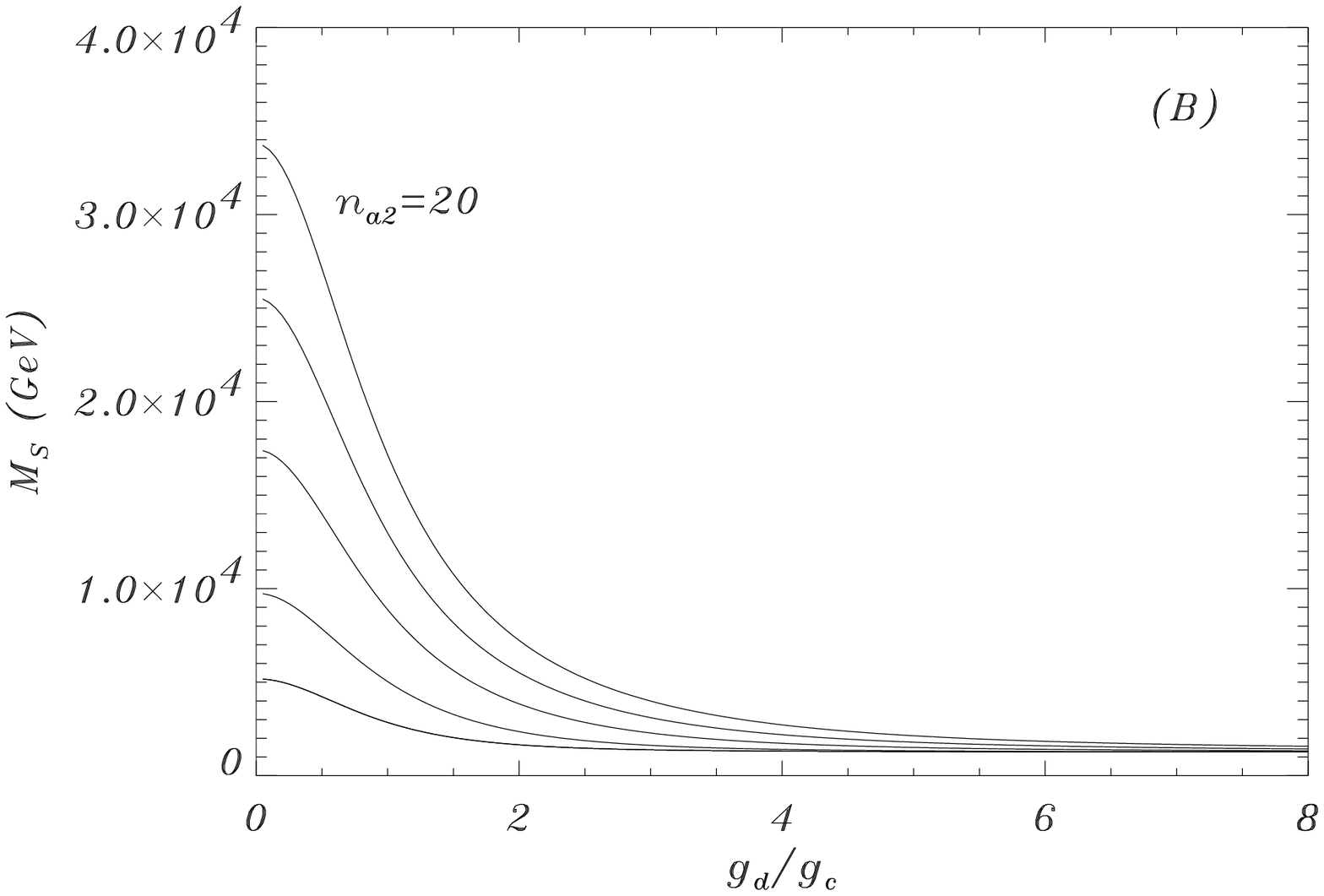,height=6.8cm,width=6.8cm}}  
\end{tabular} 
\caption{\small{D6 brane models. Class B models:
\newline\noindent
(A): The value of  the string scale, $M_S$ (GeV),
in function of the ratio $g_d/g_c$, with $n_{a2}$ fixed for each curve
and increasing upwards (step 5) from $n_{a2}=0$ (lowest curve) 
to $n_{a2}=20$. The lowest $M_S$ for which
correction to $M_Z$ is within the experimental error for the $\rho$
parameter is $\approx$ 2.5  TeV.  The model is 
defined by $\nu=1$, $\beta_1=1$, $\beta_2=1$, $n_{c1}=1$,
$n_{b1}=0$, $\theta=\pi/6$. 
\newline\noindent
(B): As for (A) but with   $\nu=1/3$ (also $\beta_1=1$, $\beta_2=1/2$, $n_{c1}=1$,
$n_{b1}=0$, $\theta=\pi/6$). The change in $\nu$ brings a range for
$M_S$ similar to that of Class A models, where this parameter has the
same value.  Lowest $M_S$ which still respects $\rho$ constraints 
is $\approx$ 1.5 TeV.}}\label{mstringb}}  

The constraints  on  the string scale obtained in this way 
for two  examples with $\nu =1,1/3$ see Table \ref{minimal}, are 
shown in  Figure \ref{mstringb}.(A) and (B). 
One may check that these figures are very similar for 
all other models of Class B. 
They show the range of bounds on $M_S$, which  in some cases
may be slightly different from Class A models.  
This  difference  is  mainly due to the value of $\nu $ parameter.
In Figure \ref{mstringb} (B),  as in Class A models,
$\nu=1/3$ and somewhat similar bounds on $M_S$ apply.
However, in Figure \ref{mstringb} (A) $\nu=1$ which gives higher 
(lower) bounds on $M_S$ than in Class A models. This is consequence 
of (\ref{newmstring}) and definition of $\xi_{21}$. 
Figure \ref{mstringb} also shows that if coupling $g_d$ is much
smaller than $g_c$, the bounds on $M_S$ are not as low as  when
they are equal or if $g_d\gg  g_c$. In the latter case these bounds 
can be as low as 2.5 TeV (Fig.\ref{mstringb} (A)) or 1.5 TeV 
(Fig.\ref{mstringb} (B)) with little dependence on $n_{a2}$. 

As a general conclusion one finds that $M_S$ may be as small as
$1.5-2.5$ TeV.
Again, we should mention that if we relax the constraint on
$\Delta\rho$ that the Higgs state have a mass equal to 115 GeV, 
the above lower bounds on $M_S$ will decrease (as discussed for  Class
A models) by a factor of $\approx 1.96$, to give bounds on $M_S$ 
as low as $\approx 1$ TeV for Figure \ref{mstringb}.(A) and (B).
In such case constraints from the value of the masses 
$M_{2,3,4}$ be larger than $\approx 500-800$ GeV  may become stronger
(but also $n_{a2}$ dependent) because
$M_S$ can be as large as 5 times $M_4$ (for $g_d\gg g_c$)
 see Figure \ref{modelb}.(F) ($n_{a2}=10$).
This would give a value for $M_S$ larger than  $2.5-4$ TeV.

\vspace{1cm}
\subsection{D5-brane models. Stringy Corrections to $M_Z$ and
bounds on $M_S$.}\label{D5AEW}
\FIGURE[hb]{
\psfig{figure=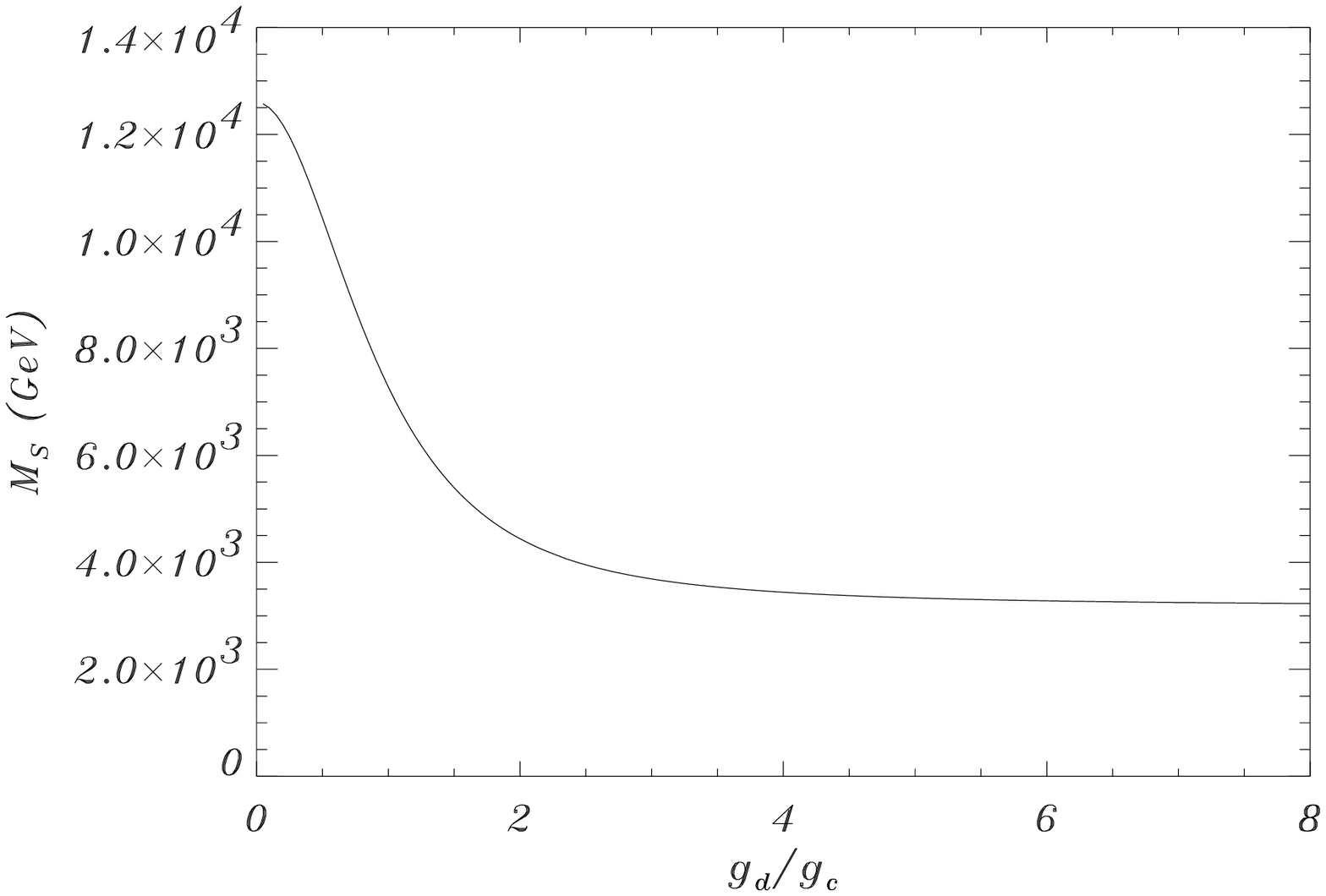,height=6.8cm,width=7cm}
\caption{\small{The string scale $M_S$ (GeV), in function of the
ratio $g_d/g_c$, for the D5-brane model of Section \ref{D5BEW}.  
Lowest $M_S$ is  $\approx 3$ TeV for $g_d/g_c\gg 1$. 
}}
\label{d5string}}

\vspace{0.3cm}
\noindent
The D5-brane model that we address here is that referred to 
in Section \ref{D5BEW} in the absence of the electroweak effects.
For this model we proceed as in  D6-brane models and search for a
solution to $M_Z$ as an expansion in $\eta=<\!\phi\!>^2/M_S^2$.
The  electroweak correction to the matrix  $M_{\alpha\beta}^2$ examined in this
model is identical to that of Class A of D6 models with
$\delta=+1$ (see eq.(\ref{5dime})) since the Higgs sector is 
similar and  contains only $H_i$. 

The mass eigenvalue equation of (\ref{soll}) has new coefficients 
$o_i$ with electroweak corrections to their
corresponding value $c_i$ of (\ref{ciD5}).  
These  corrections  contain only  terms linear in  $\eta$. 
Using an approach similar to that of previous sections
we find that
\begin{equation}
M_Z^2=\lambda_5 M_S^2= M_{Z, 0}^{2} \left[1+\eta\, \xi_{21} 
+ \eta^{2}\, \xi_{31} +\cdots\right]
\end{equation}
where $M_{Z, 0}$ is the usual SM Z boson mass and  $\xi_{21}$ is given by
\begin{eqnarray}
\xi_{21}& = & -
\left[  52 g_c^4 g_d^4 +18 g_a^2 g_c^2 g_d^2 (50 g_c^2 +6 g_d^2)+
81 g_a^4 (49 g_c^4 +12 g_c^2 g_d^2 +3 g_d^4)\right] \nonumber\\
&&\times 
\left[ 18 \sqrt{3} \left( g_c^2 g_d^2 +9 g_a^2 (g_c^2 +g_d^2)\right)^2\right]^{-1}<0
\end{eqnarray}
Thus $M_Z$ is again reduced from its Standard Model value.
Since the couplings $g_\alpha$ ($\alpha=a,..,d$) are not independent, 
but correlated to the hypercharge coupling, we also present the
expression of $\xi_{21}$ in terms of $g_y$ and\footnote{$g_y$ is 
fixed from low energy physics.}  of  the only independent
parameter $\cR=g_d/g_c$ of the model.
All these are evaluated at the string scale. With (\ref{hyperD5}) we 
have
\begin{eqnarray}\label{d5xi21}
\xi_{21} & = &  
-\left[ 72 g_a^2 g_y^2 (1+ 44 \cR^2+ 3 \cR^4)
+g_y^4 (1+4 \cR^2 +43 \cR^4) 
+ 1296 g_a^4 (49+3 \cR^2(4+\cR^2))\right] \nonumber\\
&&\times \left[ 23328 \sqrt{3} g_a^4 (1+\cR^2)^2\right]^{-1}
\end{eqnarray}
Similarly to Class A and Class B models, lower bounds on the string
scale in terms of  the ratio $\cR=g_d/g_c$  may be found from 
the experimental constraints on  $\rho$ parameter.
The relation between the string scale, $\rho$ and $\xi_{21}$,
$\xi_{31}$ found in previous cases, see eq.(\ref{msnew})
holds in this case as well. The corrections due to $\xi_{31}$
relative to the leading contribution in $\eta$ due to $\xi_{21}$ 
is less than $0.1\%$ for the string scale prediction.

In Figure \ref{d5string} the dependence of $M_S$ on 
the ratio $g_d/g_c$  shows the lower bounds on 
the string scale which still comply with  the experimental constraints 
on the parameter $\rho$. 
The bounds on $M_S$ can be as low as 3 $TeV$ for large ratio
$g_d/g_c$.  This (lower) bound derived from constraints on $\rho$ 
is stronger than those derived from using the (bounds on the) mass
of the lightest gauge boson $M_3$ (section \ref{D5BEW})
which gave  $M_S$ larger than $1.5-2.4$ TeV.
We conclude that  the lower bounds on $M_S$  are 
 similar  to those of  D6 brane models.

\vspace{0.6cm}
\noindent
We end this chapter with a  reminder of  our last remark 
of chapter 2. We have obtained in chapters 4 and 5 
 bounds on  the masses of the extra $Z$' bosons
in the specific intersecting  D6 and D5-brane models reviewed
in section 2. In translating these bounds into bounds on the string scale
$M_S$ we are setting possible volume factors to one. 
If such volume factors are very different from one, the constraints  on
$M_S$ will change accordingly.

\vspace{0.7cm}
\section{Conclusions}\label{conclusions}

The possibility of detecting additional bosons $Z'$  when exploring 
energies beyond the Standard Model scale has attracted much attention
over the years, partly because it is one of the simplest extensions
of  the Standard Model and also because of the clean experimental 
signatures. String theory further increased this interest in the
mid 1980's by the natural appearance of extra $U(1)$'s inside $E_6$, 
for instance. A thorough analysis has been performed in the past 
on the low energy effects  of that particular class of $U(1)$'s.

In this paper we have studied a very well-defined class of additional
massive $Z$ bosons that appear very naturally in D-brane models. 
Two of them have triangle anomalies which are cancelled by
a Green-Schwarz mechanism. The third additional (Abelian) gauge boson 
is familiar from left-right symmetric extensions of the SM.
We may say that the existence of these extra $U(1)$'s 
is a generic prediction of D-brane models. If the string
scale is low, these $Z'$s should be detected very soon. Furthermore,
as we have discussed in the text, these correspond to very familiar
and natural $U(1)$'s that have not only a compelling motivation from
D-brane models, but also play an important role in explaining for 
instance the stability of the proton.

We would like to emphasize that string theory provides a mechanism in
which these gauge bosons can acquire a mass independent of the Higgs
mechanism.  Their mass can be understood in terms of two dual Stuckelberg 
mechanisms in which either the massless gauge boson ``eats'' a scalar or
the stringy antisymmetric tensor ``eats'' the massless gauge boson. The
important point is that for this to happen
it is not necessary for a scalar field to acquire
a vev and therefore the origin of their mass is not due to a Higgs
field. In addition, the mixing of the $U(1)$'s implies that a
fraction of the mass of the Standard Model $Z^0$ boson may be due to 
this ``stringy''  mass term, which combined with the ordinary Higgs 
mechanism would provide the total mass for the $Z^0$ boson.
Such string corrections have been  investigated in this paper.

We have studied the mass matrices of these $U(1)$ fields and obtained 
general constraints from precision experiments on the value of their 
mass. Our numerical analysis has concentrated on D6- and D5-brane
intersection models, but we expect analogous results for other D-brane 
models. We have also set possible volume factors equal to one. 
We have found that generically one of the extra $Z_0$'s tends to 
be more massive, with mass of approximately 10 times larger than the  
string scale, typically the  multi-TeV region.  The second and  the
third boson may be  light, within a factor varying from 1 to 10 
smaller  than  the string  scale.  
In some cases  these states may be even lighter, 
depending on the exact parameters of the model 
(large $n_{a2}$).  These states mix in a non-negligible manner with 
the physical $Z_0$.

Given the generic nature of these gauged $U(1)$'s in D-brane models, 
we think that the experimental search of  their production at
colliders  is of great interest. Further, their presence could 
eventually be detected  in precision electroweak data, as we have 
discussed  in this paper. It would be amusing if the first hint of 
string theory could come from electroweak precision measurements.

\vspace{1.5cm}
\centerline{\bf Acknowledgements}
\vspace{0.5cm}
\noindent
We are grateful to G. Aldazabal, D. Cremades, S. Groot Nibbelink,
G. Honecker, C. Kokorelis, F. Marchesano, J. Moreno 
 and   A.~Uranga for useful
discussions.
The  work of L.I. is partially supported by CICYT (Spain) and the
European Commission (RTN contract HPRN-CT-2000-00148).
The work of D.M.G. and F.Q. was  supported by  PPARC (U.K.).

\newpage
\def\theequation{A-\arabic{equation}}
\appendix

\section*{Appendix I:\  The ${\bf Z}$ Boson Eigenvector}

In this Appendix we show how to compute the Z-boson eigenvector
in intersecting $D$-brane models.
We start with a comment on the ``Standard Model limit".
In the Standard Model case, to restore the electroweak symmetry limit,
 one has to take the limit $v\rightarrow 0$ in the mass formulae
\be m_W^2={1\over 4}(g_L)^2 v^2\ee
\be m_Z^2={1\over 4}(g_L^2+g_y^2) v^2\ee
and at the same time the limit  
$g_L\rightarrow 0$ in the formulae involving mixing:
\be \cos{\theta}_W={g_L\over {\sqrt{g_L^2+g_y^2}}}\ee
\be A_{\gamma}= \cos{\theta}_WB_h - \sin{\theta}_WW^3 \label{phot}\ee
\be A_{Z_0}= \sin{\theta}_WB_h + \cos{\theta}_WW^3.\label{Z}\ee
In this limit  the Z boson wavefunction
approaches the hypercharge wavefunction:
\be A_{Z_0}\rightarrow B_h.\ee
This is because in the unbroken phase $SU(2)$ is restored
and the $U(1)$ inside the $SU(2)$ does not mix anymore with
hypercharge.

Thus, one possible limit that one can take in our formulae to follow
is  $v\rightarrow 0$, as long as we also take $g_L\rightarrow 0$.
This takes us to the regime where electroweak
symmetry is not broken and therefore hypercharge is still a good
symmetry. Another interesting limit is  $M_S\rightarrow \infty$.
This decouples the heavy states and one should arrive in this limit
to the broken phase of the Standard model.

We  first derive a perturbative solution to the eigenvector of Z boson.
We denote this eigenvector ${\cal Z}$.
From our perturbative solution to the eigenvalue problem,
we found that the mass matrix
${\cal M}$ has one zero eigenvalue corresponding to the photon
with eigenvector
\be \gamma={1\over |\gamma|}
({1\over {3g_a}},0,-{1\over {g_c}},{1\over {g_d}},-{2\over {g_L}}),\ee
where
\be {1\over |\gamma|}={1\over 2}{g_yg_L\over {\sqrt{g_L^2+g_y^2}}}.\ee
Then, the photon
eigenstate can be written as
\be A_{\gamma}={g_L\over {\sqrt{g_L^2+g_y^2}}}{g_y\over 2}\Bigl[
{1\over {3g_a}}A_a-{1\over {g_c}}A_c-{1\over {g_d}}A_d\Bigr]   
-{g_y\over {\sqrt{g_L^2+g_y^2}}}W^3,\ee
with $W^3$ the third component of the $SU(2)$ gauge field.
We can now write this in a more
familiar form, namely
\be A_{\gamma}=\cos{\theta_W}B_h - \sin{\theta_W}W^3,\ee
where we have defined the hypercharge state
\be B_h={g_y\over 2}\Bigl[
{1\over {3g_a}}A_a-{1\over {g_c}}A_c+{1\over {g_d}}A_d\Bigr]\label{hyp}\ee
and the Weinberg angle
\be \cos^2{\theta_W}={g_L^2\over {g_L^2+g_y^2}}\ee   
as in the Standard Model. One immediate consistency check
is that (\ref{hyp}) corresponds precisely to the hypercharge   
eigenvector
\be h={g_y\over 2}({1\over {3g_a}},0,-{1\over {g_c}},
{1\over {g_d}})\label{hypeig}.\ee
The (light) eigenvalue corresponding to $Z$ boson is
\be e_Z={1\over 4}(g_L^2+g_y^2)v^2(1+\eta \xi_{21}+
{\cal O}(\eta^2)),\ee
We denote its associated eigenvector by  
\be {\cal Z}={1\over {|{\cal Z}|}}(z_1,z_2,z_3,z_4,z_5).\ee
The remaining three eigenvalues (large)  and associated  
eigenvectors can be approximated by their values 
 in the absence of electroweak effects,
the latter having in this case   negligible effects.

To compute the eigenvector ${\cal Z}$ one 
solves  the equation
\be {\cal M}^2\cdot {\cal Z} = e_Z \cdot {\cal Z}.\ee
This is a $5\!\times \!5$  linear system with
$det({\cal M}^2)=det(M^2+\Delta )=0$,
which implies that one of the equations is redundant.   
Removing the last  of them,
the  system that we have to solve is
\be \pmatrix {z_1\cr z_2\cr z_3\cr z_4}=
\eta Y\cdot \pmatrix {0\cr {1\over 2} g_b g_L
\cr {1\over 2}g_cg_L\cr 0} z_5,\ee
where
\be Y\equiv\Bigl[{ M}_{(4)}^2+\Delta _{(4)}-
{e_Z\over M_S^2}\cdot {\bf 1}_{(4)}\Bigr]^{-1}\ee
and
${M}_{(4)}^2$ is the upper four by four non vanishing sub-block of
${M}^2$ and
$\Delta _{(4)}$ is the corresponding sub block of
$\Delta $ (with the units divided out).
The solution to this can be written as
\be {z_1}=-{1\over 2}\eta(Y_{12}g_b +Y_{13}g_c),\ee
\be {z_2}=-{1\over 2}\eta(Y_{22}g_b +Y_{23}g_c),\ee
\be {z_3}=-{1\over 2}\eta(Y_{32}g_b +Y_{33}g_c),\ee
\be {z_4}=-{1\over 2}\eta(Y_{42}g_b +Y_{43}g_c),\ee
where we have normalized $z_5=-{1\over g_L}$.
We find the solution
\be z_1={1\over 6g_a}{g_y^2\over
{g_y^2-(g_L^2+g_y^2)(1+\eta \xi_{21})}}+
{(g_L^2+g_y^2)(1+\eta \xi_{21})\over {[g_y^2-(g_L^2+g_y^2)
(1+\eta \xi_{21})]^2}}\cdot p_1\cdot \eta+{\cal O}(\eta^2)\ee
\be z_2={(g_L^2+g_y^2)(1+\eta \xi_{21})\over {[g_y^2-(g_L^2+g_y^2)
(1+\eta \xi_{21})]^2}}\cdot p_2\cdot \eta+{\cal O}(\eta^2)\ee
\be z_3=-{1\over 2g_c}{g_y^2\over
{g_y^2-(g_L^2+g_y^2)(1+\eta \xi_{21})}}+
{(g_L^2+g_y^2)(1+\eta \xi_{21})\over {[g_y^2-(g_L^2+g_y^2)
(1+\eta \xi_{21})]^2}}{g_y^2\over g_c}\cdot
p_3\cdot \eta+{\cal O}(\eta^2)\ee
\be z_4={1\over 2g_d}{g_y^2\over
{g_y^2-(g_L^2+g_y^2)(1+\eta \xi_{21})}}+
{(g_L^2+g_y^2)(1+\eta \xi_{21})\over {[g_y^2-(g_L^2+g_y^2)
(1+\eta \xi_{21})]^2}}\cdot p_4\cdot \eta+{\cal O}(\eta^2),\ee
where $p_1,\cdots ,p_4$ are some functions of the model parameters
which we do not present explicitly.
To check our result we take the two limits that we mentioned
in the beginning.
\begin{itemize}
\item $M_S\rightarrow \infty$.

All terms proportional to $\eta$ vanish and we are left with
a simple eigenvector which however has to be properly normalized first.
Once we do so, after the limit we arrive at the Z eigenstate
\be A_{Z_0}={g_y\over \sqrt{g_L^2+g_y^2}}\Bigl[{g_y\over 2}
({1\over {3g_a}}A_a-{1\over {g_c}}A_c+{1\over {g_d}}A_d)\Bigr]+
{g_L\over \sqrt{g_L^2+g_y^2}}W^3,\label{Zlimit}\ee
which is
\be A_{Z_0}= \sin{\theta}_WB + \cos{\theta}_WW^3,\ee
precisely as in (\ref{Z}).
To further take the limit $v\rightarrow 0$, we can proceed  as in the
SM case.
\item $v\rightarrow 0$.

Again, all terms proportional to $\eta$ vanish.
Before we take the $g_L\rightarrow 0$ limit, we have to normalize
the eigenvector. If we do so, after the limit we arrive at the eigenvector
\be h={g_y\over 2}({1\over {3g_a}},0,-{1\over {g_c}},{1\over {g_d}},0),\ee
which is just the hypercharge eigenvector (\ref{hypeig})
found previously.
Actually, since the two limits commute, this is just the eigenvector that
corresponds to the
$g_L\rightarrow 0$ limit of (\ref{Zlimit}).
\end{itemize}
We can write the eigenvector in a simpler form if we expand  
and keep  only terms to ${\cal O}(\eta)$:
\be {\cal Z}={1\over {|{\cal Z}|}}\Bigl[{\cal Z}_0
+\eta \cdot {\cal Z}'\Bigr],\ee
where the first part is the Standard Model part (see (\ref{Zlimit}))
\be {\cal Z}_0 = (z_{1_0},z_{2_0},z_{3_0},z_{4_0},z_{5_0})
= ({g_y^2\over {3g_a}},
0,-{g_y^2\over {g_c}},{g_y^2\over {g_d}},2g_L),\ee
and the perturbation is
\be {\cal Z}'= (z_1',z_2',z_3',z_4',z_5')\ee
with
\be z_1'={1\over g_L^2\cos^2{\theta_W}}\bigl[g_y^2\xi_{21}+p_1\bigr]\ee
\be z_2'={1\over g_L^2\cos^2{\theta_W}}p_2\ee
\be z_3'={1\over g_L^2\cos^2{\theta_W}}\bigl[g_y^2\xi_{21}+
{g_y^2\over g_c}p_3\bigr]\ee
\be z_4'={1\over g_L^2\cos^2{\theta_W}}\bigl[g_y^2\xi_{21}+p_4\bigr]\ee
\be z_5'=0.\ee
Now we  determine the norm to order $\eta$.
Let
\be {|\cal Z|}=N+\eta N',\ee
where
\be N= 2 g_y \sqrt{g_L^2+g_y^2}.\ee
The requirement that to order $\eta$ the norm is equal to one
fixes $N'$ to
\be N'=z_{1_0}z_1'+\cdots +z_{5_0}z_5'.\ee
Finally we can write the Z boson eigenvector as the
normalized to one SM eigenvector,
plus a perturbation due to the presence of the extra $U(1)$'s,
to order $\eta$:
\be {\cal Z}={1\over N}{\cal Z}_0-\eta{1\over N}
({N'\over N}{\cal Z}_0-{\cal Z}')+{\cal O}(\eta^2).\ee
This corresponds to the eigenstate:
\be {A}_Z= A_{Z_0}-\eta{1\over N}
\Bigl[N'A_{Z_0}-(z_1'A_a+z_2'A_b+z_3'A_c+z_4'A_d)\Bigr]
+{\cal O}(\eta^2).\ee
It is possible to   express
the correction in terms of
the new basis $A_1, A_2, A_3, A_4$ instead of the $A_a,A_b,A_c,A_d$ basis.
In fact, since we are already at ${\cal O}(\eta)$ we can
express the old basis in terms of the new basis to zero order.
We finally find:
\begin{eqnarray}
 {A}_Z&=& A_{Z_0}(1-\eta \frac{N'}{N}) +\eta{1\over N}
\Bigl(
z_1'({\cal F})^{-1}_{a i}A_i+
z_2'({\cal F})^{-1}_{b i}A_i+
z_3'({\cal F})^{-1}_{c i}A_i+
z_4'({\cal F})^{-1}_{d i}A_i\Bigr)\nonumber\\
&+&{\cal O}(\eta^2)
\end{eqnarray}
with sum over i understood.

\section*{Appendix II:
Explicit formulae of $\xi_{31}$ of Class A $\&$ B of D6-brane models.}\label{form}

\setcounter{equation}{0}
For Class A models, the expression of $\xi_{31}$ of eq.(\ref{xi31}) is
$$\xi_{31}=
\left(16 \beta_1^4
\beta_2^4 \ge^4 g_b^4 a_3^4 n_{c1}^4\right)^{-1}
\left[2 a_0^2 a_1^2 + 3 a_0 a_1 a_2 a_3 
-\beta_1^2 \beta_2^4 g_b^2 a_0^2 a_3 a_4
+a_3^2 (\beta_1^2 \beta_2^4 g_b^2 a_0 a_5+a_2^2 ) \right]
$$ 
where  we used the following notation:
\begin{eqnarray}
a_0&=& g_b^2 g_c^2 g_d^2 +9 g_a^2 (g_c^2 g_d^2 + g_b^2
(g_c^2+g_d^2)),\nonumber\\
\nonumber\\
a_1&=&
 {{\beta_2}^4}\,\left[ (9 g_a^2+ g_d^2 )
\left( g_c^2 n_{c1}^2 +9 g_b^2 n_{b1}^2 \nu^2\right)
+9 g_a^2 g_d^2 n_{c1}^2\right]   \nonumber\\
&&+  {g_b^2} \beta_1^2 \left[
 \beta_2^2 (9 g_a^2+g_d^2)(4 \beta_2^2\ge^2 +n_{a2}^2 \nu^2)
 -   4 {{\beta_1}} {\beta_2} {g_d^2}\,
       n_{a2} {n_{c1}} + 4 {{\beta_1}^2}\left( {g_c^2} + 
         {g_d^2} \right) {n_{c1}^2} \right] {{\nu}^2},\nonumber\\
\nonumber\\
a_2&=&{{\beta_2}^4}\left[- 4\,{{\beta_1}^2}\,{{{\ge}}^2}
       {g_b^2} \left({g_b^2} + {g_c^2} \right) 
       \left( 9 {g_a^2} + {g_d^2} \right) - \left( 2 {g_b^2} {g_c^2}
    {g_d^2} + 9 {g_a^2} \left( {g_c^2} {g_d^2} +  2 {g_b^2}
     \left( {g_c^2} + {g_d^2} \right) \right)  \right)
\,{{n_{c1}}^2} \right]  \nonumber\\
&&\!\!\!\!\!\!\!\!\!\!\!\!\!
+   \, 6\,{{\beta_2}^4}\,{g_b^2}\,{g_c^2}\,\left( 9\,{g_a^2} + 
       {g_d^2} \right) \, n_{b1}\,n_{c1}\,\nu\, \delta
 -  {g_b^2}\,\left[ {{\beta_2}^2}
       \left( {g_b^2} + {g_c^2} \right) \,
   \left( 9\,{g_a^2} + {g_d^2} \right) \,\left( {{\beta_1}^2}\,{{{n_{a2}}}^2} + 
         9\,{{\beta_2}^2}\,{{n_{b1}}^2} \right) \right.\nonumber\\
&&-  \left. 4\,{{\beta_1}^3} \beta_2\,\left( {g_b^2} + {g_c^2} \right) \,
       {g_d^2}\,{n_{a2}} n_{c1} + 4\,{{\beta_1}^4}\,\left( {g_c^2}\,
     {g_d^2} + {g_b^2}\,\left( {g_c^2} + {g_d^2} \right)  \right) 
        {{n_{c1}}^2} \right] \,{{\nu}^2},\nonumber\\
\nonumber\\
a_3&=&g_c^2 g_d^2 +9 g_a^2(g_c^2+g_d^2),\nonumber\\
\nonumber\\
a_4&=&n_{c1}^2\left[
4 \beta_2^4 \ge^2 (9 g_a^2+g_d^2)+ \beta_2^2n_{a2}^2 \nu^2 (9
g_a^2+g_d^2)
-4 \beta_1 \beta_2 g_d^2 n_{a2} n_{c1} \nu^2\right.\nonumber\\
&&\left.+4 \beta_1^2 \nu^2 \left(4 \beta_1^2\ge^2 g_b^2+(g_c^2+g_d^2)n_{c1}^2
+9 g_b^2 n_{b1}^2 \nu^2\right)\right],\nonumber\\
\nonumber\\
a_5&=& {{n_{c1}}^2}\,\left\{ 4\,{{\beta_2}^4}\,
      {{{\ge}}^2}\,\left( 2\,{g_b^2} + 
    {g_c^2} \right) \,\left( 9\,{g_a^2} + {g_d^2} \right)  + 
     {{\beta_2}^2}\,\left( 2\,{g_b^2} + {g_c^2} \right) \,
      \left( 9\,{g_a^2} + {g_d^2} \right) \,
      {{{n_{a2}}}^2}\,{{\nu}^2}\right. \nonumber\\
&& -   4\,\beta_1\,\beta_2\,\left( 2\,{g_b^2} + {g_c^2} \right) \,
      {g_d^2}\,{n_{a2}}\,n_{c1}\,{{\nu}^2} + 4\,{{\beta_1}^2}\,{{\nu}^2}\,
      \left[ 4\,{{\beta_1}^2}\,{{{\ge}}^2}\,{g_b^2}\,\left( {g_b^2} + 
           {g_c^2} \right)      \right.\nonumber\\
&& + \left.\left.   \left( {g_c^2}\,{g_d^2} + 
           2\,{g_b^2} \left( {g_c^2} + {g_d^2} \right) \right) {{n_{c1}}^2} - 
        6\,{g_b^2}\,{g_c^2}\,n_{b1} n_{c1}\nu \,\delta+ 9\,{g_b^2}\,
         \left( {g_b^2} + {g_c^2} \right) \,
         {{n_{b1}}^2} {{{\nu}}^2} \right]  \right\}\nonumber
\end{eqnarray}

\noindent
For Class B we have for $\xi_{31}$ of eq.(\ref{newxi31}):
$$
\xi_{31}=g_c^2 \left[  b_4 +b_7-b_5+b_6 \cos(4
\theta)\right]\left( 32 \beta_1^4 \beta_2^4 \epsilon^4 g_b^2 
n_{c1}^4 b_1^4\right)^{-1}
$$
where we used the following notation:
\begin{eqnarray}
b_7&=&2 \beta_1^2 \beta_2^4 g_b^2 g_c^2 b_2^2 \left[ - 8 \beta_1^2 \epsilon^2
(9g_a^2 +g_d^2) b_0 +b_1^2 n_{c1}^2 \right] \nu^2\nonumber\\
\nonumber\\
b_6&=&\beta_2^4 b_1^2 n_{c1}^2 \left[\beta_2^4 \left( 8 \beta_1^2
\epsilon^2 g_b^2 g_c^2 (9 g_a^2 +g_d^2)^2-9 g_a^2 g_d^2 b_1
n_{c1}^2\right)+2 \beta_1^2 g_b^2 g_c^2 b_1 \nu^2\right]\nonumber\\
\nonumber\\
b_5&=& 2 \beta_1^4 g_c^2 g_b^2 b_2 \nu^4 
\left[\beta_2^2 (9 g_a^2 +g_d^2) b_0 n_{a2}^2-4 \beta_1 \beta_2 g_d^2
b_0 n_{a2} n_{c1}+ 4 \beta_1^2 n_{c1}^2 b_3\right]\nonumber\\
\nonumber\\
b_4&=& \beta_2^8 \left [8 \beta_1^2 \epsilon^2 g_b^2 g_c^2 (9 g_a^2+g_d^2)^2 (-4
\beta_1^2 \epsilon^2 (9 g_a^2+g_d^2) b_0+b_1^2 n_{c1}^2)-9 g_a^2 g_d^2
b_1^3 n_{c1}^4\right]\nonumber\\
\nonumber\\
b_3&=&-g_c^4 g_d^4 +g_b^2 g_c^2 g_d^2 (g_c^2 +g_d^2)+9 g_a^2
(g_c^2+g_d^2) [g_c^2 g_d^2 +g_b^2 (g_c^2+g_d^2)]\nonumber\\
\nonumber\\
b_2&=&\left[ \beta_2 (9 g_a^2+g_d^2) n_{a2}-2 \beta_1 g_d^2
n_{c1}\right]^2\nonumber\\
\nonumber\\
b_1&=&g_c^2 g_d^2 +9 g_a^2 (g_c^2+g_d^2)\nonumber\\
\nonumber\\
b_0&=&g_c^2 g_d^2 (g_b^2-g_c^2) +9 g_a^2 [-g_c^4+g_c^2
g_d^2+g_b^2(g_c^2+g_d^2)]\nonumber
\end{eqnarray}

\end{document}